\documentclass[aps,notitlepage,nofootinbib]{revtex4-1}

\usepackage{amstext,amsmath,amssymb,amsfonts,bbm}
\usepackage[latin1]{inputenc}
\usepackage{epsfig}
\usepackage{hyperref}
\usepackage{amsthm}
\usepackage{subfigure}
\usepackage{color}
\usepackage{multirow}

\newcommand{\bea}{\begin{eqnarray}}	
\newcommand{\eea}{\end{eqnarray}}
\newcommand{\be}{\begin{equation}}	
\newcommand{\ee}{\end{equation}}

\newcommand{\I}{\mathbb{I}}
\newcommand{\inter}{\text{int\,}}
\newcommand{\kin}{\text{kin\,}}

\newtheorem{lemma}{Lemma}

\newtheorem{definition}{Definition}
\newtheorem{theorem}{Theorem}

\begin{document}

\title{Ward-Takahashi identities for the colored Boulatov model}

\author{\bf Joseph Ben Geloun}\email{jbengeloun@perimeterinstitute.ca}

\medskip

\affiliation{Perimeter Institute for Theoretical Physics, 31 Caroline
St, Waterloo, ON N2L 2Y5, Canada \\
International Chair in Mathematical Physics and Applications, 
ICMPA-UNESCO Chair, 072BP50, Cotonou, Rep. of Benin}

\begin{abstract}
\noindent 
Ward-Takahashi identities of the colored Boulatov model are derived 
using a generic unitary field transformation. In a specific
instance, this generic  transformation turns out 
to be a symmetry of the interaction so that  
particular classes of reduced Ward-Takahashi identities 
for that symmetry are consequently identified
and interpreted.
\medskip

\noindent Pacs numbers: 04.60.Gw, 05.10.Cc, 11.10.Gh\\
Key words: Group field theory, Ward-Takahashi identities\\
\end{abstract}

\hspace{16cm} pi-qg-213

\hspace{13.9cm}  ICMPA/MPA/04/2011

\maketitle

\section{Introduction}

Group field theories (GFTs) are generalization of matrix models 
as higher rank tensor quantum field theories
over a  group manifold \cite{GFT,GFT2}. They provide 
a relevant framework for the quantization of gravity 
\cite{laurentgft, oriti, quantugeom2, oriti2, Gielen:2011dg}
as well as possess, as recently shown for the particular class of colored  GFT models \cite{color, PolyColor, sing},
an equivalent formulation of a large $1/N$ limit expansion \cite{Gurau:2010ba, Gurau:2011aq, Gurau:2011xq}
and exhibit a critical behavior
\cite{Bonzom:2011zz, Gurau:2011tj}. 

It should be emphasized that the presence of colored fields
in GFT plays an increasing role in the search of symmetry of
these theories.
Indeed, some investigations 
pertaining to the symmetry aspects of such colored GFTs have been led recently
\cite{Girelli:2010ct, gftnoncom, Baratin:2011tg, Carrozza:2011jn, Geloun:2011cz} and
one proved that, to mention a few, the colored theories  are endowed
with a genuine quantum 
group symmetry \cite{Girelli:2010ct} 
which encodes  a notion of diffeomorphism
symmetry in GFTs \cite{Baratin:2011tg}   and 
by direct computations in group space, they prove 
to possess a conserved quantity associated with
translation symmetry \cite{Geloun:2011cz}. 
All these properties
have been only proved for the colored case and there is,
at the moment, no hint of how they could be
implemented in the non colored situation.

Let us remind that the partition functions of GFT models in the sense of Boulatov-Ooguri \cite{GFT,GFT2} are defined through a
nonlocal interaction and a Gaussian measure with a degenerate covariance.
This covariance is indeed made of a group averaging and product of 
delta functions which in fact projects onto the gauge invariant
sector of the space of square integrable functions $L^2(G^D,\mathbb{C})$, for a $D$ dimensional GFT. Hence the kinetic term in the Lagrange formulation of the action, can be seen as trivial (this term is of the mass kind when restricting
field on the gauge invariant sector)
or even inexistent from the pure quantum field theory point of view. 
Besides, it is well-known that the ordinary notion of symmetry in a presence of nonlocal interaction becomes more involved or even
could break down.
These peculiar features, in returns, render unclear the 
ordinary definition of classical symmetry and the corresponding notion of Noether theorem (the notion of Noether currents 
 in a trivial dynamical theory
may only reduce to the Lagrangian density itself for translations for instance).
At the current stage of investigations on symmetries of
GFTs, there are three ways to address these issues: either to deal with a
quantum group symmetry and making use of Hopf algebra techniques \cite{Girelli:2010ct}, or
to introduce a nontrivial kinetic term \cite{Geloun:2011cz} (motivated indeed by renormalization requirements in \cite{Geloun:2011cy} \cite{quantugeom2}), or, finally,
to state directly the Noether theorem for a given symmetry at the quantum level.
The latter is well-known to be related with 
the identification of Ward-Takahashi (WT) identities  associated with a particular
field symmetry. To shed more light on  the last aspect 
is the purpose of the present work.

A review some basic but important facts
of WT identities in general quantum field theory 
and precisions on what can be investigated in our situation is in order. 
WT identities are consequences
of a symmetry of a system at the quantum level (for instance in
path integral formalism) in the form
of relations between correlation functions. These identities
should hold even after renormalization. 
As an illustration, in quantum electrodynamics with its $U(1)$ gauge symmetry, the WT identities relate
the full three-point function with the two-point function hence the wave function renormalization with the vertex renormalization.
It is nevertheless true that, in an all-embracing view, 
a given field transformation without requiring any action symmetry
induces relations between correlation functions that can be 
legitimately called WT identities. 
A known example of these occurs in the context of noncommutative quantum field theory (also a nonlocal theory) 
mapped onto non identically distributed matrix models \cite{Grosse:2004yu}. The dynamics explicitly breaks the ordinary $U(N)$
symmetry of matrix models and, at the quantum level,
produces WT identities. These latter play a crucial role in the proof of asymptotic safety at all orders
of perturbation theory \cite{Disertori:2006nq, Geloun:2008zk}.
Hence, having or not a symmetry of an action, some relations 
between correlation functions can be unraveled
provided one has a well defined field transformation. 
These relations yield a weaker sense of equations
of motion which in return can be related to modified 
(quantum) conservation law
for currents associated with that transformation. 
WT identities therefore encode the obstructions 
of a theory to be invariant at the quantum level,
obstructions which are useful in the renormalization program
and also from the point of view of generalizing the
sense of symmetry of a given system. 
Let us come back to our situation dealing 
with a strict quantum tensor field theory 
without classical dynamics and possessing a nonlocal interaction.
It becomes obvious that for reasons invoked
so far (the general concept of symmetry and its influence 
on the renormalization program) investigating WT identities 
is certainly significant in the GFT understanding. 
More specifically, one may 
ask: can we make a (whatsoever general) sense of symmetries with conserved currents by stating them directly at the quantum level using WT identities ? 
This work will give a partial but positive answer to that question
provided one uses colored fields. 

In this paper, we define a generic unitary field
transformation from which are studied WT identities
of the colored Boulatov model \cite{GFT}. 
This general unitary field transformation turns out to be 
a symmetry for the Boulatov action provided a specific
way that one chooses to act on the field arguments.
Associated with that symmetry,  reduced WT equations 
satisfied by the correlations functions are determined
and analyzed.   
We find that these reduced WT identities can be interpreted 
as quantum equation of motions for an underlying classical 
theory with a trivial kinetic term. Hence, bearing in mind 
that the field equation of motion is closely related to 
the conservation law of any kind of symmetry, it is not hopeless
that WT relations obtained in this nonlocal theory could describe the same
idea of modified conservation law stated at the quantum level
once one makes a sense of classical conserved charges
(this question will be not treated here).
Moreover, the various kind of identities carried out
should be useful for in-depth perturbative and nonperturbative renormalization programs of GFTs \cite{Rivasseau:2011xg, Carrozza:2011jn, FreiGurOriti, sefu1, sefu2, sefu3, Bonzom:2010ar, Geloun:2010vj, Geloun:2011cy}.
Finally, the results achieved 
can be extended in any GFT dimension.

The paper is organized as follows: Section 2 reviews 
the basics of the Boulatov model and discusses the
main strategy to obtain WT identities from a field symmetry 
and the \emph{universal} feature that they will assume for
any tensor models. 
Section 3 introduces
the unitary field transformation that will define the variations of the different parts entering
in the definition of the partition function. Mainly, 
one can define a general unitary transformation 
and one more specific unitary with the property 
that it preserves the gauge invariance of fields and
is a symmetry of the interaction term. For the
general and the more specific symmetry 
the study of WT identities are discussed in Section 4. 
A summary and outlook of the results are given in Section 5
and, finally, a detailed appendix provides basic facts
on the theory, on unitary field transformations and
other precisions on derivations used in the text.

\section{The Colored Boulatov Model}\label{sec:model}

Let $G$ be some compact multiplicative Lie group, denote
$e$ its unit, and $\int dh=:\int_h$ the integral with respect to the Haar measure.
Consider $\bar \varphi^i,\varphi^i$,  $i=0,1,2,3$ some complex 
scalar fields over three copies of $G$, $\varphi^i:G^3 
\rightarrow \mathbb{C}$. We denote $\delta^N(h)$ the 
delta function over $G$ that we cutoff such that 
$\delta^N(e)$ is finite, but diverges as $N$ goes to infinity. 
In the specific situation of 
$G=SU(2)$ which will be focus of this paper 
(respectively $G=U(1)$), one has 
\bea
  \delta^N(h)\Big{|}_{G=SU(2)} = \sum_{j=0}^{N} d_j\, \chi^{j}(h) \qquad
\Big(\text{resp.}\qquad 
\delta^N(\varphi)\Big{|}_{G=U(1)} = \sum_{p=-N}^{N} e^{\imath p \varphi} \Big)\;,
\eea
 where $\chi^j(h)$ stands for 
the character of group element $h$ in the representation $j$  
and $d_j =2j+1$.

The following path integral defines the partition function of the colored Boulatov model \cite{color} over $G$:
\bea\label{eq:part}
  Z(\lambda,\bar\lambda)= e^{-F(\lambda,\bar\lambda)} = 
\int \prod_{i=0}^4 d\mu_C(\bar \varphi,\varphi) \; e^{-S^{\inter}(\bar \varphi^i,\varphi^i)} \; ,
\eea
with normalized Gaussian measure 
$d\mu_C(\bar \varphi,\varphi) $ of covariance $C$ (denoting
henceforth $\varphi_{hpq}:=\varphi(h,p,q)$):
\bea
C^{kj}_{h_{0}h_{1}h_{2} ;\; h_{0}'h_{1}'h_{2}'} = 
\int d\mu_C(\bar \varphi,\varphi) \; \bar\varphi^k_{h_{0}h_{1}h_{2}}
  \varphi^j_{h_{0}'h_{1}'h_{2}'} 
 = \delta^{kj}\int_h \;  \delta^{N}\bigl( h_{0} h (h_{0}')^{-1} \bigr)
\delta^{N}\bigl( h_{1} h (h_{1}')^{-1} \bigr)
\delta^{N}\bigl( h_{2} h (h_{2}')^{-1} \bigr) \; ,
\label{gaugeproj}
\eea
and interaction  wisely chosen 
\bea
\label{eq:interaction}
S^{\inter}&=& \frac{\lambda}{\sqrt{\delta^N(e)}} 
\int_{h_{ij}} \varphi^0_{h_{03}h_{02}h_{01}} 
 \bar \varphi^1_{h_{10}h_{13}h_{12}} \varphi^2_{h_{21}h_{20}h_{23}} 
 \bar \varphi^3_{h_{32}h_{31}h_{30}} \crcr
&&+\frac{\bar\lambda}{\sqrt{\delta^N(e)}} 
\int_{h_{ij}} \bar \varphi^0_{h^{03}h^{02}h^{01}} 
 \varphi^1_{h^{10}h^{13}h^{12}} \bar \varphi^2_{h^{21}h^{20}h^{23}} 
  \varphi^3_{h^{32}h^{31}h^{30}} \; ,
\eea
where one should identify $h_{ij}=h_{ji}$ and the symbol $\int_{h_{ij}}$
means that the integral is performed over all variables labelled by $h_{ij}$
(here above six copies of $G$). The resulting pairing of field arguments
of this nonlocal interaction can be dually and graphically 
associated with a tetrahedron (each argument for each edge) 
and Feynman graphs, in this theory, are noting but a collection
of tetrahedra (simplicial complexes) glued along one of their faces (triangle) following the
covariance rule. More precisions on the GFT diagrammatics
can be found in \cite{oriti}. For a colored theory, there
is an additional gluing constraint enforcing that only 
colors of the same index can be glued together (hence the presence of $\delta^{jk}$ in the covariance).
One also notes that, in the ordinary 
colored GFT \cite{color}, the interaction with coupling constant $\lambda$ is of the form
$\varphi^0\varphi^1\varphi^2\varphi^3$ and the one with 
coupling $\bar\lambda$ can be chosen as its complex conjugate. Here, we have just renamed $\varphi^{1,3}$ as $\bar\varphi^{1,3}$ (and vice-versa) in order to have a correct notion of field transformation with respect to  colors. 
Hence,  the formalism that we will develop holds without
loss of generality in the ordinary colored GFT, with another field transformation.

As an operator, $C$ can act also onto functions. We 
will use indifferently the compact notations, when no confusion may
occur
\bea
&&
[C \varphi]^i_{[g]} = C^{ij}_{[g][\tilde g]} \, \varphi^j_{[\tilde g]} 
=\int_{[\tilde g]} C^{ij}_{[g][\tilde g]} \, \varphi^j_{[\tilde g]} 
:= \int_{  \tilde g_\alpha} C^{ij}_{g_1g_2g_3;\; \tilde g_1\tilde g_2\tilde g_3}
\varphi^j_{\tilde g_1\tilde g_2\tilde g_3}  \;,\crcr
&&
\bar\varphi C \varphi = \bar\varphi^j_{[g]} [C \varphi]^j_{[g]}
= \int_{[g]} \bar\varphi^j_{[g]} [C \varphi]^j_{[g]}
:=\int_{ g_\alpha \tilde g_\alpha} \bar\varphi^j_{ g_1 g_2 g_3} 
C^{ji}_{g_1g_2g_3;\; \tilde g_1\tilde g_2\tilde g_3} \varphi^i_{\tilde g_1\tilde g_2\tilde g_3} \;.
\eea

We sketch here the main steps of our program that
will allow us to extract the WT identities. 
Foremost, we should stress that albeit the entire study 
is realized in three dimensions, 
the same analysis can be easily carried out for arbitrary dimension
with not so much effort: for instance, any symbol $[g]$ 
will be just adjusted to be $D$ copies of group elements
(note that most of the appendices discusses the $D$
dimensional situation).
The only issue there will be to trace in the interaction the correct
group element with respect to which the transformation
is realized and then to adapt again our notations for that
context. 
Furthermore, in the next paragraph, we discuss the method 
for getting WT identities and their main features   
which should be valid in full generality, 
for any kind of field transformation $\varphi \to \tilde{\varphi}$
(save the identity for which everything obviously trivializes), 
for any tensor model and in arbitrary dimension.
We will emphasize the specificities of 
GFT models at each major step. 

As initial data, consider a partition function $Z$
defined with a Gaussian measure $d\mu_C(\varphi)$ 
of covariance $C$ (which could be the identity, any invertible operator or
even projector as it occurs in  GFTs), an interaction $S^{\inter}$ 
and also source terms $J\varphi$. 
Second, consider a field transformation yielding at the
infinitesimal level a nontrivial variation 
$\delta_\epsilon \varphi \neq 0$,
with $\epsilon$ the infinitesimal parameter associated with
that transformation. 
Collect the variations of the above quoted 
constituents of the partition function, 
namely $\delta_\epsilon S^{\inter}$, $\delta (J\varphi) 
= J \delta_\epsilon \varphi$ (assumed to be never vanishing)
and last, more interestingly, 
$\delta_\epsilon d \mu_C(\varphi)$.
Then, many cases can occur: $\delta_\epsilon S^{\inter} =0$
or not, and $\delta_\epsilon d \mu_C(\varphi) =0$ 
or not. Whenever $\delta_\epsilon S^{\inter} =0$
then the transformation is a symmetry of the
interaction part. Meanwhile $\delta_\epsilon d \mu_C(\varphi) =0$
means that the transformation does not affect
the kinetic term. 
As an example, considering any 
independent identically distributed (i.i.d.) tensor model with 
$C$ is the identity operator, 
an unitary symmetry will exactly produce a result of this kind
whereas all GFT models endowed with a gauge projector as
(\ref{gaugeproj}), the analogous part will have a nontrivial contribution.
The next stage is to differentiate of the free
energy $\ln Z$ (generating function of connected components) with respect to the infinitesimal parameter $\epsilon$
and to equate obviously the result to zero.
If $\delta_\epsilon S^{\inter}$ or $\delta_\epsilon  d \mu_C(\varphi)$
are non zero, then one expects to get {\it a priori} 
relevant contributions. Pursuing the analysis, one
differentiates the equation 
with respect to source fields $J$ a given number of time  and then puts these sources at zero. The result is a generalized WT identity
which can be decomposed in two parts: a contribution
coming from  $J \delta_\epsilon \varphi$
and other contributions issued from $\delta_\epsilon S^{\inter}$ 
and $\delta_\epsilon  d \mu_C(\varphi)$.
In an ordinary local theory, 
these last mentioned contributions are nothing but 
expectation values of field equations of motion
which can be related to a modified conservation property
of the currents associated with the transformation.
Thus, WT identities can be interpreted as a quantum version
of the local conservation of these currents.  
In our tensor situation, 
the sense of conserved charges is less clear without true dynamics.
But still, as we will show after some manipulations, 
the WT identities find a relation with equations of motion
of an auxiliary GFT theory.
Going back to the general setting, 
a final stage is to analyze and to reduce these WT identities
and to interpret them  graphically. 
We can comment that for i.i.d. models with an invariant interaction under some symmetry, 
WT identities only give relations between $N$-point functions
(of the same species, i.e. with $N$ external legs)
because the only contributions to the differentiation 
will come from the source terms $J \delta_\epsilon \varphi$. In GFT models implemented with an invariant interaction, 
due to the contribution of $\delta_\epsilon  d \mu_C(\varphi)$,
WT equations will definitely mix different kinds of $N$-point functions
henceforth worthy to be scrutinized. 
Then, the bottom line is the definition of 
a field transformation and this is the purpose of the
next section.

\section{Unitary transformations}
\label{sect:Unit}

\subsection{General unitary transformations}
\label{subsect:inftransgener}

From now on, we will restrict to the situation where $G=SU(2)$
though most of the ensuing developments may found an extension 
for an arbitrary Lie group. 

Consider the partition function $Z$ including source terms 
and the associated free energy $W$ for the Boulatov model, namely
\bea
e^{-W(\bar \eta,\eta)}=Z(\bar \eta, \eta) 
= \int d\mu_C (\bar\varphi,\varphi) e^{-S^{\inter}(\varphi,\bar \varphi) + \bar \varphi \eta+ \varphi \bar \eta} \;,
\eea
where notations remain the same as earlier and the source term
 incorporates color indices: 
\bea
\bar \varphi \eta+ \varphi \bar \eta  : = 
\sum_i \int_{g_\alpha} \left(
\bar \varphi^i_{g_1g_2g_3} \eta^i_{g_1g_2g_3} +
\bar \eta^i_{g_1g_2g_3} \varphi^i_{g_1g_2g_3} \right) .
\eea
In order to avoid confusion, we will
use different notations for a single index $a$ and
a triple index $[a]=(a_1,a_2,a_3)$. 
Hence the composition of the covariance and 
a field will be written henceforth 
$C_{[a][b]}\varphi_{[b]}$ and an operator $T$ acting
onto a single group element of fields will be denoted
as $T_{ab} \varphi_b$.  

Let us consider now a general unitary operator $U$ 
satisfying a composition law  $U^{ij}_{[a][b]}  \bar U^{i'j}_{[c][b]} = \delta^{ii'}\delta_{[a][c]}$, with $\delta_{[a][c]}$ the kernel 
of the unit operator, such that the fields transform under $U$ as
\bea
^U\,  \varphi^i_{[a]} = U^{ij}_{[a][b]} \varphi^j_{[b]} \;, \qquad
^U\, \bar \varphi^i_{[a]} = \bar\varphi^j_{[b]} \bar U^{ij}_{[a][b]}\;, \qquad
\frac{\delta}{\delta \,  ^U \varphi^i_{[a]}} = \frac{\delta}{\delta \varphi^j_{[b]}} (U^{-1})^{ji}_{[b][a]}\;, \qquad 
\frac{\delta}{\delta\,  ^U \bar\varphi^i_{[a]}} = U^{ij}_{[a][b]}\frac{\delta}{\delta \bar \varphi^j_{[b]}}
\label{eq:generunita0}
\eea 
(one can prove that such 
 operators exist, see  Appendices \ref{sect:unittransf} and \ref{sect:leftinvunit}).
Let us keep at the moment these as formal expressions. 
  
Under $U$, $S^{\inter}$ transforms as
\bea
^U \, S^{\inter} &=&\frac{\lambda}{\sqrt{\delta^N(e)}} \int_{h_{ij}}\,
^U \varphi^0_{h_{03}h_{02}h_{01}} \,
^U  \bar \varphi^1_{h_{10}h_{13}h_{12}} \,  ^U\,  \varphi^2_{h_{21}h_{20}h_{23}} \,
^U  \bar \varphi^3_{h_{32}h_{31}h_{30}} \crcr
&&+\frac{\bar\lambda}{\sqrt{\delta^N(e)}} 
\int_{h_{ij}} \  ^U \bar \varphi^0_{h^{03}h^{02}h^{01}} \,
 ^U  \varphi^1_{h^{10}h^{13}h^{12}} \,
^U  \bar \varphi^2_{h^{21}h^{20}h^{23}} \, 
^U   \varphi^3_{h^{32}h^{31}h^{30}} \; , \crcr
& =&\frac{\lambda}{\sqrt{\delta^N(e)}} \int_{h_{ij}}\,\int_{[a][b][c][d]}
U^{0i_0}_{h_{03}h_{02}h_{01}\,[a]} \varphi^{i_0}_{[a]} \,
 \bar \varphi^{i_1}_{[b]} \bar U^{1i_1}_{h_{10}h_{13}h_{12}\;[b]} \,  U^{2i_2}_{h_{21}h_{20}h_{23}\,[c]}\,  \varphi^{i_2}_{[c]} \,
\bar \varphi^{i_3}_{[d]} \bar U^{3i_3}_{h_{32}h_{31}h_{30}\,[d]}\crcr
&&+\frac{\bar\lambda}{\sqrt{\delta^N(e)}} 
\int_{h_{ij}}\,\int_{[a][b][c][d]} \ 
\bar \varphi^{i_0}_{[a]}  \bar U^{0i_0}_{h^{03}h^{02}h^{01}[a]} 
  \varphi^{i_1}_{[b]} U^{1i_1}_{h^{10}h^{13}h^{12} [b]}\,
\bar \varphi^{i_2}_{[c]}   \bar U^{2i_2}_{h^{21}h^{20}h^{23}\,[c]}  \, 
 U^{3i_3}_{h^{32}h^{31}h^{30}\,[d]} \varphi^{i_3}_{[d]}  \; .
\eea 
The partition function under (\ref{eq:generunita0})
can be mapped onto (see Eq.(\ref{eq:partiuu}) of Appendix \ref{app:inftransgener})
\bea
 Z(\bar \eta,\eta) = \int d\mu_{UCU^{-1}} (\bar \varphi, \varphi ) \; 
e^{-S^{\inter}(U \varphi ,  \bar\varphi  U^{-1}) +
\bar \varphi  U^{-1}  \eta  +  
\bar \eta U \varphi  
 } \;.
\eea
By Lemma \ref{lem:sum} (see Appendix \ref{app:gauss}), 
we  write this modified partition function as
\bea
 Z(\bar \eta,\eta) = \int d\mu_{C} (\bar \varphi, \varphi ) \;
  e^{\delta_{\varphi} (UCU^{-1} -C) \delta_{\bar \varphi}}\;
 e^{-S^{\inter}( U\varphi, \bar\varphi U^{-1}) +
\bar \varphi  U^{-1}\eta  +  \bar \eta U \varphi 
 } \;.
\label{eq:partiinterm0}
\eea
Close to the identity, we can set $U^{ij}_{[a][b]} = \delta^{ij}\delta_{[a][b]}+ \imath B^{ij}_{[a][b]}$, where $B$ is a Hermitian kernel in the 
sense that  $\bar B^{ji}_{[b][a]} = B^{ij}_{[a][b]}$. At first order in $B$, the variation of the interaction part is of the form 
(see Appendix \ref{app:inftransgener}  for precisions
concerning the notations and derivations of the following
infinitesimal variations)
\bea
&&
\delta_B S^{\inter} := \imath \lambda\Big[ [B\varphi]^0 \bar\varphi^1 \varphi^2 \bar\varphi^3 
- \varphi^0 [\bar\varphi B]^1 \varphi^2 \bar\varphi^3  
+\varphi^0 \bar\varphi^1 [B\varphi]^2 \bar\varphi^3 
- \varphi^0 \bar\varphi^1 \varphi^2 [\bar\varphi  B ]^3    \Big] 
 \;+\; \imath \bar\lambda \{\bar\varphi\}\;, \crcr
&&
 \bar\lambda \{\bar\varphi\}=  
- [\bar\varphi^0 B] \varphi^1 \bar\varphi^2 \varphi^3 
+ \bar\varphi^0 [B\varphi]^1 \bar\varphi^2 \varphi^3  
- \bar\varphi^0\varphi^1 [\bar\varphi B]^2 \varphi^3 
+ \bar\varphi^0 \varphi^1 \bar\varphi^2 [B\varphi]^3 \;.
\label{eq:generinterinf0} 
\eea
Meanwhile, the source terms have the infinitesimal
transformations
\bea
\delta_B (\bar\eta\varphi + \bar\varphi \eta)
=  \imath \sum_{i} 
\int_{[g] [a]} 
 \left(   -\bar \varphi^i_{[a]} B^{ij}_{[a][g]} \eta^j_{[g]} +
 \bar \eta^i_{[g]} B^{ij}_{[g][a]} \varphi^j_{[a]} \right) 
= :  \imath (-\bar \varphi B \eta  + \bar \eta B \varphi )
\;,
\label{eq:genersourceinf0}
\eea
and varying the covariance, we get
\bea
[U C U^{-1}]^{ij}_{[a][b]}  - C^{ij}_{[a][b]}
= \imath  [  B^{li}_{[c][a]} C^{lj}_{[c][b]}-C^{il'}_{[a][c']} B^{jl'}_{[b][c']}]
=: \imath  [ B C-CB ]^{ij}_{[a][b]}\;.
\label{eq:genercov0}
\eea

\subsection{ Right invariant unitary transformations}
\label{subsect:infvarright}

Among unitary operators, there exists a particular
class that we propose also to study. This class includes
unitary operators called 
\emph{right invariant unitaries} for having the particular
property to preserve the gauge invariance of
fields\footnote{ GFTs can be indeed defined with gauge invariant fields.
Hence, the mentioned transformation will preserve this property of fields.} and let $S^{\inter}$ invariant.
A detailed discussion about these right invariant unitaries 
can be found in Appendix \ref{sect:lrinvunit} .

Working with a right invariant unitary $U$ in the sense of the 1-action of Appendix \ref{sect:leftinvunit} Eq. (\ref{eq:notlefinv}), 
we get a change of field variables such that 
\bea
&&
^U \varphi^{0}_{a.} = U_{ab}  \varphi^0_{b.}\, \; , \qquad ^U \bar \varphi^0_{a.} = \,\bar  \varphi^0_{b.} (U^{-1})_{ba}
=\bar{ U}_{ab}\bar \varphi^0_{b.}\, \; , \qquad
\frac{\delta}{\delta \,  ^U \varphi^0_{a.}} = \frac{\delta}{\delta \varphi^0_{b.}} (U^{-1})_{ba}\;, \qquad 
\frac{\delta}{\delta\,  ^U \bar\varphi^0_{a.}} = U_{ab}\frac{\delta}{\delta \bar \varphi^0_{b.}}\ \;, \crcr
&&
^U \varphi^{1}_{.a} = U_{ab}  \varphi^1_{.b}\, \; , \qquad ^U \bar \varphi^1_{.a} = \,\bar  \varphi^1_{.b} (U^{-1})_{ba}
=\bar{ U}_{ab}\bar \varphi^1_{.b}\, \; , \qquad
\frac{\delta}{\delta  \,  ^U\varphi^1_{.a}} = \frac{\delta}{\delta \varphi^1_{.b}} (U^{-1})_{ba}\;, \qquad 
\frac{\delta}{\delta\,  ^U \bar\varphi^1_{.a}} =U_{ab}\frac{\delta}{\delta \bar \varphi^1_{.b}}\ \;,
\label{eq:unitarinv}
\eea
and colors $2$ and $3$ transform like $0$ and $1$, respectively.
The subscript $a$ or $b$ refers to 
a unique group element  and the dot
notifies the position of the remaining arguments.
Thus $0$ and $2$ are transformed with respect to their
first argument whereas $1$ and $3$ to their 
last argument.

$S^{\inter}$ remains invariant under this transformation
(see Appendix \ref{app:inftransright} for details of
the identities in the remaining of this section)
\bea
^U \, S^{\inter}  = S^{\inter}\;.
\eea
In fact, for the colored Boulatov model there are six such 
right invariant unitaries, namely 
one for each couple of arguments in the interaction, 
leaving the colored GFT interaction invariant. More generally,
a $D$ dimensional colored GFT will be invariant under $D(D+1)/2$ of such basic transformations that one can think as 
\emph{minimal} symmetries. In the following, we will use one of 
these minimal symmetry in order to simplify some general WT identities
and, consequently, to prove that the WT identities derived in this
work have a nontrivial content.

Under (\ref{eq:unitarinv}), the partition function takes the form
\bea
 Z(\bar \eta,\eta) = \int d\mu_{UCU^{-1}} (\bar \varphi, \varphi ) e^{-S^{\inter}(U \varphi ,  \bar\varphi  U^{-1}) +
\bar \varphi  U^{-1}  \eta  +  
\bar \eta U \varphi  
 } \;,
\eea
where the action of $U$ on the covariance is defined by 
\bea
\bar U_{h_{0}'c'}  C^{ii=0,2}_{h_{0}h_{1}h_{2} ;\; h_{0}'h_{1}'h_{2}'}
U_{h_{0}c}
& =& \int d\mu_{C} (\,^U\bar \varphi,\, ^U\varphi )\int_{ab h_{0} h'_{0}} U_{h_{0}c}  \bar U_{h_{0}a}  \bar U_{h_{0}'c'} U_{h_{0}'b}
 \,\bar\varphi^{i=0,2}_{ah_{1}h_{2}}\,\varphi^{i=0,2}_{bh_{1}'h_{2}'} 
:= [U C U^{-1}]^{ii=0,2}_{ch_{1}h_{2} ;\; c'h_{1}'h_{2}'}  \;, \cr\cr
\bar U_{h_{2}' c'}  C^{ii=1,3}_{h_{0}h_{1}h_{2} ;\; h_{0}'h_{1}'h_{2}'} 
U_{h_{2}c}
& =&\int d\mu_{C} (\,^U\bar \varphi,\, ^U\varphi )\int_{ab h_{2} h_{2}'} U_{h_{2}c} \bar U_{h_{2}a} 
\bar U_{h_{2}' c'} U_{h_{2}'b}
\bar\varphi^{i=1,3}_{h_{0}h_{1}a} 
\varphi^{i=1,3}_{h_{0}'h_{1}'b}  := [U C U^{-1}]^{ii=1,3}_{h_{0}h_{1}c ;\; h_{0}'h_{1}'c'} \;.
\eea
Turning the discussion to infinitesimal transformations,
we have $\delta_B S^{\inter} =0$, 
\bea
[U C U^{-1}]^{ii=0,2}_{[a][b]}  - C^{ii=0,2}_{[a][b]}
&=& \imath  [-C^{ii=0,2}_{[a];\;c'b_1b_2} B_{b_0c'} +  B_{ca_0} C^{ii=0,2}_{ca_1a_2;\; [b]}]
=: \imath  [ B C-CB ]^{ii=0,2}_{[a][b]}  \;, 
\cr\cr
[U C U^{-1}]^{ii=1,3}_{[a][b]}  - C^{ii=1,3}_{[a][b]}
&=& \imath  [-C^{ii=1,3}_{[a];\;b_0b_1c'} B_{b_2c'} +  B_{ca_2} C^{ii=1,3}_{a_0a_1c;\; [b]}]
=: \imath  [ B C-CB ]^{ii=1,3}_{[a][b]} \;,
\label{eq:infcovrinv}
\eea
whereas varying source terms  yields
\bea
&&
\delta_B (\bar\eta\varphi + \bar\varphi \eta)
= \sum_i \int_{[g]} \; 
  \left(  ^ U \bar \varphi^i_{[g]} \eta^i_{[g]} +
 \bar \eta^i_{[g]} \  ^U \varphi^i_{[g]} \right)  -  (\bar\eta\varphi + \bar\varphi \eta) \crcr
&& = 
 \imath\int_{g_i a} \Big{\{}
  \sum_{i=0,2} \left(   -\bar \varphi^i_{a.} B_{ag_{0}} \eta^i_{g_{0}.} +
 \bar \eta^i_{g_{0}.} B_{g_{0}a} \varphi^i_{a.} \right) 
+ \sum_{i=1,3} \left(   -\bar \varphi^i_{.a} B_{ag_{2}} \eta^i_{.g_{2}} +
 \bar \eta^i_{.g_{2}} B_{g_{2}a} \varphi^i_{.a} \right) 
\Big{\}}\crcr
&&
= :  \imath (-\bar \varphi B \eta  + \bar \eta B \varphi )
\;.
\label{eq:infsourrinv}
\eea
Having collected all infinitesimal terms, we are in position 
to study the WT identities of the model
and that will be the focus of the rest of this work. 

\section{Ward-Takahashi Identities for the Colored Boulatov Model}
\label{sect:Ward}

\subsection{ General unitary transformation}
\label{subsect:gener}

Our starting point is the partition function (\ref{eq:partiinterm0}).
Considering the infinitesimal transformations
(\ref{eq:generinterinf0}), (\ref{eq:genersourceinf0})
and (\ref{eq:genercov0}), $Z$ may be written
\bea
&&
 Z(\bar \eta,\eta) = \int d\mu_{C} (\bar \varphi, \varphi ) 
  \bigl( 1+ \imath \delta_{\varphi} [ BC-CB] \delta_{\bar \varphi} \bigr) 
e^{-S^{\inter}( \varphi, \bar\varphi ) -\delta_B S^{\inter} +\bar \varphi \eta  +  \bar \eta \varphi + \delta_B (\bar \varphi \eta  +  \bar \eta \varphi )
 } \crcr
&&  = \int d\mu_{C} (\bar \varphi, \varphi ) 
  \bigl( 1+ \imath \delta_{\varphi} [ BC-CB] \delta_{\bar \varphi} \bigr) e^{-S^{\inter}( \varphi, \bar\varphi )  +\bar \varphi \eta  +  \bar \eta \varphi 
 }\times
\crcr
&& 
\Bigl\{1- \imath \bar \varphi  B \eta  +  \bar \eta  B \varphi 
- \imath \lambda\Big[ [B\varphi]^0 \bar\varphi^1 \varphi^2 \bar\varphi^3 
- \varphi^0 [\bar\varphi B]^1 \varphi^2 \bar\varphi^3  
+\varphi^0 \bar\varphi^1 [B\varphi]^2 \bar\varphi^3 
- \varphi^0 \bar\varphi^1 \varphi^2 [\bar\varphi  B ]^3    \Big] 
 \;-\; \imath \bar\lambda \{\bar\varphi\}\Bigl\}\;.
\eea
Being interesting only on connected functions, 
we now derivate the free energy with respect to the infinitesimal parameter $B^{ij}_{[\mu][\nu]}$:
\bea
&&  
 \frac{\delta \ln Z(\eta,\bar \eta)}{\imath\delta B^{ij}_{[\mu][\nu]}}=0  = \frac{1}{ Z(\eta,\bar \eta)}\int d\mu_{C} (\bar \varphi, \varphi ) 
\Bigl\{ \delta_{\varphi^j_{[\nu]}} C^{il}_{[\mu][\alpha]}\delta_{\bar \varphi^l_{[\alpha]}}   
-\delta_{\varphi^l_{[\alpha]}} C^{lj}_{[\alpha][\nu]}
\delta_{\bar \varphi^i_{[\mu]}} - \bar \varphi^i_{[\mu]} \eta^j_{[\nu]}  +  \bar \eta^i_{[\mu]}   \varphi^j_{[\nu]} 
\label{eq:freeexpan0}\\
&&
- \lambda\Big[ \delta^{i0}\varphi^{j}_{[\nu]} [\bar\varphi^1\varphi^2\bar\varphi^3]_{[\mu]}
-\delta^{j1} \bar\varphi^i_{[\mu]} 
[\varphi^0 \varphi^2 \bar\varphi^3]_{[\nu]}  
+ \delta^{i2}\varphi^{j}_{[\nu]}
[\varphi^0\bar\varphi^1\bar\varphi^3]_{[\mu]}
-\delta^{j3}\bar\varphi^{i}_{[\mu]} [\varphi^0\bar\varphi^1\varphi^2]_{[\nu]}    \Big] \crcr
&&
 \;-\;  \bar\lambda \{\bar\varphi\}\Bigl\}
e^{-S^{\inter}( \varphi, \bar\varphi )  +\bar \varphi \eta  +  \bar \eta \varphi 
 }\;,
\nonumber
\eea
where the new notations mean
\bea
&&
[\bar\varphi^1\varphi^2\bar\varphi^3]_{[\mu]} 
:= \int_{h_{ij}} \; \bar\varphi^1_{\mu_{10}h_{13}h_{12}}
\varphi^2_{h_{21} \mu_{20} h_{23} } 
\bar\varphi^3_{h_{32}h_{31}\mu_{30}}\;,
\crcr
&&
[\varphi^0 \varphi^2 \bar\varphi^3]_{[\nu]}   
:= \int_{h_{ij}} \; \bar\varphi^1_{ h_{03}h_{02}\nu_{01}}
\varphi^2_{\nu_{21} h_{20} h_{23} } \bar\varphi^3_{h_{32}\nu_{31}h_{30}}
\;,\crcr
&&
[\varphi^0\bar\varphi^1\bar\varphi^3]_{[\mu]} :=  \int_{h_{ij}}  \varphi^0_{h_{03}\mu_{02}h_{01}} \bar\varphi^1_{h_{10}h_{13}\mu_{12}}   \bar\varphi^3_{\mu_{32}h_{31}h_{30}} \;,\crcr
&& 
[\varphi^0\bar\varphi^1\varphi^2]_{[\nu]}
:=   \int_{h_{ij}} 
\varphi^0_{\nu_{03}h_{02}h_{01}} \bar\varphi^1_{h_{10}\nu_{13}h_{12}} \varphi^2_{h_{21}h_{20}\nu_{23}}\;.
\label{eq:hidden}
\eea
After some algebra (the details of which are collected 
in Appendix \ref{app:freereduce}), the variation 
of the free energy function (\ref{eq:freeexpan0}) can be recast as
\bea
 \frac{\delta \ln Z(\eta,\bar \eta)}{\imath\delta B^{ij}_{[\mu][\nu]}}=0&  =& \frac{1}{ Z(\eta,\bar \eta)}\int d\mu_{C} (\bar \varphi, \varphi ) 
\Bigl\{ 
- [C\bar \varphi]^i_{[b]} \, [\eta C^\dag]^j_{[a]}  
+  [C\bar \eta]^i_{[b]}\,   [\varphi C^\dag]^j_{[a]} 
\cr\cr
&&
- \lambda\Big[  C^{i0}_{[b][\mu]}[\varphi C^\dag]^j_{[a]} [\bar\varphi^1\varphi^2\bar\varphi^3]_{[\mu]} 
- \bar C^{j1}_{[a][\nu]}[C\bar\varphi]^i_{[b]}
[\varphi^0\varphi^2\bar\varphi^3]_{[\nu]}  \cr\cr
&&
+ C^{i2}_{[b][\mu]} [ \varphi C^\dag]^j_{[a]}
[\varphi^0\bar\varphi^1\bar\varphi^3]_{[\mu]} 
- \bar C^{j3}_{[a][\nu]}[C\bar\varphi]^{i}_{[b]}
[\varphi^0\bar\varphi^1\varphi^2]_{[\nu]}     \Big] 
 \;-\;  \bar\lambda \{\bar\varphi\}\cr\cr
&& +C^{ii'}_{[b][\mu]}  \varphi^{i'}_{[\mu]} \bar\varphi^j_{[a]} 
- \bar C^{jj'}_{[a][\nu]} \bar\varphi^{j'}_{[\nu]}\varphi^i_{[b]}\Bigl\}
\;
e^{-S^{\inter}( \varphi, \bar\varphi )  +\bar \varphi \eta  +  \bar \eta \varphi 
 }\;.
\label{eq:genericstartward}
\eea
\medskip 

\noindent{\bf WT identity for two-point functions -}
The next stage is to differentiate the expression (\ref{eq:genericstartward}) using the
operator\footnote{The indices $p,k$ are fixed in $\partial_{\eta^p}\partial_{\bar\eta^k}(\cdot)|_{\eta^p=\bar\eta^k=0}$.  Moreover, omitting for a moment these indices, $\partial_{\eta^p}\partial_{\bar\eta^k}(\cdot)|_{\eta^p=\bar\eta^k=0}$  will be denoted by $\partial_{\eta}\partial_{\bar\eta}$ in the following.}  
 $\partial_{\eta^p}\partial_{\bar\eta^k}(\cdot)|_{\eta^p=\bar\eta^k=0}$ 
for getting the connected components of the
correlation functions (we shall denote $J = \bar \varphi \eta  +  \bar \eta \varphi$ $= \sum_i \bar \varphi^i \eta^i  +  \bar \eta^i \varphi^i$)
\bea
&&
0=\langle \partial_{\eta}\partial_{\bar\eta}\Bigl[- [C\bar \varphi]^i_{[b]} \, [\eta C^\dag]^j_{[a]}  
+  [C\bar \eta]^i_{[b]}\,   [\varphi C^\dag]^j_{[a]} 
\cr\cr
&&
- \lambda\Big[  C^{i0}_{[b][\mu]}[\varphi C^\dag]^j_{[a]} [\bar\varphi^1 \varphi^2 \bar\varphi^3]_{[\mu]} 
- \bar C^{j1}_{[a][\nu]} [C\bar\varphi]^i_{[b]}
[\varphi^0\varphi^2 \bar\varphi^3]_{[\nu]}  \cr\cr
&&
+ C^{i2}_{[b][\mu]} [ \varphi C^\dag]^j_{[a]}
[\varphi^0\bar\varphi^1\bar\varphi^3]_{[\mu]} 
- \bar C^{j3}_{[a][\nu]}[C\bar\varphi]^{i}_{[b]}
[\varphi^0\bar\varphi^1\varphi^2]_{[\nu]}     \Big] 
 \;-\;  \bar\lambda \{\bar\varphi\}\cr\cr
&& +C^{ii'}_{[b][\mu]}  \varphi^{i'}_{[\mu]} \bar\varphi^j_{[a]} 
- \bar C^{jj'}_{[a][\nu]} \bar\varphi^{j'}_{[\nu]}\varphi^i_{[b]}\Bigr] 
e^{J }|_{\eta=\bar\eta=0} \rangle_c \;.
\eea
A direct computation  yields at first
\bea
&&
\langle \Bigl[ 
 C^{ii'}_{[b][\mu]}  \varphi^{i'}_{[\mu]} \bar\varphi^j_{[a]} 
- [C^\dag]^{j'j}_{[\nu][a]} \bar\varphi^{j'}_{[\nu]}\varphi^i_{[b]}  \Bigr] \partial_{\eta}(\bar\varphi\eta)\partial_{\bar\eta}(\bar\eta\varphi) |_{\eta=\bar\eta=0} \rangle_c  \crcr
&&
- \lambda\,
\langle \Bigl[ C^{i0}_{[b][\mu]}[\varphi C^\dag]^j_{[a]}  [\bar\varphi^1 \varphi^2 \bar\varphi^3]_{[\mu]} 
  \Bigr] \partial_{\eta} (\bar\varphi\eta)\partial_{\bar\eta}(\bar\eta\varphi)  |_{\eta=\bar\eta=0} \rangle_c  
+ \lambda\,\langle  \Bigl[
  [C^\dag]^{1j}_{[\nu][a]} [C\bar\varphi]^i_{[b]}
[\varphi^0\varphi^2 \bar\varphi^3]_{[\nu]} 
  \Bigr] \partial_{\eta} (\bar\varphi\eta)\partial_{\bar\eta}(\bar\eta\varphi)  |_{\eta=\bar\eta=0} \rangle_c  \crcr
&&
- \lambda\,\langle \Bigl[ 
   C^{i2}_{[b][\mu]} [ \varphi C^\dag]^j_{[a]}
[\varphi^0\bar\varphi^1\bar\varphi^3]_{[\mu]}  \Bigr] \partial_{\eta} (\bar\varphi\eta)\partial_{\bar\eta}(\bar\eta\varphi) |_{\eta=\bar\eta=0} \rangle_c  
+ \lambda\,\langle\Bigl[ 
  [C^\dag]^{3j}_{[\nu][a]} [C\bar\varphi]^{i}_{[b]} 
[\varphi^0\bar\varphi^1\varphi^2]_{[\nu]}  \Bigr] \partial_{\eta} (\bar\varphi\eta)\partial_{\bar\eta}(\bar\eta\varphi)  |_{\eta=\bar\eta=0} \rangle_c  \cr\cr
&&
- \bar\lambda\{\bar\varphi\}= 
\langle \Bigl[ 
 \partial_{\eta}[C\bar \varphi]^i_{[b]} \, [\eta C^\dag]^j_{[a]}  \partial_{\bar\eta}(\bar\eta\varphi) 
-\partial_{\bar\eta} [C\bar \eta]^i_{[b]}\,   [\varphi C^\dag]^j_{[a]} \partial_{\eta} (\bar\varphi\eta) 
  \Bigr] |_{\eta=\bar\eta=0} \rangle_c  \;.
\eea
Performing the explicit differentiation with respect to $\eta^p_{[m]}$ and $\bar\eta^k_{[n]}$, one gets
\bea
&&
\langle \Bigl[
[C\bar \varphi]^i_{[b]}  [C^\dag]^{pj}_{[m][a]} \varphi^k_{[n]} 
-   [\varphi C^\dag]^{j}_{[a]}  C^{ik}_{[b][n]}\bar\varphi^p_{[m]}
   \Bigr] \rangle_c =\cr\cr
&&
\langle 
\Bigl[ 
 C^{ii'}_{[b][\mu]}  \varphi^{i'}_{[\mu]} \bar\varphi^j_{[a]} 
- [C^\dag]^{j'j}_{[\nu][a]} \bar\varphi^{j'}_{[\nu]}\varphi^i_{[b]}   \Bigr] \bar\varphi^p_{[m]}\varphi^k_{[n]} \rangle_c  \cr\cr
&&
- \lambda\,
\langle \Bigl[  C^{i0}_{[b][\mu]}[\varphi C^\dag]^j_{[a]} [\bar\varphi^1\varphi^2\bar\varphi^3]_{[\mu]} 
  \Bigr] \bar\varphi^p_{[m]}\varphi^k_{[n]} \rangle_c  
+ \lambda\,\langle  \Bigl[
   [C^\dag]^{1j}_{[\nu][a]} [C\bar\varphi]^i_{[b]} [\varphi^0\varphi^2\bar\varphi^3]_{[\nu]} 
  \Bigr] \bar\varphi^p_{[m]}\varphi^k_{[n]} \rangle_c\cr\cr
&&
- \lambda\,\langle \Bigl[ 
 C^{i2}_{[b][\mu]}[ \varphi C^\dag]^j_{[a]}
[\varphi^0\bar\varphi^1  \bar\varphi^3]_{[\mu]}   \Bigr] \bar\varphi^p_{[m]}\varphi^k_{[n]} \rangle_c
+ \lambda\,\langle\Bigl[ 
   [C^\dag]^{3j}_{[\nu][a]}[C\bar\varphi]^{i}_{[b]} 
[\varphi^0 \bar\varphi^1\varphi^2]_{[\nu]}  \Bigr] \bar\varphi^p_{[m]}\varphi^k_{[n]} \rangle_c
- \bar\lambda\{\bar\varphi\} \;,
\eea
so that summing over $[\mu]$ and $[\nu]$,  the following statement holds:
\begin{theorem}
Two-point functions of the colored Boulatov under a generic unitary field transformation satisfy the relation,
for $i,j,p,k=0,1,2,3$,
\bea
&&
\langle
 [C\bar \varphi]^i_{[b]}  [C^\dag]^{pj}_{[m][a]} \varphi^k_{[n]} \rangle_c
- \langle   [\varphi C^\dag]^{j}_{[a]}  C^{ik}_{[b][n]}\bar\varphi^p_{[m]}
   \rangle_c =\cr\cr
&&
\langle 
\bar\varphi^j_{[a]} [C \varphi]^i_{[b]} \bar\varphi^p_{[m]}\varphi^k_{[n]} \rangle_c  
- \langle [\bar\varphi C^\dag]^j_{[a]}\varphi^i_{[b]}   \bar\varphi^p_{[m]}\varphi^k_{[n]} \rangle_c \cr\cr
&&
- \lambda\,
\langle [\varphi C^\dag]^j_{[a]}\; ^{i0}[C\cdot\bar\varphi^1\varphi^2\bar\varphi^3]_{[b]} \, \bar\varphi^p_{[m]}\varphi^k_{[n]} \rangle_c  
+ \lambda\,\langle  
[C\bar\varphi]^i_{[b]}\;
^{1j}[\varphi^0\varphi^2\bar\varphi^3 \cdot C^\dag]_{[a]}\,
  \bar\varphi^p_{[m]}\varphi^k_{[n]} \rangle_c\cr\cr
&&
- \lambda\,\langle[ \varphi C^\dag]^j_{[a]}\;
  ^{i2} [C \cdot \varphi^0\bar\varphi^1 \bar\varphi^3]_{[b]}  \, \bar\varphi^p_{[m]}\varphi^k_{[n]} \rangle_c
+ \lambda\,\langle [C\bar\varphi]^i_{[b]}\;
^{3j}[\varphi^0\bar\varphi^1\varphi^2 \cdot C^\dag]_{[a]}  \, \bar\varphi^p_{[m]}\varphi^k_{[n]} \rangle_c 
- \bar\lambda\{\bar\varphi\} \;,
\eea
 where we introduced the notations
\bea
&&
^{i0}[C\cdot\bar\varphi^1\varphi^2\bar\varphi^3]_{[a]} := 
 C^{i0}_{[a][\mu]} [\bar\varphi^1\varphi^2\bar\varphi^3]_{[\mu]} \;,\qquad
^{1j}[\varphi^0\varphi^2\bar\varphi^3 \cdot C^\dag]_{[a]}
:= 
 [C^\dag]^{1j}_{[\nu][a]}
[\varphi^0\varphi^2\bar\varphi^3]_{[\nu]} \;,\crcr
&&
^{i2} [C \cdot \varphi^0\bar\varphi^1 \bar\varphi^3]_{[a]}:=
 C^{i2}_{[a][\mu]}
[\varphi^0 \bar\varphi^1\bar\varphi^3]_{[\mu]} \;,\qquad
^{3j}[\varphi^0\bar\varphi^1\varphi^2 \cdot C^\dag]_{[a]}
:= 
 [C^\dag]^{3j}_{[\nu][a]}
[\varphi^0 \bar\varphi^1 \varphi^2]_{[\nu]} \;,
\eea
for which repeated indices are summed.

\end{theorem}

\noindent{\bf The case of four external colors -} Let us consider that the external color labels $i,j,p,k$ are pairwise
distinct. For definiteness, let us assume that $i=0$, $j=1$,
$p=2$ and $k=3$,  then the WT identity becomes
\bea
&&
0=
\langle 
\bar\varphi^1_{[a]} [C \varphi]^0_{[b]} \bar\varphi^2_{[m]}\varphi^3_{[n]} \rangle_c  
- \langle [\bar\varphi C^\dag]^1_{[a]}\varphi^0_{[b]}   \bar\varphi^2_{[m]}\varphi^3_{[n]} \rangle_c \cr\cr
&&
- \lambda\,
\langle [\varphi C^\dag]^1_{[a]}\; ^{00}[C\cdot\bar\varphi^1\varphi^2\bar\varphi^3]_{[b]} \, \bar\varphi^2_{[m]}\varphi^3_{[n]} \rangle_c  
+ \lambda\,\langle  
[C\bar\varphi]^0_{[b]}\;
^{11}[\varphi^0\varphi^2\bar\varphi^3 \cdot C^\dag]_{[a]}\,
  \bar\varphi^2_{[m]}\varphi^3_{[n]} \rangle_c - \bar\lambda\{\bar\varphi\} \;.
\eea
This is an identity for a four-point function 
with four external color which might be useful in the study of the
coupling constant renormalization.

\noindent{\bf The case of two external colors -} 
Let us assume now that $i=k=0$ and $j=p=1$
\bea
&&
\langle
 [C\bar \varphi]^0_{[b]}  [C^\dag]^{11}_{[m][a]} \varphi^0_{[n]} \rangle_c
- \langle   [\varphi C^\dag]^{1}_{[a]}  C^{00}_{[b][n]}\bar\varphi^1_{[m]}
   \rangle_c =\cr\cr
&&
\langle 
\bar\varphi^1_{[a]} [C \varphi]^0_{[b]} \bar\varphi^1_{[m]}\varphi^0_{[n]} \rangle_c  
- \langle [\bar\varphi C^\dag]^1_{[a]}\varphi^0_{[b]}   \bar\varphi^1_{[m]}\varphi^0_{[n]} \rangle_c \cr\cr
&&
- \lambda\,
\langle [\varphi C^\dag]^1_{[a]}\; ^{00}[C\cdot\bar\varphi^1\varphi^2\bar\varphi^3]_{[b]} \, \bar\varphi^1_{[m]}\varphi^0_{[n]} \rangle_c  
+ \lambda\,\langle  
[C\bar\varphi]^0_{[b]}\;
^{11}[\varphi^0\varphi^2\bar\varphi^3 \cdot C^\dag]_{[a]}\,
  \bar\varphi^1_{[m]}\varphi^0_{[n]} \rangle_c
- \bar\lambda\{\bar\varphi\} \;.
\eea

\noindent{\bf WT identities for four-point functions -}
To obtain higher order point functions, we derivate again
the free energy.
Derivating twice $\ln Z$ for computing the four-point function identities,
we have
\bea
&&
\langle \Bigl[ 
 C^{ii'}_{[b][\mu]}  \varphi^{i'}_{[\mu]} \bar\varphi^j_{[a]} 
- [C^\dag]^{j'j}_{[\nu][a]} \bar\varphi^{j'}_{[\nu]}\varphi^i_{[b]}  \Bigr] \partial_{\eta_2}J\partial_{\bar\eta_2}J\partial_{\eta_1}J\partial_{\bar\eta_1}J |_{\eta=\bar\eta=0} \rangle_c  \crcr
&&
- \lambda\,
\langle \Bigl[ C^{i0}_{[b][\mu]}[\varphi C^\dag]^j_{[a]} [\bar\varphi^1\varphi^2\bar\varphi^3]_{[\mu]} 
  \Bigr] \partial_{\eta_2}J\partial_{\bar\eta_2}J\partial_{\eta_1}J\partial_{\bar\eta_1}J  |_{\eta=\bar\eta=0} \rangle_c  \crcr
&&
+ \lambda\,\langle  \Bigl[
 [C^\dag]^{1j}_{[\nu][a]}[C\bar\varphi]^i_{[b]} 
[\varphi^0\varphi^2\bar\varphi^3]_{[\nu]} 
  \Bigr] \partial_{\eta_2}J\partial_{\bar\eta_2}J\partial_{\eta_1}J\partial_{\bar\eta_1}J  |_{\eta=\bar\eta=0} \rangle_c  \crcr
&&
- \lambda\,\langle \Bigl[ 
  C^{i2}_{[b][\mu]} [ \varphi C^\dag]^j_{[a]}
[\varphi^0\bar\varphi^1\bar\varphi^3]_{[\mu]}  \Bigr] \partial_{\eta_2}J\partial_{\bar\eta_2}J\partial_{\eta_1}J\partial_{\bar\eta_1}J|_{\eta=\bar\eta=0} \rangle_c  \crcr
&&
+\lambda\,\langle\Bigl[ 
 [C^\dag]^{3j}_{[\nu][a]}[C\bar\varphi]^{i}_{[b]} 
[\varphi^0\bar\varphi^1\varphi^2]_{[\nu]} \Bigr] \partial_{\eta_2}J\partial_{\bar\eta_2}J\partial_{\eta_1}J\partial_{\bar\eta_1}J  |_{\eta=\bar\eta=0} \rangle_c - \bar\lambda\{\bar\varphi\} \cr\cr
&&
= 
\langle \Bigl[ 
 [C\bar \varphi]^i_{[b]} \partial_{\eta_1}[\eta C^\dag]^j_{[a]} \partial_{\bar\eta_1}J 
-  \partial_{\bar\eta_1}[C\bar \eta]^i_{[b]}   [\varphi C^\dag]^j_{[a]} \partial_{\eta_1} J
  \Bigr]\partial_{\bar\eta_2} J \partial_{\eta_2} J |_{\eta=\bar\eta=0} \rangle_c  + (1 \leftrightarrow 2)\;.
 \eea
 Fixing the indices of $(\eta_1)^k_{[m]}$,$ (\bar \eta_1)^l_{[n]}$
$(\eta_2)^t_{[p]}$ and $(\bar\eta_2)^s_{[q]}$, the differentiations
yield
\bea
&&
\langle \Bigl[ 
C^{ii'}_{[b][\mu]}  \varphi^{i'}_{[\mu]} \bar\varphi^j_{[a]} 
- [C^\dag]^{j'j}_{[\nu][a]} \bar\varphi^{j'}_{[\nu]}\varphi^i_{[b]}  \Bigr] 
\bar\varphi^{t}_{[p]}\varphi^s_{[q]} \bar\varphi^k_{[m]}
\varphi^l_{[n]} \rangle_c  \crcr
&&
- \lambda\,
\langle \Bigl[ C^{i0}_{[b][\mu]}[\varphi C^\dag]^j_{[a]} [\bar\varphi^1\varphi^2\bar\varphi^3]_{[\mu]} 
  \Bigr]\bar\varphi^{t}_{[p]}\varphi^s_{[q]} \bar\varphi^k_{[m]}
\varphi^l_{[n]}  \rangle_c
\crcr
&&  
+ \lambda\,\langle  \Bigl[
  [C^\dag]^{1j}_{[\nu][a]} [C\bar\varphi]^i_{[b]}
[\varphi^0\varphi^2 \bar\varphi^3]_{[\nu]} 
  \Bigr]\bar\varphi^{t}_{[p]}\varphi^s_{[q]} \bar\varphi^k_{[m]}
\varphi^l_{[n]}   \rangle_c  \crcr
&&
- \lambda\,\langle \Bigl[ 
 C^{i2}_{[b][\mu]} [ \varphi C^\dag]^j_{[a]}
[\varphi^0 \bar\varphi^1 \bar\varphi^3]_{[\mu]}  \Bigr]
\bar\varphi^{t}_{[p]}\varphi^s_{[q]} \bar\varphi^k_{[m]}
\varphi^l_{[n]} \rangle_c  \crcr
&&
+ \lambda\,\langle\Bigl[ 
 [C^\dag]^{3j}_{[\nu][a]}[C\bar\varphi]^{i}_{[b]} 
[\varphi^0\bar\varphi^1\varphi^2]_{[\nu]} \Bigr]
\bar\varphi^{t}_{[p]}\varphi^s_{[q]} \bar\varphi^k_{[m]}
\varphi^l_{[n]} \rangle_c  
- \bar\lambda\{\bar\varphi\} \crcr
&&
 = \langle \Bigl[ 
 [C\bar \varphi]^i_{[b]} [C^\dag]^{kj}_{[m][a]} \varphi^l_{[n]} 
- C^{il}_{[b][n]}   [\varphi C^\dag]^j_{[a]} \bar\varphi^k_{[m]}
  \Bigr]\bar\varphi^{t}_{[p]}\varphi^s_{[q]} \rangle_c 
 + [(k,l) \;\leftrightarrow \; (t,s)]\;. 
\eea
Summing over $[\mu]$ and $[\nu]$,  on this last expression rests our
\begin{theorem}
Four-point functions of the colored Boulatov model under a generic unitary field transformation satisfy 
the relation, for $i,j,k,l,s,t=0,1,2,3$,
\bea
&&
\langle [C\bar \varphi]^i_{[b]}   \varphi^l_{[n]}[C^\dag]^{kj}_{[m][a]}\bar\varphi^t_{[p]}\varphi^s_{[q]}\rangle_c 
-  \langle   [\varphi C^\dag]^j_{[a]}  C^{il}_{[b][n]}\bar\varphi^k_{[m]}\bar\varphi^t_{[p]}\varphi^s_{[q]}\rangle_c 
+ [(k,l) \;\leftrightarrow \; (t,s)]= 
\cr\cr&&
\langle 
\bar\varphi^j_{[a]}  [C\varphi]^i_{[b]}  \bar\varphi^{t}_{[p]}\varphi^s_{[q]} \bar\varphi^k_{[m]}
\varphi^l_{[n]} \rangle_c
-  \langle [\bar\varphi C^\dag]^j_{[a]}\varphi^i_{[b]}   \bar\varphi^{t}_{[p]}\varphi^s_{[q]} \bar\varphi^k_{[m]}
\varphi^l_{[n]} \rangle_c  \cr\cr
&&
- \lambda\,
\langle [\varphi C^\dag]^j_{[a]} \;
^{i0}[C\cdot\bar\varphi^1\varphi^2\bar\varphi^3]_{[b]} \,
\bar\varphi^{t}_{[p]}\varphi^s_{[q]} \bar\varphi^k_{[m]}
\varphi^l_{[n]}  \rangle_c  
+ \lambda\,\langle 
 [C\bar\varphi]^i_{[b]}\; ^{1j}[\varphi^0\varphi^2\bar\varphi^3 \cdot C^\dag]_{[a]}\,\bar\varphi^{t}_{[p]}\varphi^s_{[q]} \bar\varphi^k_{[m]}
\varphi^l_{[n]}  \rangle_c\cr\cr
&&
- \lambda\,\langle [ \varphi C^\dag]^j_{[a]}\; 
^{i2} [C \cdot \varphi^0\bar\varphi^1 \bar\varphi^3]_{[b]}\, \bar\varphi^{t}_{[p]}\varphi^s_{[q]} \bar\varphi^k_{[m]}
\varphi^l_{[n]} \rangle_c
+ \lambda\,\langle [C\bar\varphi]^i_{[b]}\;
^{3j}[\;\varphi^0\bar\varphi^1\varphi^2 \cdot C^\dag]_{[a]} \, \bar\varphi^{t}_{[p]}\varphi^s_{[q]} \bar\varphi^k_{[m]}
\varphi^l_{[n]} \rangle_c
- \bar\lambda\{\bar\varphi\}\;.
\eea
\end{theorem}

\medskip
  
\noindent{\bf WT identities for even-point functions -}
The WT identities for $[n=2p\geq 2]$-point functions
can be deduced by simple recursion from the aforementioned equations. 
We first need to introduce some notations
\bea
&&
\mathcal F = [C\bar \varphi]^i_{[b]} \, [\eta C^\dag]^j_{[a]}  
- [C\bar \eta]^i_{[b]}\,   [\varphi C^\dag]^j_{[a]} \;,\qquad
\mathcal H = \mathcal F  e^{J} \;,\cr\cr
&&
\partial_{\bar \eta_k} \partial_{\eta_k} \mathcal H = 
\left(\mathcal F_{k} + (\partial_{\bar \eta_k} J) (\partial_{\eta_k}  J)\,\mathcal F \right)e^{J}\;,\qquad
\mathcal F_{k}  
= 
 [C \bar \varphi ]^i_{[b]} \partial_{\eta_k}  [\eta C^{\dagger} ]^j_{[a]}   (\partial_{\bar\eta_k} J)
- \partial_{\bar\eta_k}  [C\bar \eta]^i_{[b]} [\varphi C^{\dagger}]^j_{[a]}  \,
    (\partial_{\eta_k} J) \;.
\eea
It is by simple recursion that one proves that
\bea
&&
\Bigl[\prod_{l=1}^n \partial_{\eta_l}\partial_{\bar\eta_l} \Bigr]
\mathcal H    = \Bigl[\sum_{k=1}^n 
 \mathcal F_k \prod_{l\neq k}^n\left[ 
\partial_{\eta_l} J \partial_{\bar\eta_l} J \right]
+  
\prod_{l=1}^n  \left[\partial_{\eta_l} J \partial_{\bar\eta_l} J \right]
\mathcal F
\Bigr]\, e^J \;,  \\
&& 
\Bigl[\prod_l^n \partial_{\eta_l}\partial_{\bar\eta_l} \Bigr]
\mathcal H  |_{\eta=\bar\eta=0}  = \sum_{k=1}^n 
 \mathcal F_k \prod_{l\neq k}^n\left[ 
\partial_{\eta_l} J \partial_{\bar\eta_l} J \right]|_{\eta=\bar\eta=0} 
\;.
\label{eq:deriv0}
\eea
From the last line (\ref{eq:deriv0}), we explicitly obtain by
fixing the derivative with respect to the indices such that
$(\eta_l)^{\alpha_l}_{[a_l]}$ and $(\bar\eta_l)^{\beta_l}_{[b_l]}$:
\bea
&&
\Bigl[\prod_{l=1}^n \partial_{\eta_l}\partial_{\bar\eta_l} \Bigr]
\mathcal H  |_{\eta=\bar\eta=0}  = \sum_{k=1}^n 
\left( [C \bar \varphi ]^i_{[b]}  [C^{\dagger} ]^{\alpha_k j}_{[a_k][a]}    \varphi^{\beta_k}_{[b_k]}
- C^{i \beta_k}_{[b][b_k]} [\varphi C^{\dagger}]^j_{[a]} \bar\varphi^{\alpha_k}_{[a_k]} \, \right)
\prod_{l\neq k}^n\left[ 
\bar\varphi^{\alpha_l}_{[a_l]} \varphi^{\beta_l}_{[b_l]} \right] \;.
\eea
The generalized WT identity for an even-point function
can be written as
\bea
&&
\Big\langle \Bigl[ 
 C^{ii'}_{[b][\mu]}  \varphi^{i'}_{[\mu]} \bar\varphi^j_{[a]} 
- [C^\dag]^{j'j}_{[\nu][a]} \bar\varphi^{j'}_{[\nu]}\varphi^i_{[b]} \Bigr] 
\prod_{l =1}^n \left[ 
\bar\varphi^{\alpha_l}_{[a_l]} \varphi^{\beta_l}_{[b_l]} \right] \Big\rangle_c  \crcr
&&
- \lambda\,
\Big\langle \Bigl[ C^{i0}_{[b][\mu]}[\varphi C^\dag]^j_{[a]} [\bar\varphi^1\varphi^2\bar\varphi^3]_{[\mu]} 
  \Bigr]\prod_{l =1}^n \left[ \bar\varphi^{\alpha_l}_{[a_l]} \varphi^{\beta_l}_{[b_l]}  \right]   \Big\rangle_c
\crcr
&&  
+ \lambda\,\Big\langle  \Bigl[
    [C^\dag]^{1j}_{[\nu][a]} [C\bar\varphi]^i_{[b]} 
[\varphi^0\varphi^2\bar\varphi^3]_{[\nu]} 
  \Bigr]\prod_{l =1}^n \left[ \bar\varphi^{\alpha_l}_{[a_l]} \varphi^{\beta_l}_{[b_l]} \right]  \Big\rangle_c  \crcr
&&
- \lambda\,\Big\langle \Bigl[ 
  C^{i2}_{[b][\mu]}[ \varphi C^\dag]^j_{[a]}
[\varphi^0\bar\varphi^1\bar\varphi^3]_{[\mu]}  \Bigr]
\prod_{l =1}^n \left[ \bar\varphi^{\alpha_l}_{[a_l]} \varphi^{\beta_l}_{[b_l]}  \right]\Big\rangle_c  \crcr
&&
+ \lambda\,\Big\langle\Bigl[ 
   [C^\dag]^{3j}_{[\nu][a]}[C\bar\varphi]^{i}_{[b]}  
[\varphi^0 \bar\varphi^1 \varphi^2]_{[\nu]} \Bigr]
\prod_{l =1}^n \left[\bar\varphi^{\alpha_l}_{[a_l]} \varphi^{\beta_l}_{[b_l]}  \right] \Big\rangle_c  
- \bar\lambda\{\bar\varphi\} \crcr
&&
 = \sum_{k=1}^n 
\Big\langle 
\Big[  [C \bar \varphi ]^i_{[b]}  [C^{\dagger} ]^{\alpha_k j}_{[a_k][a]}    \varphi^{\beta_k}_{[b_k]}
- C^{i \beta_k}_{[b][b_k]} [\varphi C^{\dagger}]^j_{[a]} \bar\varphi^{\alpha_k}_{[a_k]} 
\Big]
\prod_{l\neq k}^n\left[ 
\bar\varphi^{\alpha_l}_{[a_l]} \varphi^{\beta_l}_{[b_l]}  \right] \Big\rangle_c \;.
\eea
Summing over the remaining 
running indices $[\mu]$ and $[\nu]$,  we have, in the same anterior notations, the following statement:
\begin{theorem}
\label{theogener}
Even $n$-point functions of the colored Boulatov model under a generic unitary field transformation satisfy the relation
\bea
&&
 \sum_{k=1}^n \Big\{  
\Big\langle  [C \bar \varphi ]^i_{[b]}  [C^{\dagger} ]^{\alpha_k j}_{[a_k][a]}    \varphi^{\beta_k}_{[b_k]}
\prod_{\ell\neq k}^n\left[ 
\bar\varphi^{\alpha_\ell}_{[a_\ell]} \varphi^{\beta_\ell}_{[b_\ell]} \right] \Big\rangle_c
- \Big\langle  C^{i \beta_k}_{[b][b_k]} [\varphi C^{\dagger}]^j_{[a]} \bar\varphi^{\alpha_k}_{[a_k]} 
\prod_{\ell\neq k}^n\left[ 
\bar\varphi^{\alpha_\ell}_{[a_\ell]} \varphi^{\beta_\ell}_{[b_\ell]} \right] \Big\rangle_c \Big\} \crcr
& = &
\Big\langle
 \bar\varphi^j_{[a]} [C\varphi]^i_{[b]} \prod_{\ell =1}^n \left[ \bar\varphi^{\alpha_\ell}_{[a_\ell]} \varphi^{\beta_\ell}_{[b_\ell]}  \right]\Big\rangle_c 
-  \Big\langle [\bar\varphi C^\dag]^j_{[a]}\varphi^i_{[b]} \prod_{\ell =1}^n \left[\bar\varphi^{\alpha_\ell}_{[a_\ell]} \varphi^{\beta_\ell}_{[b_\ell]}  \right]\Big\rangle_c 
 \crcr
&-& 
\lambda\,
\Big\langle  [\varphi C^\dag]^j_{[a]} \;
^{i0}[C\cdot\bar\varphi^1\varphi^2\bar\varphi^3]_{[b]} 
\prod_{\ell =1}^n \left[\bar\varphi^{\alpha_\ell}_{[a_\ell]} \varphi^{\beta_\ell}_{[b_\ell]}  \right]   \Big\rangle_c
+ 
\lambda\,\Big\langle  
[C\bar\varphi]^i_{[b]}\; ^{1j}[\varphi^0\varphi^2\bar\varphi^3 \cdot C^\dag]_{[a]}
\prod_{\ell =1}^n \left[ \bar\varphi^{\alpha_\ell}_{[a_\ell]} \varphi^{\beta_\ell}_{[b_\ell]} \right]  \Big\rangle_c  \crcr
&-&
 \lambda\,\Big\langle   [ \varphi C^\dag]^j_{[a]}\; 
^{i2} [C \cdot \varphi^0\bar\varphi^1 \bar\varphi^3]_{[b]}\,
\prod_{\ell =1}^n \left[\bar\varphi^{\alpha_\ell}_{[a_\ell]} \varphi^{\beta_\ell}_{[b_\ell]}  \right]\Big\rangle_c  
+ 
\lambda\,\Big\langle
\, [C\bar\varphi]^i_{[b]}\;
^{3j}[\;\varphi^0\bar\varphi^1\varphi^2 \cdot C^\dag]_{[a]} \,
\prod_{\ell =1}^n \left[\bar\varphi^{\alpha_\ell}_{[a_\ell]} \varphi^{\beta_\ell}_{[b_\ell]} \right] \Big\rangle_c  
- \bar\lambda\{\bar\varphi\}  \;.
\label{eq:genericWARD}
\eea
\end{theorem}
The WT identity (\ref{eq:genericWARD}) is valid in full generality regarding
a generic unitary field transformation without requiring
that transformation to be a symmetry of the action. 
We can make the following striking observation:
the equation (\ref{eq:genericstartward}) (and hence the
general WT identity (\ref{eq:genericWARD})) generates without ambiguity  the WT identities associated with 
independent non identically distributed matrix models with 
invertible covariances as the models studied in \cite{Disertori:2006nq,Geloun:2008zk}. 
Indeed, forgetting the color index and restricting all
tensors to matrices, 
 one has just to use the facts that, on one hand, 
these models are covariant under a unitary symmetry 
in order to cancel 
the interaction terms in $\lambda,\bar\lambda$ and, 
on the other hand, to invert two extra covariances such that
(\ref{eq:genericstartward}) and (\ref{eq:genericWARD}) generate
the corresponding WT equations for these more simple cases.  
Hence, it is often useful to specify which kind of WT 
identities could be inferred from the same reasoning
with now a well defined symmetry of the model. It is the main 
purpose of the remaining of this paper.

\subsection{Ward-Takahashi identities for the 1-action symmetry}
\label{subsect:1act}

We begin with the partition function (\ref{eq:partiinterm0})
  under infinitesimal transformations given by
$\delta_B S^{\inter} =0$, 
(\ref{eq:infcovrinv}) and (\ref{eq:infsourrinv}) that
we symbolically write 
\bea
 Z(\bar \eta,\eta) = \int d\mu_{C} (\bar \varphi, \varphi ) 
  \Bigl( 1- \imath \bar \varphi  B\eta +  \bar \eta B \varphi  + \imath \delta_{\varphi} [BC-CB] \delta_{\bar \varphi} \Bigr) 
e^{-S^{\inter}( \varphi, \bar\varphi ) +\bar \varphi \eta  +  \bar \eta \varphi
 }\;.
\eea
Derivating the free energy with respect to $\imath B_{\mu\nu}$  yields  
\bea
 \frac{\ln Z(\eta,\bar\eta)}{\imath\delta B_{\mu\nu}}
&=&0 =  \frac{1}{Z(\eta,\bar\eta)}
\int d\mu_{C} (\bar \varphi, \varphi ) 
\int_{g_\alpha h_\beta}
\Bigl( -\bar \varphi^{0,2}_{\mu.} \eta^{0,2}_{\nu.} +  \bar \eta^{0,2}_{\mu.}  \varphi^{0,2}_{\nu.}  
 - \bar \varphi^{1,3}_{.\mu} \eta^{1,3}_{.\nu} +  \bar \eta^{1,3}_{.\mu}  \varphi^{1,3}_{.\nu}    \crcr
&&
   + \frac{\delta}{\delta \varphi^{0,2}_{\nu.}} C^{0,2}_{\mu. ;\;h_1h_2h_3}
\frac{\delta}{\delta\bar \varphi^{0,2}_{h_0h_1h_2} }
 - \frac{\delta}{\delta \varphi^{0,2}_{h_0h_1h_2}} 
C^{0,2}_{h_0h_1h_2;\; \nu.}
\frac{\delta}{\delta\bar \varphi^{0,2}_{\mu.} } 
\crcr
&&
 + \frac{\delta}{\delta \varphi^{1,3}_{.\nu}} C^{1,3}_{.\mu ;\;h_1h_2h_3}
\frac{\delta}{\delta\bar \varphi^{1,3}_{h_0h_1h_2} }
 - \frac{\delta}{\delta \varphi^{1,3}_{h_0h_1h_2}} 
C^{1,3}_{h_0h_1h_2;\;  .\nu}
\frac{\delta}{\delta\bar \varphi^{1,3}_{.\mu} } 
 \Bigr) 
e^{-S^{\inter}( \varphi, \bar\varphi ) +\bar \varphi \eta  +  \bar \eta \varphi  }\;,
\label{eq:intermed}
\eea
where repeated indices and $(\cdot)$ arguments (called henceforth 
dot arguments) are summed.
The latter expression can be computed to
(see Appendix \ref{app:freereduc1action} for derivations and 
notations)
\bea
 0&=&  \frac{1}{Z(\eta,\bar\eta)}  \int d\mu_{C} (\bar \varphi, \varphi )    \Bigl[ 
- \bar \varphi^{0,2}_{\mu.} \eta^{0,2}_{\nu.} +  \bar \eta^{0,2}_{\mu.}  \varphi^{0,2}_{\nu.}  
 - \bar \varphi^{1,3}_{.\mu} \eta^{1,3}_{.\nu} +  \bar \eta^{1,3}_{.\mu}  \varphi^{1,3}_{.\nu}    
+\varphi^{0,2}_{\mu.}\bar\eta^{0,2}_{\nu.}
-\bar\varphi^{0,2}_{\nu.} \eta^{0,2}_{\mu.}
+
\varphi^{1,3}_{.\mu}\bar\eta^{1,3}_{.\nu}
-
\bar\varphi^{1,3}_{.\nu}\eta^{1,3}_{.\mu} \crcr
&&
- \lambda \;\varphi^{0,2}_{\mu.} \cdot [\bar\varphi^{1}\varphi^{2,0}\bar \varphi^{3}]_{\nu.} 
-\bar\lambda\; \varphi^{1,3}_{.\mu}\cdot [\bar\varphi^{0}\bar \varphi^{2}\varphi^{3,1}]_{.\nu}
+\lambda\;
\bar\varphi^{0,2}_{\nu.}
\cdot [\varphi^{1}\bar\varphi^{2,0} \varphi^{3}]_{\mu.}
+
\bar\lambda \;\bar\varphi^{1,3}_{.\nu}
\cdot[ 
\varphi^{0} \varphi^{2}\bar\varphi^{3,1}]_{.\mu}
 \Bigr]
e^{-S^{\inter}( \varphi, \bar\varphi ) +\bar \varphi \eta  +  \bar \eta \varphi  } \crcr
&&
\label{reducfreeener1act}
\eea
and therefore, differentiating by an even product
of $\eta^{\alpha_\ell}_{[a_\ell]} \bar\eta^{\beta_\ell}_{[b_\ell]}$, we can readily identify the WT identities as given by the equation
\bea
&&
 \sum_{k=1}^n \Big\{  
\Big\langle  \bar \varphi ^{0,2}_{\mu.}\,  [\delta ]^{\alpha_k ,0,2}_{[a_k]\nu.}\,    \varphi^{\beta_k}_{[b_k]}
\prod_{\ell\neq k}^n\left[ 
\bar\varphi^{\alpha_\ell}_{[a_\ell]} \varphi^{\beta_\ell}_{[b_\ell]} \right] \Big\rangle_c
- \Big\langle  [\delta]^{\beta_k,0,2}_{ [b_k]\mu.} \,\varphi ^{0,2}_{\nu.} \bar\varphi^{\alpha_k}_{[a_k]} 
\prod_{\ell\neq k}^n\left[ 
\bar\varphi^{\alpha_\ell}_{[a_\ell]} \varphi^{\beta_\ell}_{[b_\ell]} \right] \Big\rangle_c\crcr
&&
-\Big\langle   \varphi ^{0,2}_{\mu.} \, [\delta ]^{\beta_k ,0,2}_{[b_k]\nu.}  \, \bar \varphi^{\alpha_k}_{[a_k]}
\prod_{\ell\neq k}^n\left[ 
\bar\varphi^{\alpha_\ell}_{[a_\ell]} \varphi^{\beta_\ell}_{[b_\ell]} \right] \Big\rangle_c
+ \Big\langle  [\delta]^{\alpha_k,0,2}_{[a_k]\mu. }\, \bar\varphi ^{0,2}_{\nu.} \varphi^{\beta_k}_{[b_k]} 
\prod_{\ell\neq k}^n\left[ 
\bar\varphi^{\alpha_\ell}_{[a_\ell]} \varphi^{\beta_\ell}_{[b_\ell]} \right] \Big\rangle_c \crcr
&&
+\Big\langle  \bar \varphi ^{1,3}_{.\mu}  \,[\delta ]^{\alpha_k ,1,3}_{[a_k].\nu}  \,  \varphi^{\beta_k}_{[b_k]}
\prod_{\ell\neq k}^n\left[ 
\bar\varphi^{\alpha_\ell}_{[a_\ell]} \varphi^{\beta_\ell}_{[b_\ell]} \right] \Big\rangle_c
- \Big\langle  [\delta]^{ \beta_k,1,3}_{.\mu [b_k]} \,\varphi ^{1,3}_{.\nu} \bar\varphi^{\alpha_k}_{[a_k]} 
\prod_{\ell\neq k}^n\left[ 
\bar\varphi^{\alpha_\ell}_{[a_\ell]} \varphi^{\beta_\ell}_{[b_\ell]} \right] \Big\rangle_c\crcr
&&
-\Big\langle   \varphi ^{1,3}_{.\mu} \, [\delta ]^{\beta_k ,1,3}_{[b_k].\nu}   \,\bar \varphi^{\alpha_k}_{[a_k]}
\prod_{\ell\neq k}^n\left[ 
\bar\varphi^{\alpha_\ell}_{[a_\ell]} \varphi^{\beta_\ell}_{[b_\ell]} \right] \Big\rangle_c
+ \Big\langle  [\delta]^{\alpha_k,1,3}_{.\mu [a_k]}\, \bar\varphi ^{1,3}_{.\nu} \varphi^{\beta_k}_{[b_k]} 
\prod_{\ell\neq k}^n\left[ 
\bar\varphi^{\alpha_\ell}_{[a_\ell]} \varphi^{\beta_\ell}_{[b_\ell]} \right] \Big\rangle_c \Big\} \crcr
& = &
- 
\lambda\,
\Big\langle \varphi^{0,2}_{\mu.} \cdot [\bar\varphi^{1}\varphi^{2,0}\bar \varphi^{3}]_{\nu.} 
\prod_{\ell =1}^n \left[\bar\varphi^{\alpha_\ell}_{[a_\ell]} \varphi^{\beta_\ell}_{[b_\ell]}  \right]   \Big\rangle_c
-\bar\lambda\,\Big\langle  
 \varphi^{1,3}_{.\mu}\cdot [\bar\varphi^{0}\bar \varphi^{2}\varphi^{3,1}]_{.\nu}
\prod_{\ell =1}^n \left[ \bar\varphi^{\alpha_\ell}_{[a_\ell]} \varphi^{\beta_\ell}_{[b_\ell]} \right]  \Big\rangle_c  \crcr
&+&
 \lambda\,\Big\langle \bar\varphi^{0,2}_{\nu.} \cdot [\varphi^{1}\bar\varphi^{2,0}\varphi^{3}]_{\mu.} 
\prod_{\ell =1}^n \left[\bar\varphi^{\alpha_\ell}_{[a_\ell]} \varphi^{\beta_\ell}_{[b_\ell]}  \right]\Big\rangle_c  
+
\bar\lambda \,\Big\langle
\bar\varphi^{1,3}_{.\nu}
\cdot[ 
\varphi^{0} \varphi^{2}\bar\varphi^{3,1}]_{.\mu}\,
\prod_{\ell =1}^n \left[\bar\varphi^{\alpha_\ell}_{[a_\ell]} \varphi^{\beta_\ell}_{[b_\ell]} \right] \Big\rangle_c  
 \;,
\eea
where $[\delta]^{\alpha,i,j}_{[a][b]} = (\delta^{\alpha,i} + \delta^{\alpha,j}) \delta_{[a][b]}$. 
This leads us to the following statement.
\begin{theorem}
\label{coro1action}
Even $n$-point functions of the colored Boulatov model under a right invariant unitary transformation satisfy the relation
\bea
&&
 \sum_{k=1}^n \Big\{ \sum_{\alpha=0,2} \Big[ 
 \delta ^{\alpha_k \alpha} \delta_{a_{k1}\nu} \Big\langle  \bar \varphi ^{\alpha}_{\mu a_{k2} a_{k3}}    \varphi^{\beta_k}_{[b_k]}
\prod_{\ell\neq k}^n\left[ 
\bar\varphi^{\alpha_\ell}_{[a_\ell]} \varphi^{\beta_\ell}_{[b_\ell]} \right] \Big\rangle_c
-  \delta^{\alpha \beta_k}  \delta_{\mu b_{k1}} \Big\langle \varphi ^{\alpha}_{\nu b_{k2}b_{k3}} \bar\varphi^{\alpha_k}_{[a_k]} 
\prod_{\ell\neq k}^n\left[ 
\bar\varphi^{\alpha_\ell}_{[a_\ell]} \varphi^{\beta_\ell}_{[b_\ell]} \right] \Big\rangle_c\crcr
&&
- \delta ^{\alpha \beta_k} \delta_{b_{k1}\nu}\Big\langle   \varphi ^{\alpha}_{\mu b_{k2}b_{k3}}   \bar \varphi^{\alpha_k}_{[a_k]}
\prod_{\ell\neq k}^n\left[ 
\bar\varphi^{\alpha_\ell}_{[a_\ell]} \varphi^{\beta_\ell}_{[b_\ell]} \right] \Big\rangle_c
+ \delta^{\alpha \alpha_k} \delta_{\mu a_{k1}} \Big\langle  \bar\varphi ^{\alpha}_{\nu a_{k2} a_{k3}} \varphi^{\beta_k}_{[b_k]} 
\prod_{\ell\neq k}^n\left[ 
\bar\varphi^{\alpha_\ell}_{[a_\ell]} \varphi^{\beta_\ell}_{[b_\ell]} \right] \Big\rangle_c \Big]\crcr
&&
+ \sum_{\alpha=1,3}\Big[\delta ^{\alpha_k \alpha} \delta_{a_{k3}\nu}  \Big\langle  \bar \varphi ^{\alpha}_{a_{k1}a_{k2}\mu}     \varphi^{\beta_k}_{[b_k]}
\prod_{\ell\neq k}^n\left[ 
\bar\varphi^{\alpha_\ell}_{[a_\ell]} \varphi^{\beta_\ell}_{[b_\ell]} \right] \Big\rangle_c
-  \delta^{\alpha \beta_k}\delta_{\mu b_{k3}} \Big\langle \varphi ^{\alpha}_{b_{k1}b_{k2}\nu} \bar\varphi^{\alpha_k}_{[a_k]} 
\prod_{\ell\neq k}^n\left[ 
\bar\varphi^{\alpha_\ell}_{[a_\ell]} \varphi^{\beta_\ell}_{[b_\ell]} \right] \Big\rangle_c\crcr
&&
- \delta ^{\alpha\beta_k }\delta_{b_{k3}\nu}  \Big\langle   \varphi ^{\alpha}_{b_{k1}b_{k2}\mu}  \bar \varphi^{\alpha_k}_{[a_k]}
\prod_{\ell\neq k}^n\left[ 
\bar\varphi^{\alpha_\ell}_{[a_\ell]} \varphi^{\beta_\ell}_{[b_\ell]} \right] \Big\rangle_c
+  \delta^{\alpha \alpha_k} \delta_{\mu a_{k3}} \Big\langle \bar\varphi ^{\alpha}_{a_{k1}a_{k2}\nu} \varphi^{\beta_k}_{[b_k]} 
\prod_{\ell\neq k}^n\left[ 
\bar\varphi^{\alpha_\ell}_{[a_\ell]} \varphi^{\beta_\ell}_{[b_\ell]} \right] \Big\rangle_c \Big] \Big\} \crcr
& = &
\lambda \sum_{\alpha=0,2}\Big[
-\Big\langle \varphi^{\alpha}_{\mu.} \cdot [\bar\varphi^{1}\varphi^{\bar\alpha}\bar \varphi^{3}]_{\nu.} 
\prod_{\ell =1}^n \left[\bar\varphi^{\alpha_\ell}_{[a_\ell]} \varphi^{\beta_\ell}_{[b_\ell]}  \right]   \Big\rangle_c
 +
\Big\langle \;\bar\varphi^{\alpha}_{\nu.} \cdot [\varphi^{1}\bar\varphi^{\bar\alpha}\varphi^{3}]_{\mu.} 
\prod_{\ell =1}^n \left[\bar\varphi^{\alpha_\ell}_{[a_\ell]} \varphi^{\beta_\ell}_{[b_\ell]}  \right]\Big\rangle_c  
\Big] \crcr
&+&
\bar\lambda  \sum_{\alpha=1,3}\Big[
-\Big\langle  
\; \varphi^{\alpha}_{.\mu}\cdot [\bar\varphi^{0}\bar \varphi^{2}\varphi^{\bar\alpha}]_{.\nu}
\prod_{\ell =1}^n \left[ \bar\varphi^{\alpha_\ell}_{[a_\ell]} \varphi^{\beta_\ell}_{[b_\ell]} \right]  \Big\rangle_c  
+
\Big\langle
\bar\varphi^{\alpha}_{.\nu}
\cdot[ 
\varphi^{0} \varphi^{2}\bar\varphi^{\bar\alpha}]_{.\mu}\,
\prod_{\ell =1}^n \left[\bar\varphi^{\alpha_\ell}_{[a_\ell]} \varphi^{\beta_\ell}_{[b_\ell]} \right] \Big\rangle_c  \Big],
\label{eq:wardident1act}
\eea
where $\bar \alpha=2,0$ if $\alpha =0,2$, respectively,
and $\bar\alpha=3,1$ if $\alpha=1,3$, respectively, and
in the right hand side of the equality, 
the notations explicitly mean
\bea
&&
\varphi^{0}_{\mu.} \cdot [\bar\varphi^{1}\varphi^{2}\bar \varphi^{3}]_{\nu.} :=\int_{hgh_{ij}} 
\varphi^{0}_{\mu hg} \bar\varphi^{1}_{gh_{13}h_{12} }\varphi^{2}_{h_{21} h h_{23}}\bar \varphi^{3}_{h_{32}h_{31}\nu} \;,\crcr
&& 
\varphi^{2}_{\mu.} \cdot [\bar\varphi^{1}\varphi^{0}\bar \varphi^{3}]_{\nu.} :=
\int_{hgh_{ij}} 
\varphi^{0}_{h_{03}hh_{01}} \bar\varphi^{1}_{h_{10}h_{13}\nu}
\varphi^{2}_{\mu h g}\bar \varphi^{3}_{gh_{31}h_{30}} \;,
\crcr
&&
 \varphi^{1}_{.\mu}\cdot [\bar\varphi^{0}\bar \varphi^{2}\varphi^{3}]_{.\nu} := 
\int_{hgh_{ij}} 
\bar\varphi^{0}_{h_{03}h_{02}h} \varphi^{1}_{hg\mu} 
\bar \varphi^{2}_{\nu h_{20}h_{23} } \varphi^{3}_{ h_{32}gh_{30} } 
\;,\crcr
&&
 \varphi^{3}_{.\mu}\cdot [\bar\varphi^{0}\bar \varphi^{2}\varphi^{1}]_{.\nu} :=
\int_{hgh_{ij}} 
\bar\varphi^{0}_{\nu h_{02}h_{01}} \varphi^{1}_{h_{10}gh_{12}} 
\bar \varphi^{2}_{ h_{21}h_{20}h }\varphi^{3}_{hg\mu}\;,
\label{integrals0}
\eea
and the analogous for
$\bar\varphi^{0,2}_{\nu.} \cdot [\varphi^{1}\bar\varphi^{2,0}\varphi^{3}]_{\mu.} $
and 
$
\bar\varphi^{1,3}_{.\nu}
\cdot[\varphi^{0} \varphi^{2}\bar\varphi^{3,1}]_{.\mu}$.
\end{theorem}
At the first sight, one may wonder why 
Theorem \ref{theogener} looks simpler 
than Theorem \ref{coro1action}. This is really an illusion 
because, in the second case, the symmetry constrains
much more the equality and fewer terms will survive.
From the general WT identity (\ref{eq:wardident1act}), we can derive
some more specific relations characterizing
particular graphs. 
Let us for instance discuss the case of two-point 
graphs. For this category of graphs, (\ref{eq:wardident1act})
simplifies to
\bea
&&
 \sum_{\alpha=0,2} \Big[ 
 \delta ^{\alpha_0 \alpha} \delta_{a_1\nu} \Big\langle  \bar \varphi ^{\alpha}_{\mu a_2 a_{3}}    \varphi^{\beta_0}_{[b]}
 \Big\rangle_c
-  \delta^{\alpha \beta_0}  \delta_{\mu b_{1}} \Big\langle \varphi ^{\alpha}_{\nu b_{2}b_{3}} \bar\varphi^{\alpha_0}_{[a]} 
 \Big\rangle_c\crcr
&&
- \delta ^{\alpha \beta_0} \delta_{b_{1}\nu}\Big\langle   \varphi ^{\alpha}_{\mu b_{2}b_{3}}   \bar \varphi^{\alpha_0}_{[a]}
 \Big\rangle_c
+ \delta^{\alpha \alpha_0} \delta_{\mu a_{1}} \Big\langle  \bar\varphi ^{\alpha}_{\nu a_{2} a_{3}} \varphi^{\beta_0}_{[b]} 
 \Big\rangle_c \Big]\crcr
&&
+ \sum_{\alpha=1,3}\Big[\delta ^{\alpha_0 \alpha} \delta_{a_{3}\nu}  \Big\langle  \bar \varphi ^{\alpha}_{a_{1}a_{2}\mu}     \varphi^{\beta_0}_{[b]}
 \Big\rangle_c
-  \delta^{\alpha \beta_0}\delta_{\mu b_{3}} \Big\langle \varphi ^{\alpha}_{b_{1}b_{2}\nu} \bar\varphi^{\alpha_0}_{[a]} 
 \Big\rangle_c\crcr
&&
- \delta ^{\alpha\beta_0 }\delta_{b_{3}\nu}  \Big\langle   \varphi ^{\alpha}_{b_{1}b_{2}\mu}  \bar \varphi^{\alpha_0}_{[a]}
 \Big\rangle_c
+  \delta^{\alpha \alpha_0} \delta_{\mu a_{3}} \Big\langle \bar\varphi ^{\alpha}_{a_{1}a_{2}\nu} \varphi^{\beta_0}_{[b]} 
\Big\rangle_c \Big]\crcr
& = &
- 
\lambda\,
\Big\langle \varphi^{0,2}_{\mu.} \cdot [\bar\varphi^{1}\varphi^{2,0}\bar \varphi^{3}]_{\nu.} \,
\bar\varphi^{\alpha_0}_{[a]} \varphi^{\beta_0}_{[b]}  \Big\rangle_c
-\bar\lambda\,\Big\langle  
\; \varphi^{1,3}_{.\mu}\cdot [\bar\varphi^{0}\bar \varphi^{2}\varphi^{3,1}]_{.\nu}\,
 \bar\varphi^{\alpha_0}_{[a]} \varphi^{\beta_0}_{[b]}   \Big\rangle_c  \crcr
&+&
 \lambda\,\Big\langle \;\bar\varphi^{0,2}_{\nu.} \cdot [\varphi^{1}\bar\varphi^{2,0}\varphi^{3}]_{\mu.} \,
\bar\varphi^{\alpha_0}_{[a]} \varphi^{\beta_0}_{[b]}  \Big\rangle_c  
+
\bar\lambda \,\Big\langle
\bar\varphi^{1,3}_{.\nu}
\cdot[ 
\varphi^{0} \varphi^{2}\bar\varphi^{3,1}]_{.\mu}\,
\bar\varphi^{\alpha_0}_{[a]} \varphi^{\beta_0}_{[b]}  \Big\rangle_c  
 \;.
\label{eq:wardident1act3}
\eea
where a sum is performed repeated color indices and 
on dot arguments whereas $[a]$ and $[b]$ are kept fixed.
Using a minimal symmetry, $\varphi^0 \to\; ^U \varphi^0$
and $\varphi^3 \to \; ^U\varphi^3$, which makes again
$S^{\inter} =\; ^U\, S^{\inter}$, the whole analysis 
gets simplified further. We then assume that only remains terms involving
$\alpha = 0$ and $3$, then the above WT identity can be recast in the following way
\bea
&&
 \delta ^{\alpha_0 0} \delta_{a_1\nu} \Big\langle  \bar \varphi ^{0}_{\mu a_2 a_{3}}    \varphi^{\beta_0}_{[b]}
 \Big\rangle_c
-  \delta^{0 \beta_0}  \delta_{\mu b_{1}} \Big\langle \varphi ^{0}_{\nu b_{2}b_{3}} \bar\varphi^{\alpha_0}_{[a]} 
 \Big\rangle_c
- \delta ^{0 \beta_0} \delta_{b_{1}\nu}\Big\langle   \varphi ^{0}_{\mu b_{2}b_{3}}   \bar \varphi^{\alpha_0}_{[a]}
 \Big\rangle_c
+ \delta^{0 \alpha_0} \delta_{\mu a_{1}} \Big\langle  \bar\varphi ^{0}_{\nu a_{2} a_{3}} \varphi^{\beta_0}_{[b]} 
 \Big\rangle_c \crcr
&&
+ \delta ^{\alpha_0 3} \delta_{a_{3}\nu}  \Big\langle  \bar \varphi ^{3}_{a_{1}a_{2}\mu}     \varphi^{\beta_0}_{[b]}
 \Big\rangle_c
-  \delta^{3 \beta_0}\delta_{\mu b_{3}} \Big\langle \varphi ^{3}_{b_{1}b_{2}\nu} \bar\varphi^{\alpha_0}_{[a]} 
 \Big\rangle_c
- \delta ^{3\beta_0 }\delta_{b_{3}\nu}  \Big\langle   \varphi ^{3}_{b_{1}b_{2}\mu}  \bar \varphi^{\alpha_0}_{[a]}
 \Big\rangle_c
+  \delta^{3 \alpha_0} \delta_{\mu a_{3}} \Big\langle \bar\varphi ^{3}_{a_{1}a_{2}\nu} \varphi^{\beta_0}_{[b]} 
\Big\rangle_c \crcr
& = &
- 
\lambda\,
\Big\langle \varphi^{0}_{\mu.} \cdot [\bar\varphi^{1}\varphi^{2}\bar \varphi^{3}]_{\nu.} \,
\bar\varphi^{\alpha_0}_{[a]} \varphi^{\beta_0}_{[b]}  \Big\rangle_c
-\bar\lambda\,\Big\langle  
\; \varphi^{3}_{.\mu}\cdot [\bar\varphi^{0}\bar \varphi^{2}\varphi^{1}]_{.\nu}\,
 \bar\varphi^{\alpha_0}_{[a]} \varphi^{\beta_0}_{[b]}   \Big\rangle_c  \crcr
&+&
 \lambda\,\Big\langle \;\bar\varphi^{0}_{\nu.} \cdot [\varphi^{1}\bar\varphi^{2}\varphi^{3}]_{\mu.} \,
\bar\varphi^{\alpha_0}_{[a]} \varphi^{\beta_0}_{[b]}  \Big\rangle_c  
+
\bar\lambda \,\Big\langle
\bar\varphi^{3}_{.\nu}
\cdot[ 
\varphi^{0} \varphi^{2}\bar\varphi^{1}]_{.\mu}\,
\bar\varphi^{\alpha_0}_{[a]} \varphi^{\beta_0}_{[b]}  \Big\rangle_c  
 \;.
\label{eq:wardident1act4}
\eea

\begin{figure}[h]
\begin{center}
\includegraphics[width=150mm]{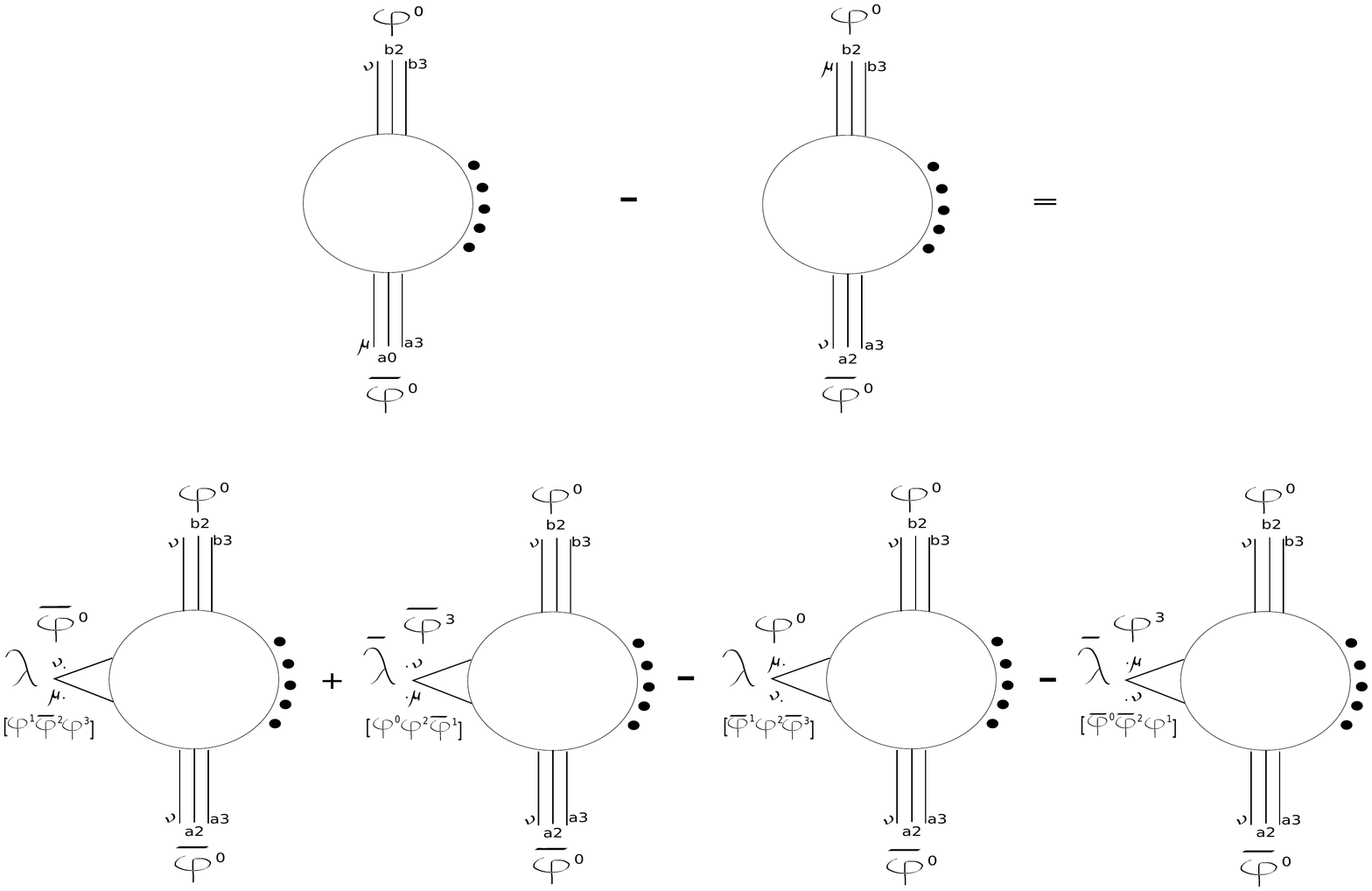}
\caption{Ward-Takahashi identity for the 1-action symmetry for colors $0$ and $3$ }
\label{figwii}
\end{center}
\end{figure}

From the fact that (see \cite{sefu2} Lemma 2.1) in a color model, 
 an even-point function with a color missing on the external legs 
has external colors appearing always in pairs $(\bar\varphi\varphi)$, 
therefore, assuming further that two-point functions
do not vanish, we can require
that $\alpha_0=\beta_0=0$.\footnote{This is without loss of generality since the case $\alpha_0=\beta_0=3$ can be inferred by symmetry and
will lead to the similar conclusion.} 
Taking into account these assumptions, the relation (\ref{eq:wardident1act4}) becomes
\bea
&&
  \delta_{a_1\nu} \Big\langle  \bar \varphi ^{0}_{\mu a_2 a_{3}}    \varphi^{0}_{[b]}
 \Big\rangle_c
-    \delta_{\mu b_{1}} \Big\langle \bar\varphi^{0}_{[a]} \varphi ^{0}_{\nu b_{2}b_{3}} 
 \Big\rangle_c
-\delta_{b_{1}\nu}\Big\langle   \bar \varphi^{0}_{[a]} \varphi ^{0}_{\mu b_{2}b_{3}}  
 \Big\rangle_c
+  \delta_{\mu a_{1}} \Big\langle  \bar\varphi ^{0}_{\nu a_{2} a_{3}} \varphi^{0}_{[b]} 
 \Big\rangle_c \crcr
& = &
- 
\lambda\,
\Big\langle \varphi^{0}_{\mu.} \cdot [\bar\varphi^{1}\varphi^{2}\bar \varphi^{3}]_{\nu.} 
\bar\varphi^{0}_{[a]} \varphi^{0}_{[b]}  \Big\rangle_c
-\bar\lambda\,\Big\langle  
\; \varphi^{3}_{.\mu}\cdot
 [\bar\varphi^{0}\bar \varphi^{2}\varphi^{1}]_{.\nu}
 \bar\varphi^{0}_{[a]} \varphi^{0}_{[b]}   \Big\rangle_c  \crcr
&+&
 \lambda\,\Big\langle \;\bar\varphi^{0}_{\nu.} \cdot [\varphi^{1}\bar\varphi^{2}\varphi^{3}]_{\mu.} 
\bar\varphi^{0}_{[a]} \varphi^{0}_{[b]}  \Big\rangle_c  
+
\bar\lambda \,\Big\langle
\bar\varphi^{3}_{.\nu}
\cdot[ 
\varphi^{0} \varphi^{2}\bar\varphi^{1}]_{.\mu}\,
\bar\varphi^{0}_{[a]} \varphi^{0}_{[b]}  \Big\rangle_c  
 \;.
\label{eq:wardident1act5}
\eea
The following cases could be studied

\noindent{\bf Case 1:} $\mu=\nu=a_1=b_1$
yielding a trivial relation
\bea
&&
 0=
- 
\lambda\,
\Big\langle \varphi^{0}_{\mu.} \cdot [\bar\varphi^{1}\varphi^{2}\bar \varphi^{3}]_{\mu.} 
\bar\varphi^{0}_{\mu a_2 a_3} \varphi^{0}_{\mu b_2 b_3}  \Big\rangle_c
-\bar\lambda\,\Big\langle  
\; \varphi^{3}_{.\mu}\cdot
 [\bar\varphi^{0}\bar \varphi^{2}\varphi^{1}]_{.\mu}
 \bar\varphi^{0}_{\mu a_2 a_3} \varphi^{0}_{\mu b_2 b_3}   \Big\rangle_c  \crcr
&+&
 \lambda\,\Big\langle \;\bar\varphi^{0}_{\mu.} \cdot [\varphi^{1}\bar\varphi^{2}\varphi^{3}]_{\mu.} 
\bar\varphi^{0}_{\mu a_2 a_3} \varphi^{0}_{\mu b_2 b_3}  \Big\rangle_c  
+
\bar\lambda \,\Big\langle
\bar\varphi^{3}_{.\mu}
\cdot[ 
\varphi^{0} \varphi^{2}\bar\varphi^{1}]_{.\mu}\,
\bar\varphi^{0}_{\mu a_2 a_3} \varphi^{0}_{\mu b_2 b_3}  \Big\rangle_c  .
\label{eq:wardident1act6}
\eea
\noindent{\bf Case 2:} $\mu\neq \nu$, and
$a_1= \nu$ and $b_1= \mu$. These assumptions lead to
\bea
&&
 \Big\langle  \bar \varphi ^{0}_{\mu a_2 a_{3}}   
 \varphi^{0}_{\mu b_1 b_2}
 \Big\rangle_c
-     \Big\langle \bar\varphi^{0}_{\nu a_2 a_{3}} \varphi ^{0}_{\nu b_{2}b_{3}} 
 \Big\rangle_c =0
\crcr
0& = &
- 
\lambda\,
\Big\langle \varphi^{0}_{\mu.} \cdot [\bar\varphi^{1}\varphi^{2}\bar \varphi^{3}]_{\nu.} 
\bar\varphi^{0}_{\nu a_2 a_{3}} \varphi^{0}_{\mu b_1 b_2}  \Big\rangle_c
-\bar\lambda\,\Big\langle  
\; \varphi^{3}_{.\mu}\cdot
 [\bar\varphi^{0}\bar \varphi^{2}\varphi^{1}]_{.\nu}
 \bar\varphi^{0}_{\nu a_2 a_{3}} \varphi^{0}_{\mu b_1 b_2}   \Big\rangle_c  \crcr
&+&
 \lambda\,\Big\langle \bar\varphi^{0}_{\nu.} \cdot [\varphi^{1}\bar\varphi^{2}\varphi^{3}]_{\mu.} 
\bar\varphi^{0}_{\nu a_2 a_{3}} \varphi^{0}_{\mu b_1 b_2}  \Big\rangle_c  
+
\bar\lambda \,\Big\langle
\bar\varphi^{3}_{.\nu}
\cdot[ 
\varphi^{0} \varphi^{2}\bar\varphi^{1}]_{.\mu}\,
\bar\varphi^{0}_{\nu a_2 a_{3}} \varphi^{0}_{\mu b_1 b_2}  \Big\rangle_c  
 \;,
\label{eq:wardident1act7}
\eea
where one should use the fact that any correlation including a beginning and an
end-point with the same index (for instance,
$\Big\langle  \bar \varphi ^{0}_{\mu a_2 a_{3}}   
 \varphi^{0}_{\mu b_1 b_2}\Big\rangle_c$ involving 
a open strand with amplitude $ \delta(\mu (\prod h) \mu^{-1})$) 
does not depend on that point  (say $\mu$). Again,
this WT identity is trivial.

\noindent{\bf Case 3:} $\mu\neq \nu$ and assume that
$a_1= b_1= \nu$, these restrictions give
\bea
&&
 \Big\langle  \bar \varphi ^{0}_{\mu a_2 a_{3}}    \varphi^{0}_{\nu b_2 b_3}
 \Big\rangle_c
-\Big\langle   \bar \varphi^{0}_{\nu a_2 a_3} \varphi ^{0}_{\mu b_{2}b_{3}}  
 \Big\rangle_c
 \crcr
& = &
- 
\lambda\,
\Big\langle \varphi^{0}_{\mu.} \cdot [\bar\varphi^{1}\varphi^{2}\bar \varphi^{3}]_{\nu.} 
\bar\varphi^{0}_{\nu a_2 a_3} \varphi^{0}_{\nu b_2 b_3}  \Big\rangle_c
-\bar\lambda\,\Big\langle  
 \varphi^{3}_{.\mu}\cdot
 [\bar\varphi^{0}\bar \varphi^{2}\varphi^{1}]_{.\nu}
 \bar\varphi^{0}_{\nu a_2 a_3} \varphi^{0}_{\nu b_2 b_3}   \Big\rangle_c  \crcr
&+&
 \lambda\,\Big\langle \bar\varphi^{0}_{\nu.} \cdot [\varphi^{1}\bar\varphi^{2}\varphi^{3}]_{\mu.} 
\bar\varphi^{0}_{\nu a_2 a_3} \varphi^{0}_{\nu b_2 b_3}  \Big\rangle_c  
+
\bar\lambda \,\Big\langle
\bar\varphi^{3}_{.\nu}
\cdot[ 
\varphi^{0} \varphi^{2}\bar\varphi^{1}]_{.\mu}\,
\bar\varphi^{0}_{\nu a_2 a_3} \varphi^{0}_{\nu b_2 b_3}  \Big\rangle_c  
\label{eq:wardident1act8}
\eea
which is a nontrivial relation.
This WT identity and the kind with more external legs have been illustrated in Figure 1 where a field is graphically 
represented by three parallel strands each of which are associated 
with a field argument.

Let us find a meaning
of the last equations (\ref{eq:wardident1act6}), 
(\ref{eq:wardident1act7}) and (\ref{eq:wardident1act8}).
The whole analysis in the paper 
has been performed by directly stating all computations at the 
quantum level by totally ignoring a possible 
kinetic part in some hypothetic classical action. 
A crucial fact is happening in the derivations
having a closer look on equations (\ref{integrals0}) and (\ref{reducfreeener1act}). 
Let us digress a little bit in order to explain the meaning of this
claim by considering instead 
a $3D$  colored GFT classical model with, now, fields projected onto the gauge
invariant sector of $L^2(G^3,\mathbb{C})$ 
as it is in ordinary assumed in literature. 
Such a GFT possesses a trivial kinetic term (and so, at the quantum level,
becomes an i.i.d. model) which 
can be written 
\bea
S^{\kin} = \sum_{a=0}^{3} \int_{[h]} 
\bar\varphi^a_{[h]} \varphi^a_{[h]}
\label{kinect}
\eea
and an interaction $S^{\inter}$ still of the form  (\ref{eq:interaction}).
Within this framework, the equations of motion for fields are
obtained by simple variation of the action.  For instance, 
the equation of $\varphi^0$ can be written as (omitting 
the $1/\sqrt{\delta^N(e)}$)
\bea
0  = \bar\varphi^0_{h_{03}h_{02}h_{01}} + 
\lambda\int_{h_{ij}}
\bar\varphi^1_{h_{10} h_{13}h_{12}} 
\varphi^2_{h_{21} h_{20}h_{23}}
\bar\varphi^3_{h_{32} h_{31}h_{30}}\;, 
\label{eom}
\eea
the rest of equations for other fields can be easily inferred.
Assuming that all these equations of motion for all
colors hold, it is striking that 
 all interaction terms perfectly cancel: 
\bea
- \lambda \;\varphi^{0,2}_{\mu.} \cdot [\bar\varphi^{1}\varphi^{2,0}\bar \varphi^{3}]_{\nu.} 
+\lambda\;
\bar\varphi^{0,2}_{\nu.}
\cdot [\varphi^{1}\bar\varphi^{2,0} \varphi^{3}]_{\mu.}
= - \lambda \;\varphi^{0,2}_{\mu.} \cdot \bar\varphi^{0,2}_{\nu.} +\lambda\;
\bar\varphi^{0,2}_{\nu.}
\cdot \varphi^{0,2}_{\mu.}
 = 0\;,
\eea
so that (\ref{reducfreeener1act}) reduces to simple contributions
with sources. Under this light, equations (\ref{eq:wardident1act6})
and (\ref{eq:wardident1act7}) are of course totally trivial whereas 
(\ref{eq:wardident1act8}) just teaches us that the full two-point function
is symmetric under strand index permutation $\mu \leftrightarrow \nu$
which is true for an i.i.d model since the
strand index should be conserved along an open face of amplitude $\delta(\mu\nu^{-1})$. 
Hence, the non trivial relation
(\ref{eq:wardident1act8}) corresponds to a quantum 
version of the classical equation of motion (\ref{eom}) living
in the gauge projected theory defined by (\ref{kinect}). It is at least in that 
sense that one recovers what is ordinary expected
from the point of view of the WT identity formalism. 
From this point,
we may reasonably think that the identities (\ref{eq:wardident1act8})
and more generally the non trivial contributions of 
(\ref{eq:wardident1act}) encode 
quantum modified versions of equations of 
motion for colored fields 
together with a weaker version of the conservation of 
some conserved quantities (in the gauge invariant formulation). 
This last statement is the purpose
of a current investigation \cite{Geloun}.

\section{Conclusion}

Let us recapitulate. 
First, the notion of unitary gauge field symmetry 
in (colored or not) tensor theories has been 
identified: it amounts to 
multiply the modes of the tensor field expansion via Peter-Weyl theorem
by unitary matrices, and if the field possess a color (vector) index
then this index could also undergo a transformation. 
 Then, using this generic unitary field transformation, 
we have succeeded in finding an extension of the ordinary 
derivation of WT identities adapted in the present tensor situation
and in computing,  starting from the two-point and four-point functions,
WT identities for any even-point functions
of the colored Boulatov model. These relations have been
proved to be more general than the one obtained in another
context of non identically distributed (with invertible covariance) 
matrix models. Moreover, we show that there exists a particular class of operators (called right invariant unitary operators) under which the
  gauge invariance property of fields is preserved and the
colored GFT interaction becomes invariant. This class of operators has allowed to refine  the formalism and to identify WT identities associated with this symmetry with a nontrivial content. A main result of this
paper is that, in a colored theory, a combination of 
two-point functions can be expanded versus a peculiar class 
of six-point functions with a strand insertion. We reinterpret
these WT identities as quantum versions of equations of motion
of a GFT  fully projected onto its gauge invariant sector.
 Let us mention
that, although it is not excluded at all that 
one can define equivalent unitary transformations
for non colored fields being still a symmetry of 
the interaction in non colored GFTs (for instance by using a similar prescription as in \cite{Geloun:2011cz}), the case of non colored 
GFT will inexorably lead to far more complicated derivations 
for extracting WT identities with similar conclusions. 

This work is, as it must be acknowledged,
a prime sketch and mostly technical journey on
quantum aspects of (unitary) symmetry on correlation functions
in (non identically distributed) tensor models. 
Hence, it certainly deserves to be qualitatively refined by future prospects.
Indeed, we obtain for a tensor theory, 
both the generic and for the so-called 1-action, 
various kind of identities of which the content
remains to be clarified and, if possible, classified. 
The difficulty concerning GFT colored models here resides
 in the richness brought by 
the combination of the tensor structure and the gauge projector.
One has to systematically check the true content
of the bulk of the identities fulfilled by $N$-point functions. 
 Diagrammatics associated to each nontrivial relations 
should be also determined.  More relations 
between these identities and quantum equations
of motion and conservation of Noether
quantities associated with the interaction symmetry
are under investigations \cite{Geloun}.
It can be underlined that classical symmetries
(translations and dilatations)
have been recently investigated 
by introducing Laplacian dynamics in GFTs \cite{Geloun:2011cz}.
It has been shown that translation invariance for colored
GFTs leads to a conserved quantity associated with the
energy momentum tensor. 
It is therefore natural to ask in this dynamical context  
the same questions addressed in the ordinary case, namely 
what are the form and meaning of the WT identities
for these symmetries and their relation to the
 Noether currents ?

Another problem that one could address is 
whether or not WT identities give 
a relation between any (or specific) kind of $N$-point functions
 in terms of $N+4$
starting from the simplest colored situation of the
1-action minimal symmetry. 
In fact, this question is of relevance in attempts to solve the theory 
in the sense that to give explicit expressions for correlation functions.
Yet another way to think about this problem,
it to use other ingredients such as Schwinger-Dyson equations, so that WT identities might provide a
closed integral equation for $N$-point 
correlation functions \cite{Grosse:2009pa}.
In this latter work, the author have successfully found 
 integral equations for the full renormalized planar 
connected two and  four-point functions (the first admits perturbative solutions). 
Thus, a farther stage is to exploit the structure of 
Schwinger-Dyson in tensor models \cite{Gurau:2011xq} in 
order to get a closed form equations for a particular $N$-point
functions in the colored theory. 
Hence, the analysis performed here could be useful for both perturbative and nonperturbative renormalization.

\section*{Acknowledgements}
The author is indebted to Razvan Gurau, Valentin 
Bonzom and Tim Koslowski 
for helpful discussions at various stages of this work. 
Research at Perimeter Institute is supported by the Government of Canada through Industry 
Canada and by the Province of Ontario through the Ministry of Research and Innovation.

\section*{Appendix}
\appendix

\renewcommand{\theequation}{\Alph{section}.\arabic{equation}}
\setcounter{equation}{0}

\section{Gaussian integration for tensor models}
\label{app:gauss}

In this appendix, a series of lemmas are introduced. These pertain to the properties of the GFT Gaussian measure and are extensively used in the text.  
We will denote the fields $\varphi_{[h]} = \varphi(h_1,h_2,h_3)$,
and covariance associated with Feynman Gaussian measure as
$C_{[g][h]} = C_{g_1g_2g_3;\;h_1h_2h_3}$ and 
for all field, $C_{[g][h]} \varphi_{[h]} := \int_{[h]} C_{[g][h]} \varphi_{[h]}$. In a colored theory, fields are equipped with
an extra index $\varphi^i_{[h]}$ and the covariance
reads $C^{ij}_{[g][h]}=\delta^{ij} C_{[g][h]}$. 
The subsequent analysis admits a straightforward 
generalization in any GFT dimension. 

\begin{definition}
 A Gaussian measure of covariance $C$ is defined by its non zero correlations
  \bea
    \int d\mu_C (\bar \varphi,\varphi)\; \varphi_{[a_1]} \dots  \varphi_{[a_n]}  \bar \varphi_{[b_1]} \dots \bar \varphi_{[b_n]} = 
       \sum_{\pi } \prod_{i=1}^n C_{[a_i] [b_{\pi(i)}] }   \; ,
\label{eq:Wick}
  \eea
   where the sum is taken over all permutations $\pi$ 
of $n$ elements. For a colored theory, setting
 $d\mu_C (\bar \varphi^\ell,\varphi^\ell)=d\mu_C (\bar \varphi,\varphi)$, we have
\bea
  \int d\mu_C (\bar \varphi,\varphi)\; \varphi^{k_1}_{[a_1]} \dots  \varphi^{k_1}_{[a_n]}  \bar \varphi^{j_1}_{[b_1]} \dots \bar \varphi^{j_n}_{[b_n]} = 
       \sum_{\pi } \prod_{i=1}^n C^{k_i j_{\pi(i)}}_{[a_i][ b_{\pi(i)}] }   \; .
\label{eq:Wickcolor}
\eea
\end{definition}
In the following, the developments hold in general,
i.e. without colors using the definition (\ref{eq:Wick}).
However being interested in colored theory, we will
give the corresponding result in that particular instance. 
In order to alleviate notations, 
the source term will be denoted as
\bea
\bar\eta\varphi + \bar\varphi \eta := 
 \int_{[h_i]} (\bar\eta_{h_1h_2h_3}\,\varphi_{h_1h_2h_3} + \bar\varphi_{h_1h_2,h_3} \,\eta_{h_1h_2h_3})\;.
\eea
 \begin{lemma}
\label{lem:integsour}
We have
 \bea 
   \int d\mu_C (\bar \varphi,\varphi)\; e^{\bar\varphi \eta+ \bar\eta \varphi} = e^{\bar \eta C \eta} \;. 
 \eea
\end{lemma}
{\bf Proof:} By direct evaluation using the Wick theorem (\ref{eq:Wick}), we obtain
 \bea
  \int d\mu_C (\bar \varphi,\varphi)\; e^{\bar\varphi \eta+ \bar\eta \varphi}  = 
   \sum_{n=1}^{\infty} \frac{1}{ ( n!)^2} \sum_{\pi } \prod_{i=1}^n \int_{[a_i][b_{\pi(i)}]} 
    \bar \eta_{[a_i]} C_{[a_i] [b_{\pi(i)} ]} \eta_{[b_{\pi(i)}] } =  \sum_{n=1}^{\infty} \frac{1}{  n!} 
    \bigl[ \bar \eta C  \eta  \bigr]^n\;.
 \eea
\qed
\begin{lemma}\label{lem:parts}[Integration by parts] 
 Introducing the functional derivative 
$\delta_{\varphi_{[h]}} (\cdot) = \delta (\cdot)/\delta \varphi_{[h]} $
for any functional $H(\varphi,\bar\varphi)$, we have
 \bea
   \int d\mu_C (\bar \varphi, \varphi) \;  \Big{(} \varphi_{[a]} H(\bar\varphi,\varphi) - C_{[a][b]} \delta_{\bar \varphi_{[b]}} H(\bar\varphi,\varphi) \Big{)} =0\;.
 \eea
\end{lemma}
{\bf Proof:} This relation can be shown by first introducing
source terms. We rewrite 
\bea
  && \int d\mu_C (\bar \varphi, \varphi) \;  \Big{(}  \varphi_{[a]} H(\bar \varphi,\varphi) - 
      C_{[a][b]} \delta_{\bar \varphi_{[b]}} H(\bar \varphi, \varphi) \Big{)} \crcr
   &&= 
   \int d\mu_C (\bar \varphi, \varphi) \;  \Big{[}  \varphi_{[a]} H(\delta_{\eta}, \delta_{\bar \eta}) 
         e^{\bar\varphi \eta+ \bar\eta \varphi } |_{\bar\eta=\eta=0} - C_{[a][b]} \delta_{\bar \varphi_{[b]}} 
         \Big{(} H(\delta_{\eta}, \delta_{\bar \eta}) 
         e^{\bar\varphi \eta+ \bar\eta \varphi } |_{\bar\eta=\eta=0}  \Big{)} \Big{]} \crcr
  && = H(\delta_{\eta}, \delta_{\bar \eta})
      \Big{[} \int d\mu_C (\bar \varphi, \varphi) \;  \Big{(} \varphi_{[a]} - C_{[a][b]} \eta_{[b]} \Big{)} 
                          e^{\bar\varphi \eta+ \bar\eta \varphi } \Big{]} 
      \Big{|}_{\bar\eta=\eta=0}\;.
\eea
The latter expression can be calculated using Lemma \ref{lem:integsour} and the fact that number of fields $\varphi$
and $\bar\varphi$ should be the same in order
to give a nonvanishing amplitude:
\bea
&& \int d\mu_C (\bar \varphi, \varphi) \;  \Big{(}  \varphi_{[a]} - C_{[a][b]} \eta_{[b]} \Big{)} e^{\bar\varphi \eta+ \bar\eta \varphi } =\crcr
&&
 \int d\mu_C (\bar \varphi, \varphi) \;  \Big{(}  \varphi_{[a]} \sum_n \frac{1}{(n+1)!n!} 
(\bar\varphi \eta)^{n+1} (\bar\eta \varphi)^{n} - C_{[a][b]} \eta_{[b]}  \sum_n \frac{1}{(n!)^2} (\bar\varphi \eta)^n (\bar\eta \varphi)^n \Big{)}
\crcr
&&= \sum_n \frac{1}{(n!)^2} C_{[a][b]}  \eta_{[b]}  \; n! \; (\bar\eta C\eta)^{n} - C_{[a][b]} \eta_{[b]}  \sum_n \frac{1}{(n!)^2}\; n! \; (\bar\eta C\eta)^{n} =0\;.
\eea
\qed

Lemma \ref{lem:parts} can be translated in terms of colored
fields as
\bea
  \int d\mu_C (\bar \varphi, \varphi) \;  \Big{(} \varphi^i_{[a]} H(\bar\varphi,\varphi) - C^{ij}_{[a][b]} \delta_{\bar \varphi^j_{[b]}} H(\bar\varphi,\varphi) \Big{)} =0\;.
\eea
Indeed, following step by step the previous proof, we have
\bea
&& \int d\mu_C (\bar \varphi, \varphi) \;  \Big{(}  \varphi^i_{[a]} - C^{ij}_{[a][b]} \eta^j_{[b]} \Big{)} e^{\bar\varphi^j \eta^j+ \bar\eta^j \varphi^j } \prod_{k\neq j} \;e^{\bar\varphi^k \eta^k+ \bar\eta^k \varphi^k } =\crcr
&&
 \int d\mu_C (\bar \varphi, \varphi)  \;  \Big{(}  \varphi^i_{[a]} \sum_n \frac{1}{(n+1)!n!} 
(\bar\varphi^j \eta^j)^{n+1} (\bar\eta^j \varphi^j)^{n} - C^{ij}_{[a][b]} \eta^j_{[b]}  \sum_n \frac{1}{(n!)^2} (\bar\varphi^j \eta^j)^n (\bar\eta^j \varphi^j)^n \Big{)} \prod_{k\neq j} e^{\bar\varphi^k \eta^k+ \bar\eta^k \varphi^k } 
\crcr
&&= \int d\mu_C (\bar \varphi', \varphi')   \Big{(} \sum_n \frac{1}{(n!)^2} C^{ij}_{[a][b]}  \eta^j_{[b]}  \; n! \; (\bar\eta C\eta)^{n} - C^{ij}_{[a][b]} \eta^j_{[b]}  \sum_n \frac{1}{(n!)^2}\; n! \; (\bar\eta C\eta)^{n} =0\Big{)} \prod_{k\neq j} \; e^{\bar\varphi^k \eta^k+ \bar\eta^k \varphi^k } \;.
\eea

\begin{lemma} \label{lem:sum}
Modifying the covariance $C$ for $C+A$,
the following relation holds for any functional $H(\bar\varphi,\varphi)$:
 \bea
   \int d\mu_{C+A} ( \bar\varphi, \varphi) H ( \bar\varphi,\varphi) = \int d\mu_C (\bar\varphi, \varphi)
      \; e^{\delta_{\varphi} A \delta_{\bar \varphi}} H( \bar\varphi,\varphi)\;.
 \eea
\end{lemma}
{\bf Proof:} First, one performs the expansion using again the source
term and uses Lemma \ref{lem:integsour} in order to obtain
 \bea
   && H(\delta_{\eta}, \delta_{\bar \eta}) \Big{[}\int d\mu_{C+A} (\bar \varphi, \varphi) e^{\bar\varphi \eta+ \bar\eta \varphi } \Big{]} 
      \Big{|}_{\bar\eta=\eta=0} = 
     H(\delta_{\eta}, \delta_{\bar \eta}) \Big{[} e^{\bar \eta C \eta} e^{\bar \eta A \eta} \Big{|}_{\bar\eta=\eta=0} \crcr 
   &&=
      H(\delta_{\eta}, \delta_{\bar \eta}) \Big{[} \int d\mu_{C} (\bar \varphi, \varphi)  e^{\bar \eta A \eta}
        e^{\bar\varphi \eta+ \bar\eta \varphi } \Big{]}\Big{|}_{\bar\eta=\eta=0} =\int d\mu_{C} (\bar \varphi, \varphi) \;  e^{\delta_{\varphi} A \delta_{\bar \varphi}} H(\delta_{\eta}, \delta_{\bar \eta}) 
      \Big{[}  e^{\bar\varphi \eta+ \bar\eta \varphi } \Big{]}\Big{|}_{\bar\eta=\eta=0} \;
\eea
which is the desired result.  
\qed

\section{Unitary transformations}
\label{sect:unittransf}

\subsection{Left/Right invariant unitary operators}
\label{sect:lrinvunit}

Let us recall first some basic elements of $G=SU(2)$ representation theory.
A Wigner matrix element of an $SU(2)$ group 
element $g$  in the representation $j$  will be denoted by $D^j_{mn}(g)$. 
Note the properties of these representation matrices given by
$
 \bar D^j_{mn}(g) =D^j_{nm}(g^{-1})= (-)^{m-n} D^j_{-m-n}(g) $
and $\int dg\, D^j_{mn}(g) \bar D^{j'}_{m'n'}(g) = (1/d_j) \delta^{jj'} \delta_{mm'}\delta_{nn'}$, where $d_j =2j+1$. In the following, we will use the symbol $\int_h :=\int dh$ for denoting the Haar integral with respect to the variable $h$. 
Dumb sums like  $\sum$ without specifying the arguments 
mean that these sums are performed over all repeated discrete variables.

Any function of one variable over $G$ can be expanded  in
modes via Peter-Weyl representation theorem as 
\bea
 f(g) =\sum \sqrt{d_j} f^j_{mn} D^{j}_{mn}(g):= \sum_{j,m,n} \sqrt{d_j} f^j_{mn} D^{j}_{mn}(g) \;, \qquad f^{j}_{mn} = \sqrt{d_j} \int_g f(g) \bar D^j_{mn}(g)\;.
\eea
An operator over the one variable functions is defined by a kernel
\bea
 U(\alpha,\beta) = \sum \sqrt{d_{j_1}d_{j_2}} 
U^{j_1j_2}_{m_1n_1\; m_2n_2} D^{j_1}_{m_1n_1}(\alpha) \bar D^{j_2}_{m_2n_2}(\beta)\;, \quad \forall \alpha,\beta \in G\;.
\eea
The normalization is justified by the fact that
\bea
 U(f) (\alpha) &=& \int_{h} U(\alpha, h) f(h) = 
\sum \sqrt{d_{j_1} d_{j_2}} 
U^{j_1j_2}_{m_1n_1\; m_2n_2} D^{j_1}_{m_1n_1}(\alpha) 
\sum \sqrt{d_j} f^j_{mn} \int_h \bar D^{j_2}_{m_2n_2}(h) D^{j}_{mn}(h)\nonumber\\
&=&\sum \sqrt{d_{j_1}} \Big{[}U^{j_1j_2}_{m_1 n_1\;m_2n_2}\, f^{j_2}_{m_2n_2} \Big{]} D^{j_1}_{m_1n_1}(\alpha)\;.
\eea
For instance, the identity operator $\I$ possesses the tensor components  
$
 \I^{j_1j_2}_{m_1 n_1\;m_2 n_2} = \delta^{j_1j_2} \delta_{m_1 m_2} \delta_{n_1n_2}
$ and the kernel
\bea
 \I(\alpha,\beta)= \sum \sqrt{d_{j_1} d_{j_2}} \delta^{j_1j_2} \delta_{m_1 m_2} \delta_{n_1n_2}
 D^{j_1}_{m_1n_1}(\alpha) \bar D^{j_2}_{m_2n_2}(\beta) 
= \sum_j d_j D^{j}_{mn}(\alpha) \bar D^{j}_{mn}(\beta) = 
\sum_j d_j \chi^j(\alpha \beta^{-1}) = 
\delta(\alpha\beta^{-1}) \;.
\nonumber
\eea
 Note that we introduce
the symbol $\chi^j(g):= \sum_m D^j_{mm}(g)$ denoting
the character of the group element $g$  in the representation $j$.

The adjoint of an operator $U$ is denoted by $U^\dag$
and its kernel is defined by $[U^\dag](\alpha,\beta) = \overline{U(\beta,\alpha)}$.
Hence, an operator  $U$ is unitary if the following relation holds
\bea
[U^\dag U](\alpha,\beta) =
 [U U^\dag](\alpha,\beta) = \int_h U(\alpha, h) \overline{U( \beta, h)} = \I(\alpha,\beta)=\delta(\alpha \beta^{-1})\;.
\eea
Furthermore, by noting that
\bea 
&&\sum \sqrt{d_{j_1}d_{j_2}}  U^{j_1j_2}_{m_1 n_1\;m_2 n_2}  D^{j_1}_{m_1n_1}(\alpha)
 \sqrt{d_{j'_1} d_{j'_2}} \; \overline{U^{j'_1j'_2}_{m'_1n'_1\;m'_2 n'_2}} \bar D^{j'_1}_{m'_1n'_1}(\beta) 
\int_h\, \bar D^{j_2}_{m_2n_2}(h) D^{j'_2}_{m'_2n'_2}(h) \crcr
&&= \sum \sqrt{d_{j_1} d_{j'_1}} 
\Big{[} U^{j_1j_2}_{m_1n_1\;m_2 n_2}   \overline{U^{j'_1j_2}_{m'_1n'_1\;m_2 n_2}}  \Big{]}
D^{j_1}_{m_1n_1}(\alpha) \bar D^{j'_1}_{m'_1n'_1}(\beta)\;,
\eea
 as expected, an operator is unitary if and only if
\bea
 \sum_{j_2,m_2,n_2} 
U^{j_1j_2}_{m_1n_1\; m_2n_2}   \overline{U^{j'_1j_2}_{m'_1 n'_1\;m_2n_2}} = \delta^{j_1j_1'} \delta_{m_1m_1'}\delta_{n_1n_1'}
\;.
\eea

Among the unitary operators there exists a special class of unitaries, those invariant under left group action
\bea
\forall p, \alpha,\beta \in G\;, \qquad
A(p\alpha,p\beta)=A(\alpha,\beta) \;,\qquad \int_h A(\alpha,h) \overline{A(\beta, h)} = \delta(\alpha^{-1}\beta)\;.
\eea
Due to that invariance, we have 
\bea 
A(\alpha,\beta)&=& \int_h A(h\alpha, h\beta)
=
\sum \sqrt{d_{j_1}d_{j_2}} 
U^{j_1j_2}_{m_1m_2\; n_1n_2} 
D^{j_1}_{k_1n_1}(\alpha) \bar D^{j_2}_{k_2n_2}(\beta)
\int_h D^{j_1}_{m_1k_1}(h) \bar D^{j_2}_{m_2k_2}(h) \crcr
&&=\sum U^{j_1j_2}_{m_1m_2\; n_1n_2} 
\bar D^{j_1}_{n_1k_1}(\alpha^{-1}) \bar D^{j_2}_{k_2n_2}(\beta)
\delta^{j_1j_2} \delta_{m_1m_2} \delta_{k_1k_2} \crcr
&&
 =\sum 
\Big{[}\sum_{m_1}U^{j_1j_1}_{m_1m_1\; n_1n_2} \Big{]}
\bar D^{j_1}_{n_1n_2}(\alpha^{-1}\beta ) \equiv \sum d_j A^j_{pq}  D^{j}_{pq}(\alpha^{-1}\beta )
\eea
where we read off 
$
A^j_{pq}=[(-)^{p-q}/d_j] \sum_m U^{jj}_{mm\; -p-q}
$.
Imposing unitarity on these operators yields
\bea 
&& \int_h A(\alpha, h) \overline{A( \beta, h)} = 
\sum d_j d_{j'} A^j_{pq} \overline{A^{j'}_{p'q'}} D^j_{pr}(\alpha^{-1}) \bar D^{j'}_{p'r'}(\beta^{-1})
\int_{h} D^j_{rq}(h) \bar D^{j'}_{r'q'}(h) \;,\crcr
&&=\sum d_j \Big{[}\sum_q A^j_{pq} \overline{A^{j}_{p'q}}\Big{]} D^j_{pr}(\alpha^{-1}) \bar D^{j}_{p'r}(\beta^{-1})
=\sum d_j \Big{[}\sum_q A^j_{pq} \overline{A^{j}_{p'q}}\Big{]} D^j_{pp'}(\alpha^{-1}\beta)  \;.
\eea
Therefore the invariant unitary operators are represented by unitary matrices in each dimension $d_j$ of the representation
\bea
 \sum_q A^j_{pq} \overline{A^{j}_{p'q}} = \delta_{pp'}\;.
\label{lefinv}
\eea
We call such an $A$ as a \emph{ left invariant unitary}.
For \emph{right invariant unitary} operator, a similar definition can be given and a little computation leads  to
\bea
A(\alpha,\beta)&&= \int_h A(\alpha h, \beta h)= 
\sum 
\Big{[}\sum_{n}U^{j_1j_1}_{m_1m_2\; nn} \Big{]}
 D^{j_1}_{m_1m_2}(\alpha\beta^{-1} ) \equiv \sum d_j A^j_{pq}  D^{j}_{pq}(\alpha\beta ^{-1}) \;,
\eea
with
$
  A^j_{pq} := (1/d_j) \sum_{n}U^{jj}_{pq\; nn}
$.
Further imposing unitarity yields some conditions on the coefficients $A_{pq}^j$:
\bea
&&
\int_h A(\alpha,h)\overline{A(\beta,h)}  = 
\sum d_j d_{j'}  A^j_{pq}  \overline{A^{j'}_{p'q'}} 
\int_{h}  D^{j}_{pq}(\alpha h^{-1})  \overline{D^{j}_{p'q'}(\beta h^{-1})} \crcr
&& = \sum d_j   A^j_{pq}  \overline{A^{j}_{p'q}}  D^{j}_{pr}(\alpha)
\overline{D}^{j}_{p'r}(\beta) = 
\sum d_j [\sum_q A^j_{pq}  \overline{A^{j}_{p'q}}]  D^{j}_{pp'}(\alpha \beta^{-1})  \;,\eea
so that 
$\sum_q A^j_{pq} \overline{A^{j}_{p'q}} = \delta_{pp'}$
which is a similar  to left invariant unitary 
condition (\ref{lefinv}).
However, assuming that we impose that the
following is unitary
\bea
\int_h A(h,\alpha)\overline{A(h,\beta)}  =
\sum d_j [\sum_p A^j_{pq}  \overline{A^{j}_{pq'}}]  D^{j}_{q'q}( \beta^{-1}\alpha)  \;,\eea
one could get another condition on the $A^j_{pq}$'s that is
\bea
\sum_p A^j_{pq}  \overline{A^{j}_{pq'}} = \delta_{qq'}\;.
\eea

\subsection{Unitary transformation of fields}
\label{sect:leftinvunit}

\noindent{\bf 1-action on fields -} Consider the right invariant unitary operator as detailed in Appendix \ref{sect:lrinvunit} which is of the
form
\bea
 A(g,h) = \sum\; d_j A^j_{mn} D^{j}_{mn}(gh^{-1})\;.
\eea
The condition 
$
\sum_{n} A^j_{mn} \bar{A}^j_{m'n} =  \delta_{mm'},
$
ensures that, for all $j$, $A^j$ is a unitary matrix of rank $2j+1$, i.e. $A^j \in U(2j+1)$.

Given a $D$ dimensional GFT,
the 1-action of $A$ on a field means that the said field transforms 
with respect to its first argument, namely
\bea
^A \varphi(g_1,g_2,\dots, g_D) =\int_h\; A(g_1,h) \varphi(h,g_2,\dots,g_D)\;,\quad
^A   \bar\varphi( g_1,g_2,\dots, g_D) = \int_h\;  \overline{A(g_1,h)}\overline{\varphi(h,g_2,\dots, g_D)} \;,
\eea
this is, using the mode expansion (and equivalent
in term of matrices and tensors), 
\bea
&&
^A  \varphi(g_1,g_2,\dots, g_D) =\sum d_j  \sqrt{d_{j_1}} 
A^j_{pq}  \varphi^{j_1 j_i}_{m_1n_1\; m_{i}n_i}
\int_h  D^j_{pq}(g_1h^{-1}) D^{j_1}_{m_1n_1}(h) \prod_{i\neq 1} 
\sqrt{d_{j_i}} D^{j_i}_{m_in_i}(g_i) \crcr
&&   = \sum   
A^{j_1}_{m_1q}  \varphi^{j_1 j_i}_{qn_1\; m_{i}n_i}\; \sqrt{d_{j_1}} D^{j_1}_{m_1n_1}(g_1)
\prod_{i\neq 1} [
\sqrt{d_{j_i}} D^{j_i}_{m_in_i}(g_i)]  \;,
\cr\cr
&&
^A   \bar\varphi( g_1,g_2,\dots,g_D) =\sum d_j  \sqrt{d_{j_1}} 
\bar A^j_{pq}  \bar\varphi^{ j_ij_1}_{m_1n_1\;m_{i}n_i}
\int_h  \bar D^j_{pq}(g_1h^{-1}) \bar D^{j_1}_{m_1n_1}(h) \prod_{i\neq 1} 
\sqrt{d_{j_i}} \bar D^{j_i}_{m_in_i}(g_i) \crcr
&&   = \sum   
\bar A^{j_1}_{m_1q}  \bar \varphi^{j_ij_1 }_{qn_1\;m_{i}n_i}\; \sqrt{d_{j_1}} \bar D^{j_1}_{m_1n_1}(g_1)
\prod_{i\neq 1} [
\sqrt{d_{j_i}} \bar D^{j_i}_{m_in_i}(g_i)] \;.
\eea
Hence, the modes of the transformed field $^A \varphi$ 
can be related to the modes of the prime field as
\bea
^A \varphi^{j_1j_i}_{m_1n_1\; m_in_i} 
= \sum_p  A^{j_1}_{m_1p}  \varphi^{j_1 j_i}_{pn_1\; m_{i}n_i}\;,
\qquad
^A  \bar\varphi^{j_ij_1}_{m_in_i\;m_1n_1}
= \sum_p  \bar A^{j_1}_{m_1p}  \bar\varphi^{j_ij_1 }_{pn_1\;m_{i}n_i }\;,
\eea
with the notable feature that only the first set of labels coined by $1$, $j_1,m_1$
and $n_1$, is actually involved under this transformation. 
Returning to the group formulation, this field transformation will
be referred to the equivalent forms when no possible confusion may occur
\bea
  A_{gh} \varphi_{h.} : =  \ ^A \varphi_{g.} = \int_h A(g,h)\varphi(h,(.)) \;,\qquad
\bar\varphi_{h.} A^{-1}_{hg} : =  \ ^A \bar\varphi_{g.} = \int_h \overline{A(g,h)}\overline{\varphi(h,(.))} \;.
\label{eq:notlefinv}
\eea

\medskip

\noindent{\bf Infinitesimal transformation -}
Given a right invariant unitary $A$, its component $A^j \in U(d_j)$, and
therefore there exists $(B^j)^\dag = B^j$ a Hermitian matrix
of the same dimension $d_j$, such that
\bea
&&
A^{j}_{mn} =  \delta^{j}_{mn} + \imath B^j_{mn}  \;, \qquad
\bar B^{j}_{nm} = B^{j}_{mn}\;.
\eea
We can expand $A$ infinitesimally at first order in $B$:  
\bea
A(h,g) \simeq   \sum_j d_j\sum_{mn} (\delta^{j}_{mn} + \imath B^j_{mn}) D^{j}_{mn}(h^{-1}g) 
 =  \delta(gh^{-1}) +\imath B(g,h) \;.
\label{infinit}
\eea
$B$ is a Hermitian kernel in the sense that
\bea
\overline{B(h,g)} =  \sum_j d_j \sum_{mn} \bar B^j_{mn} \bar D^{j}_{mn}(hg^{-1}) 
 =  \sum_j d_j \sum_{mn} B^j_{nm} D^{j}_{nm}(gh^{-1})  = B(g,h)\;.
\label{hermit}
\eea

\section{Calculation of infinitesimal variations under unitary transformations}
\label{app:calcinfvar}

\subsection{General unitary transformation}
\label{app:inftransgener}

We start by considering a general unitary operator $U$
which satisfies  $U^{ij}_{[a][b]}  \bar U^{kj}_{[c][b]} =\delta^{ik} \delta_{[a][c]}$ where  $[a] := (a_1,a_2,a_3)$, and $\delta_{[a][b]}$ stands for
the kernel of the unit operator identifying each field arguments.
 Appendix \ref{sect:leftinvunit} provides 
a particular type of this unitary operator of the form
$U^{jk} = \delta^{jk} A \otimes \I \otimes \I$ that we will discuss 
in detail in the next subsection. In this appendix, we assume a 
formal and general expression for this operator and infer 
the infinitesimal variations for the action, the covariance
and source term.

We assume that,  under $U$,  the colored fields $\varphi^i$ transform as
\bea
^U\,  \varphi^i_{[a]} = U^{ij}_{[a][b]} \varphi^j_{[b]} \;, \qquad
^U\, \bar \varphi^i_{[a]} = \bar\varphi^j_{[b]} \bar U^{ij}_{[a][b]}\;, \qquad
\frac{\delta}{\delta \,  ^U \varphi^i_{[a]}} = \frac{\delta}{\delta \varphi^j_{[b]}} (U^{-1})^{ji}_{[b][a]}\;, \qquad 
\frac{\delta}{\delta\,  ^U \bar\varphi^i_{[a]}} = U^{ij}_{[a][b]}\frac{\delta}{\delta \bar \varphi^j_{[b]}}\ \;.
\label{eq:generunita}
\eea 
A sum (integration on arguments and discrete sum on colors) is understood over all repeated indices. 
This transformation
therefore mixes both colors and group arguments of the fields. 

$S^{\inter}$ becomes after this field transformation
\bea
^U \, S^{\inter}
& =&\frac{\lambda}{\sqrt{\delta^N(e)}} \int_{h_{ij}}\,\int_{[a][b][c][d]}
U^{0i_0}_{h_{03}h_{02}h_{01}\,[a]} \varphi^{i_0}_{[a]} \,
 \bar \varphi^{i_1}_{[b]} \bar U^{1i_1}_{h_{10}h_{13}h_{12}\;[b]} \,  U^{2i_2}_{h_{21}h_{20}h_{23}\,[c]}\,  \varphi^{i_2}_{[c]} \,
\bar \varphi^{i_3}_{[d]} \bar U^{3i_3}_{h_{32}h_{31}h_{30}\,[d]}\crcr
&&+\frac{\bar\lambda}{\sqrt{\delta^N(e)}} 
\int_{h_{ij}}\,\int_{[a][b][c][d]} \ 
\bar \varphi^{i_0}_{[a]}  \bar U^{0i_0}_{h^{03}h^{02}h^{01}[a]} 
  \varphi^{i_1}_{[b]} U^{1i_1}_{h^{10}h^{13}h^{12} [b]}\,
\bar \varphi^{i_2}_{[c]}   \bar U^{2i_2}_{h^{21}h^{20}h^{23}\,[c]}  \, 
 U^{3i_3}_{h^{32}h^{31}h^{30}\,[d]} \varphi^{i_3}_{[d]}  \; .
\eea 
Expanding the unitary operator around the identity, one has $U^{ij}_{[a][b]} =\delta^{ij} \delta_{[a][b]}+ \imath B^{ij}_{[a][b]}$, where $B$ is a Hermitian kernel i.e. ought to satisfy $\bar B^{ij}_{[b][a]} = B^{ji}_{[a][b]}$. At first order in $B$, the variation of the interaction part can be computed as follows: 
\bea
\delta_B S^{\inter} 
&=&
\frac{ \imath\lambda}{\sqrt{\delta^N(e)}} \int_{h_{ij}}\,\Big{[} 
-\int_{[d]}  B^{i_33}_{[d][h_{3i}]} \varphi^0_{[h_{0i}]} \,\bar \varphi^1_{[h_{1i}]}   \varphi^2_{[h_{2i}]} \,\bar \varphi^{i_3}_{[d]}
+
\int_{[a]}  B^{0i_0}_{[h_{0i}][a]} \varphi^{i_0}_{[a]} \,\bar \varphi^1_{[h_{1i}]}   \varphi^2_{[h_{2i}]} \,\bar \varphi^3_{[h_{3i}]}
   \crcr
&&\, 
-\int_{[b]} B^{i_11}_{[b][h_{1i}]} \varphi^0_{[h_{0i}]} \,\bar \varphi^{i_1}_{[b]}  \varphi^2_{[h_{2i}]} \,\bar \varphi^{3}_{[h_{3i}]} 
+ 
\int_{c}
B^{2i_2}_{[h_{2i}][c]} \, \varphi^0_{[h_{0i}]} \,\bar \varphi^1_{[h_{1i}]}  \varphi^{i_2}_{[c]} \,\bar \varphi^3_{[h_{3i}]} \Big{]}  \cr\cr
&&+\frac{ \imath\bar\lambda}{\sqrt{\delta^N(e)}} 
\int_{h^{ij}}\,
 \Big{[}\int_{[d]}
B^{3i_3}_{[h^{3i}][d]}\bar \varphi^0_{[h^{0i}]}  \varphi^1_{[h^{1i}]}  \bar \varphi^2_{[h^{2i}]} \varphi^{i_3}_{[d]} 
 - 
\int_{[a]}   B^{i_00}_{[a][h^{0i}]}
\bar \varphi^{i_0}_{[a]}  \varphi^1_{[h^{1i}]}  \bar \varphi^2_{[h^{2i}]} \varphi^3_{[h^{3i}]}  \crcr
&& + 
\int_{[b]}
 B^{1i_1}_{[h^{1i}][b]}
\bar \varphi^0_{[h^{0i}]}  \varphi^{i_1}_{[b]}  \bar \varphi^2_{[h^{2i}]} \varphi^3_{[h^{3i}]} 
-
\int_{[c]}   B^{i_22}_{[c][h^{2i}]} 
\bar \varphi^0_{[h^{0i}]}  \varphi^1_{[h^{1i}]}  \bar \varphi^{i_2}_{[c]} \varphi^3_{[h^{3i}]} \Big{]} \;,\eea
which can be denoted compactly by
\bea
&&
\delta_B S^{\inter} := \imath \lambda\Big[ [B\varphi]^0 \bar\varphi^1 \varphi^2 \bar\varphi^3 
- \varphi^0 [\bar\varphi B]^1 \varphi^2 \bar\varphi^3  
+\varphi^0 \bar\varphi^1 [B\varphi]^2 \bar\varphi^3 
- \varphi^0 \bar\varphi^1 \varphi^2 [\bar\varphi B ] ^3    \Big] 
 \;+\; \imath \bar\lambda \{\bar\varphi\} \;,\crcr
&&
 \bar\lambda \{\bar\varphi\}=  
- [\bar\varphi B]^0 \varphi^1 \bar\varphi^2 \varphi^3 
+ \bar\varphi^0 [B\varphi]^1 \bar\varphi^2 \varphi^3  
- \bar\varphi^0\varphi^1 [\bar\varphi B]^2 \varphi^3 
+ \bar\varphi^0 \varphi^1 \bar\varphi^2 [B\varphi]^3 \;.
\label{eq:generinterinf} 
\eea
Meanwhile, the source terms have the infinitesimal
variations
\bea
&&
\delta_B (\bar\eta\varphi + \bar\varphi \eta)
= \sum_i \int_{[g]} \; 
  \left(  ^ U \bar \varphi^i_{[g]} \eta^i_{[g]} +
 \bar \eta^i_{[g]} \  ^U \varphi^i_{[g]} \right)   -  (\bar\eta\varphi + \bar\varphi \eta)  \crcr
&&
= \sum_{i} \int_{[g_i][a]} 
  \left(   (\delta^{ij} \delta_{[g][a]} -\imath
\bar B^{ij}_{[g][a]}) \bar \varphi^{j}_{[a]}\eta^i_{[g]} +
 \bar \eta^i_{[g]}  (\delta^{ij} \delta_{[g][a]} +\imath
B^{ij}_{[g][a]}) \varphi^j_{[a]} \right)  -  \sum_i   \int_{[g]}(\bar\eta^i_{[g]}\varphi^i_{[g]} + \bar\varphi^i_{[g]} \eta^i_{[g]})
 \crcr
&& =  \imath \sum_{i} 
\int_{[g] [a]} 
 \left(   -\bar \varphi^j_{[a]} B^{ji}_{[a][g]} \eta^i_{[g]} +
 \bar \eta^i_{[g]} B^{ij}_{[g][a]} \varphi^{j}_{[a]} \right) 
= :  \imath (-\bar \varphi B \eta  + \bar \eta B \varphi )
\;.
\label{eq:genersourceinf}
\eea
Under (\ref{eq:generunita}), the partition function transforms according to
\bea
 Z(\bar \eta,\eta) = \int d\mu_{UCU^{-1}} (\bar \varphi, \varphi ) e^{-S^{\inter}(U \varphi ,  \bar\varphi  U^{-1}) +
\bar \varphi  U^{-1}  \eta  +  
\bar \eta U \varphi  
 } \;.
\label{eq:partiuu}
\eea
We have used the fact that the covariance varies as
\bea
 C^{kk'}_{[h][h']} &= &
\int d\mu_C(\ ^U \bar \varphi,\ ^U   \varphi) \;\int_{[a][b]} 
\bar U^{kj}_{[h][a]}\,\bar\varphi^{j}_{[a]}\,
 U^{k'j'}_{[h'][b]}\;\varphi^{j'}_{[b]} 
\crcr
\bar U^{k'l'}_{[h'][b']} 
C^{kk'}_{[h]  [h']}  U^{kl}_{[h][a']} 
& =&\int d\mu_C(\ ^U \bar \varphi,\ ^U   \varphi)\;\int_{[a][b][h][h']} U^{kl}_{[h][a']} \bar U^{kj}_{[h][a]} 
 U^{k'j'}_{[h'][b]} \bar U^{k'l'}_{[h'][b']}   \;\bar\varphi^{j}_{[a]}\varphi^{j'}_{[b]} 
\crcr
 & =&\int d\mu_C(\ ^U \bar \varphi,\ ^U   \varphi)\;\int_{ab}    
 \;\bar\varphi^{l}_{[a']}\varphi^{l'}_{[b']} 
:=\, [U C U^{-1}]^{ll'}_{[a'][b']} :=  [\, ^U\,  C]^{ll'}_{[a'][b']} \;.
\eea
Seeking the infinitesimal variation of the covariance, one finds \bea
[U C U^{-1}]^{ij}_{[a][b]}  - C^{ij}_{[a][b]}
&=& (\delta^{li}\delta_{[c][a]} +\imath B^{li}_{[c][a]} )C^{ll'}_{[c][c']} (\delta^{jl'}\delta_{[b][c']}-\imath B^{jl'}_{[b][c']} )
 -  C^{ij}_{[a][b]}  \cr\cr
&=& \imath  [-C^{il'}_{[a][c']} B^{jl'}_{[b][c']} +  B^{li}_{[c][a]} C^{lj}_{[c][b]}]
=: \imath  [ B C-CB ]^{ij}_{[a][b]}\;.
\eea

\subsection{Right invariant unitary transformation}
\label{app:inftransright}

Working with a right invariant unitary in the sense of the 1-action 
of Appendix \ref{sect:leftinvunit}, Eq. (\ref{eq:notlefinv}), 
we get a change of variables such that 
\bea
&&
^U \varphi^{0}_{a.} = U_{ab}  \varphi^0_{b.}\, \; , \qquad ^U \bar \varphi^0_{a.} = \,\bar  \varphi^0_{b.} (U^{-1})_{ba}
=\bar{ U}_{ab}\bar \varphi^0_{b.}\, \; , \qquad
\frac{\delta}{\delta \,  ^U \varphi^0_{a.}} = \frac{\delta}{\delta \varphi^0_{b.}} (U^{-1})_{ba}\;, \qquad 
\frac{\delta}{\delta\,  ^U \bar\varphi^0_{a.}} = U_{ab}\frac{\delta}{\delta \bar \varphi^0_{b.}}\ \;, \crcr
&&
^U \varphi^{1}_{.a} = U_{ab}  \varphi^1_{.b}\, \; , \qquad ^U \bar \varphi^1_{.a} = \,\bar  \varphi^1_{.b} (U^{-1})_{ba}
=\bar{ U}_{ab}\bar \varphi^1_{.b}\, \; , \qquad
\frac{\delta}{\delta  \,  ^U\varphi^1_{.a}} = \frac{\delta}{\delta \varphi^1_{.b}} (U^{-1})_{ba}\;, \qquad 
\frac{\delta}{\delta\,  ^U \bar\varphi^1_{.a}} =U_{ab}\frac{\delta}{\delta \bar \varphi^1_{.b}}\ \;,
\label{eq:unita}
\eea
meanwhile colors $2$ and $3$ transform like $0$ and $1$, respectively.
The subscripts $a,b$ should be considered here as
a unique group element (and not a triplet) and the dot
notifies the position of the remaining arguments of the field.
Hence fields $0$ and $2$ are transformed with respect to their
first argument whereas fields $1$ and $3$ to their 
last argument.

Under $U$ the term $S^{\inter}$ transforms as
\bea
&&
^U \, S^{\inter} 
 =\frac{\lambda}{\sqrt{\delta^N(e)}} \int_{h_{ij}}\,\int_{abcd}
U_{h_{03}a} \varphi^0_{ah_{02}h_{01}} \,
 \bar \varphi^1_{h_{10}h_{13}b} \bar U_{h_{12}b} \,  U_{h_{21}c}\,  \varphi^2_{ch_{20}h_{23}} \,
\bar \varphi^3_{h_{32}h_{31}d} \bar U_{h_{30}d}\crcr
&&+\frac{\bar\lambda}{\sqrt{\delta^N(e)}} 
\int_{h_{ij}}\,\int_{abcd} \  \bar U_{h^{03}a} \bar \varphi^0_{ah^{02}h^{01}} 
  \varphi^1_{h^{10}h^{13}b} U_{h^{12} b}\,
\bar U_{h^{21}c} \bar \varphi^2_{ch^{20}h^{23}}  \, 
  \varphi^3_{h^{32}h^{31}d} U_{h^{30}d} \; ,
\eea 
using the orthogonality relation the unitary operators, namely
$U_{hb}\bar U_{hc}=\delta_{bc}$, we have after a 
proper renaming of variables
\bea
^U \, S^{\inter}  = S^{\inter}\;.
\eea
Remark that this symmetry can be even decomposed
in two \emph{minimal} and independent symmetries:
one performed on the couple $(0,3)$ and another one
performed on $(1,2)$. Each of these latter symmetries
does not modify  $S^{\inter}$ and can be 
used to determine all the subsequent developments
without loss of generality. These simpler symmetries 
can be useful to reduce the generic WT identities
and to obtain particular graphical equations.

Under (\ref{eq:unita}), the partition function undergoes the following
modification: 
\bea
 Z(\bar \eta,\eta) = \int d\mu_{UCU^{-1}} (\bar \varphi, \varphi ) e^{-S^{\inter}(U \varphi ,  \bar\varphi  U^{-1}) +
\bar \varphi  U^{-1}  \eta  +  
\bar \eta U \varphi  
 } \;.
\eea
We have used the fact that the covariance transforms as follows:
\bea
C^{ii=0,2}_{h_{0}h_{1}h_{2} ;\; h_{0}'h_{1}'h_{2}'} &= &
\int d\mu_C(\ ^U \bar \varphi,\ ^U   \varphi) \;\int_{ab} 
\bar U_{h_{0}a}\,\bar\varphi^{i=0,2}_{ah_{1}h_{2}}\,
 U_{h_{0}'b}\;\varphi^{i=0,2}_{bh_{1}'h_{2}'} \crcr
\bar U_{h_{0}'c'}  C^{ii=0,2}_{h_{0}h_{1}h_{2} ;\; h_{0}'h_{1}'h_{2}'}
U_{h_{0}c}
& =&\int d\mu_C(\ ^U \bar \varphi,\ ^U   \varphi) \int_{ab h_{0} h'_{0}} U_{h_{0}c}  \bar U_{h_{0}a}  \bar U_{h_{0}'c'} U_{h_{0}'b}
 \,\bar\varphi^{i=0,2}_{ah_{1}h_{2}}\,\varphi^{i=0,2}_{bh_{1}'h_{2}'} 
:= [U C U^{-1}]^{ii=0,2}_{ch_{1}h_{2} ;\; c'h_{1}'h_{2}'}   \cr\cr\cr
C^{ii=1,3}_{h_{0}h_{1}h_{2} ;\; h_{0}'h_{1}'h_{2}'} 
&= &
\int d\mu_C(\ ^U \bar \varphi,\ ^U   \varphi) \;\int_{ab} 
\,\bar\varphi^{i=1,3}_{h_{0}h_{1}a} \bar U_{h_{2}a}\,
\;\varphi^{i=1,3}_{h_{0}'h_{1}'b}  U_{h_{2}'b} \crcr
\bar U_{h_{2}' c'}  C^{ii=1,3}_{h_{0}h_{1}h_{2} ;\; h_{0}'h_{1}'h_{2}'} 
U_{h_{2}c}
& =&\int d\mu_C(\ ^U \bar \varphi,\ ^U   \varphi) \int_{ab h_{2} h_{2}'} U_{h_{2}c} \bar U_{h_{2}a} 
\bar U_{h_{2}' c'} U_{h_{2}'b}
\bar\varphi^{i=1,3}_{h_{0}h_{1}a} 
\varphi^{i=1,3}_{h_{0}'h_{1}'b}  := [U C U^{-1}]^{ii=1,3}_{h_{0}h_{1}c ;\; h_{0}'h_{1}'c'} \;.\crcr
&&
\eea
For a small $B$, we decompose $U_{ab} = \delta_{ab}+ \imath B_{ab}$, hence at first order in $B$, we can explicitly check that the variation of the interaction part vanishes. After a straightforward computation, one has
\bea
\delta_B S^{\inter}
&=&
\frac{\imath\lambda}{\sqrt{\delta^N(e)}} \int_{h_{ij}}\, \Bigl{[}
 - \int_{d}\bar B_{h_{30}d}
\varphi^0_{h_{03}h_{02}h_{01}}\bar \varphi^1_{h_{10}h_{13}h_{12}} 
\varphi^2_{h_{21}h_{20}h_{23}} \,
\bar \varphi^3_{h_{32}h_{31}d}
\crcr
&& + 
\int_{a}  B_{h_{03}a}
\varphi^0_{ah_{02}h_{01}}\bar \varphi^1_{h_{10}h_{13}h_{12}} 
\varphi^2_{h_{21}h_{20}h_{23}} \,
\bar \varphi^3_{h_{32}h_{31}h_{30}}
 \crcr
&&
-\int_{b} \bar B_{h_{12}b} 
\varphi^0_{h_{03}h_{02}h_{01}} \bar \varphi^1_{h_{10}h_{13}b}  
\varphi^2_{h_{21}h_{20}h_{23}} \,
\bar \varphi^3_{h_{32}h_{31}h_{30}} \crcr
&&
+ \int_{c}B_{h_{21}c} \,
 \varphi^0_{h_{03}h_{02}h_{01}} \bar \varphi^1_{h_{10}h_{13}h_{12}}  
\varphi^2_{ch_{20}h_{23}} \,
\bar \varphi^3_{h_{32}h_{31}h_{30}} \Big{]}
\cr\cr
&&+\frac{\imath\bar\lambda}{\sqrt{\delta^N(e)}} 
\int_{h^{ij}}\, \Bigl{[}
 \int_{d} B_{h^{30}d}
\varphi^0_{h^{03}h^{02}h^{01}}\bar \varphi^1_{h^{10}h^{13}h^{12}} 
\varphi^2_{h^{21}h^{20}h^{23}} \,
\bar \varphi^3_{h^{32}h^{31}d}
\crcr
&& -
\int_{a}  \bar B_{h^{03}a}
\varphi^0_{ah^{02}h^{01}}\bar \varphi^1_{h^{10}h^{13}h^{12}} 
\varphi^2_{h^{21}h^{20}h^{23}} \,
\bar \varphi^3_{h^{32}h^{31}h^{30}}
 \crcr
&&
+\int_{b}  B_{h^{12}b} 
\varphi^0_{h^{03}h^{02}h^{01}} \bar \varphi^1_{h^{10}h^{13}b}  
\varphi^2_{h^{21}h^{20}h^{23}} \,
\bar \varphi^3_{h^{32}h^{31}h^{30}} \crcr
&&
- \int_{c}\bar B_{h^{21}c} \,
 \varphi^0_{h^{03}h^{02}h^{01}} \bar \varphi^1_{h^{10}h^{13}h^{12}}  
\varphi^2_{ch^{20}h^{23}} \,
\bar \varphi^3_{h^{32}h^{31}h^{30}} \Big{]} \;,
\eea
then using the Herimiticity of the kernel $\bar B_{ab} = B_{ba}$
this term cancels. Thus, $\delta_B S^{\inter}=0$. 

The infinitesimal variations of the covariance are given by
 \bea
[U C U^{-1}]^{ii=0,2}_{[a][b]}  - C^{ii=0,2}_{[a][b]}
&=& (\delta_{ca_0} +\imath B_{ca_0} )C^{ii=0,2}_{c a_1a_2;\; c' b_1b_2}  (\delta_{c'b_0} -\imath \bar B_{c'b_0} )
 -  C^{ii=0,2}_{[a][b]}  \crcr
&=& \imath  [-C^{ii=0,2}_{[a];\;c'b_1b_2} B_{b_0c'} +  B_{ca_0} C^{ii=0,2}_{ca_1a_2;\; [b]}]
=: \imath  [ B C-CB ]^{ii=0,2}_{[a][b]}  \;, 
\cr\cr
[U C U^{-1}]^{ii=1,3}_{[a][b]}  - C^{ii=1,3}_{[a][b]}
&=& (\delta_{ca_2} +\imath B_{ca_2} )C^{ii=1,3}_{a_0a_1c;\; b_0 b_1c'}  (\delta_{c'b_2} -\imath \bar B_{c'b_2} )
 -  C^{ii=1,3}_{[a][b]}  \crcr
&=& \imath  [-C^{ii=1,3}_{[a];\;b_0b_1c'} B_{b_2c'} +  B_{ca_2} C^{ii=1,3}_{a_0a_1c;\; [b]}]
=: \imath  [ B C-CB ]^{ii=1,3}_{[a][b]} \;,
\eea
whereas the source terms can be varied as follows 
\bea
&&
\delta_B (\bar\eta\varphi + \bar\varphi \eta)
= \sum_i \int_{[g]} \; 
  \left(  ^ U \bar \varphi^i_{[g]} \eta^i_{[g]} +
 \bar \eta^i_{[g]} \  ^U \varphi^i_{[g]} \right)  -  (\bar\eta\varphi + \bar\varphi \eta) \crcr
&& = 
 \imath\int_{g_i a} \Big{\{}
  \sum_{i=0,2} \left(   -\bar \varphi^i_{a.} B_{ag_{0}} \eta^i_{g_{0}.} +
 \bar \eta^i_{g_{0}.} B_{g_{0}a} \varphi^i_{a.} \right) 
+ \sum_{i=1,3} \left(   -\bar \varphi^i_{.a} B_{ag_{2}} \eta^i_{.g_{2}} +
 \bar \eta^i_{.g_{2}} B_{g_{2}a} \varphi^i_{.a} \right) 
\Big{\}}\crcr
&&
= :  \imath (-\bar \varphi B \eta  + \bar \eta B \varphi )
\;.
\eea

\section{Free energy evaluations}
\label{app:calcwardigener}

\subsection{General unitary transformation}
\label{app:freereduce}

We start by giving the variation of free energy (\ref{eq:freeexpan0})
under the infinitesimal transformation
generated by a general unitary operator\footnote{At each step of calculation and for simplicity purpose, we will not display 
the term $ \bar\lambda \{\bar\varphi\}$.
However we will provide an explicit symmetry of its analogous,
i.e. the term with coefficient $\lambda$, from which 
 $ \bar\lambda \{\bar\varphi\}$ can be determined without ambiguity.}
\bea
&&  
 \frac{\delta \ln Z(\eta,\bar \eta)}{\imath\delta B^{ij}_{[\mu][\nu]}}=0  = \frac{1}{ Z(\eta,\bar \eta)}\int d\mu_{C} (\bar \varphi, \varphi ) 
\Bigl\{ \delta_{\varphi^j_{[\nu]}} C^{il}_{[\mu][\alpha]}\delta_{\bar \varphi^l_{[\alpha]}}   
-\delta_{\varphi^l_{[\alpha]}} C^{lj}_{[\alpha][\nu]}
\delta_{\bar \varphi^i_{[\mu]}} - \bar \varphi^i_{[\mu]} \eta^j_{[\nu]}  +  \bar \eta^i_{[\mu]}   \varphi^j_{[\nu]} 
\label{eq:freeexpan}\\
&&
- \lambda\Big[ \delta^{i0}\varphi^{j}_{[\nu]} 
[\bar\varphi^1\varphi^2 \bar\varphi^3]_{[\mu]} 
-\delta^{j1} \bar\varphi^i_{[\mu]} 
[\varphi^0 \varphi^2 \bar\varphi^3]_{[\nu]}  
+ \delta^{i2}\varphi^{j}_{[\nu]}
[\varphi^0 \bar\varphi^1  \bar\varphi^3]_{[\mu]} 
-\delta^{j3} \bar\varphi^{i}_{[\mu]}  
[ \varphi^0\bar\varphi^1 \varphi^2]_{[\nu]}  \Big] \crcr
&&
 \;-\;  \bar\lambda \{\bar\varphi\}\Bigl\}\;
e^{-S^{\inter}( \varphi, \bar\varphi )  +\bar \varphi \eta  +  \bar \eta \varphi 
 }\;,
\nonumber
\eea
where the remaining arguments of fields which do not appear are integrated (see (\ref{eq:hidden})).
Using the Hermiticity of the covariance $\bar C^{ij}_{[b][a]} = C^{ji}_{[a][b]}$  (as this is the two-point correlation function and the latter is a necessarily Hermitian), 
(\ref{eq:freeexpan}) is again
\bea
&&  
 \frac{\delta \ln Z(\eta,\bar \eta)}{\imath\delta B^{ij}_{[\mu][\nu]}}=0  = \frac{1}{ Z(\eta,\bar \eta)}\int d\mu_{C} (\bar \varphi, \varphi ) 
\Bigl\{ \delta_{\varphi^j_{[\nu]}} [C\delta_{\bar \varphi}]^i_{[\mu]}   
-\delta_{\bar \varphi^i_{[\mu]}} [\delta_{\varphi} C^\dag]^j_{[\nu]}
- \bar \varphi^i_{[\mu]} \eta^j_{[\nu]}  +  \bar \eta^i_{[\mu]}   \varphi^j_{[\nu]} 
\label{intergener}\\
&&
- \lambda\Big[\delta^{i0} \varphi^j_{[\nu]}
[ \bar\varphi^1\varphi^2 \bar\varphi^3]_{[\mu]} 
- \delta^{j1}\bar\varphi^{i}_{[\mu]}
[\varphi^0 \varphi^2\bar\varphi^3]_{[\nu]}  
+\delta^{i2}   \varphi^j_{[\nu]}
[\varphi^0 \bar\varphi^1\bar\varphi^3]_{[\mu]}
-\delta^{j3} \bar\varphi^i_{[\mu]} 
[\varphi^0\bar\varphi^1\varphi^2]_{[\nu]}    \Big] \crcr
&&
 \;-\;  \bar\lambda \{\bar\varphi\}\Bigl\}
e^{-S^{\inter}( \varphi, \bar\varphi )  +\bar \varphi \eta  +  \bar \eta \varphi 
 }\;,
\nonumber
\eea
where $\bar\lambda \{\bar\varphi\}$ can be obtained 
from the term in $\lambda$ by multiplying it by $(-1)$,
the symmetry  $(j,\nu)\leftrightarrow (i,\mu)$ and complex conjugation. 

Let us multiply (\ref{intergener}) and sum over repeated
indices by $\bar C^{j'j}_{[a][\nu]}$ and $C^{i'i}_{[b][\mu]}$, we get,
after renaming $i',j'$ by $i,j$:
\bea
&&  
0  = \frac{1}{ Z(\eta,\bar \eta)}\int d\mu_{C} (\bar \varphi, \varphi ) 
\Bigl\{ 
- [C\bar \varphi]^i_{[b]} \, [\eta C^\dag]^j_{[a]}  
+  [C\bar \eta]^i_{[b]}\,   [\varphi C^\dag]^j_{[a]} 
\cr\cr
&&
- \lambda\Big[  C^{i0}_{[b][\mu]}[\varphi C^\dag]^j_{[a]} [\bar\varphi^1\varphi^2 \bar\varphi^3]_{[\mu]}
- \bar C^{j1}_{[a][\nu]}[C\bar\varphi]^i_{[b]} 
[\varphi^0\varphi^2\bar\varphi^3]_{[\nu]}  \cr\cr
&&
+ C^{i2}_{[b][\mu]} [ \varphi C^\dag]^j_{[a]}
[\varphi^0\bar\varphi^1 \bar\varphi^3]_{[\mu]} 
- \bar C^{j3}_{[a][\nu]}[C\bar\varphi]^{i}_{[b]} 
[\varphi^0 \bar\varphi^1 \varphi^2]_{[\nu]}    \Big] 
 \;-\;  \bar\lambda \{\bar\varphi\}\cr\cr
&& +C^{ii'}_{[b][\mu]} [\delta_{\varphi} C^\dag]^j_{[a]} [C\delta_{\bar \varphi}]^{i'}_{[\mu]}   
-\bar C^{jj'}_{[a][\nu]}
[C\delta_{\bar \varphi}]^i_{[b]} [\delta_{\varphi} C^\dag]^{j'}_{[\nu]} \Bigl\}
e^{-S^{\inter}( \varphi, \bar\varphi )  +\bar \varphi \eta  +  \bar \eta \varphi 
 }\;,
\eea
here, $\bar\lambda \{\bar\varphi\}$ can be obtained 
from the term in $\lambda$ by multiplication by $(-1)$,
complex conjugation of fields and symmetry  
\bea
(i,a,\nu, [\varphi C^\dag]) \leftrightarrow (j,b,\mu, [C\bar\varphi])\;.
\eea
Using Lemma \ref{lem:parts}, we have:  
\bea
&&\frac{1}{ Z(\eta,\bar \eta)}\int d\mu_{C} (\bar \varphi, \varphi ) 
\Bigl\{
C^{ii'}_{[b][\mu]} [\delta_{\varphi} C^\dag]^j_{[a]} [C\delta_{\bar \varphi}]^{i'}_{[\mu]}   
-\bar C^{jj'}_{[a][\nu]}
[C\delta_{\bar \varphi}]^i_{[b]} [\delta_{\varphi} C^\dag]^{j'}_{[\nu]} \Bigl\}
e^{-S^{\inter}( \varphi, \bar\varphi )  +\bar \varphi \eta  +  \bar \eta \varphi 
 }\crcr
&&
= \frac{1}{ Z(\eta,\bar \eta)}\int d\mu_{C} (\bar \varphi, \varphi ) 
\Bigl\{ C^{ii'}_{[b][\mu]} \bar\varphi^j_{[a]} [C\delta_{\bar \varphi}]^{i'}_{[\mu]}   
-\bar C^{jj'}_{[a][\nu]}
\varphi^i_{[b]} [\delta_{\varphi} C^\dag]^{j'}_{[\nu]} \Bigl\}
e^{-S^{\inter}( \varphi, \bar\varphi )  +\bar \varphi \eta  +  \bar \eta \varphi 
 }\;.
\eea 
Another integration by parts and Lemma \ref{lem:parts}
yield
\bea
&&  
 \frac{1}{ Z(\eta,\bar \eta)}\int d\mu_{C} (\bar \varphi, \varphi ) 
\Bigl\{ \delta_{\bar \varphi^l_{[h]}   } C^{ii'}_{[b][\mu]} \bar\varphi^j_{[a]} C^{i'l}_{[\mu][h]}  - \delta_{\bar \varphi^l_{[h]}   } [C^{ii'}_{[b][\mu]} \bar\varphi^j_{[a]} C^{i'l}_{[\mu][h]} ] \crcr
&&
-\delta_{\varphi^l_{[h']}} \bar C^{jj'}_{[a][\nu]}
\varphi^i_{[b]} \bar C^{j'l}_{[\nu][h']} + \delta_{\varphi^l_{[h']}} [\bar C^{jj'}_{[a][\nu]}
\varphi^i_{[b]} \bar C^{j'l}_{[\nu][h']}] \Bigl\}
e^{-S^{\inter}( \varphi, \bar\varphi )  +\bar \varphi \eta  +  \bar \eta \varphi 
 }\;.
\eea
Using  again the fact that $C$ is Hermitian and performing
some differentiations leads to
\bea
\frac{1}{ Z(\eta,\bar \eta)}\int d\mu_{C} (\bar \varphi, \varphi ) 
\Bigl\{ 
\delta_{\bar \varphi^l_{[h]}   } C^{ii'}_{[b][\mu]} \bar\varphi^j_{[a]} C^{i'l}_{[\mu][h]}  -  [C^2]^{ij}_{[b][a]}
-\delta_{\varphi^l_{[h']}} \bar C^{jj'}_{[a][\nu]}
\varphi^i_{[b]} \bar C^{j'l}_{[\nu][h']} +[\bar C^2]^{ji}_{[a][b]} \Bigl\}
e^{-S^{\inter}( \varphi, \bar\varphi )  +\bar \varphi \eta  +  \bar \eta \varphi  }
\label{eq:intermeward}
\eea
and since $[C^2]^{ij}_{[a][b]} = [\bar C^2]^{ji}_{[b][a]}$, these
two terms cancel and 
the last  expression (\ref{eq:intermeward})  assumes
now the form
\bea  
 \frac{1}{ Z(\eta,\bar \eta)}\int d\mu_{C} (\bar \varphi, \varphi ) 
\Bigl\{  C^{ii'}_{[b][\mu]} [C \delta_{\bar \varphi}]^{i'}_{[\mu]} \bar\varphi^j_{[a]} 
- \bar C^{jj'}_{[a][\nu]}[\delta_{\varphi}C^\dag]^{j'}_{[\nu]} 
\varphi^i_{[b]}  \Bigl\}
e^{-S^{\inter}( \varphi, \bar\varphi )  +\bar \varphi \eta  +  \bar \eta \varphi 
 }\;.
\eea
Use a third time Lemma \ref{lem:parts} in order to get
\bea
\frac{1}{ Z(\eta,\bar \eta)}\int d\mu_{C} (\bar \varphi, \varphi ) 
\Bigl\{  C^{ii'}_{[b][\mu]}  \varphi^{i'}_{[\mu]} \bar\varphi^j_{[a]} 
- \bar C^{jj'}_{[a][\nu]} \bar\varphi^{j'}_{[\nu]}\varphi^i_{[b]}  \Bigl\}
e^{-S^{\inter}( \varphi, \bar\varphi )  +\bar \varphi \eta  +  \bar \eta \varphi 
 }\;.
\eea
Plugging this into the variation of the free energy,
one obtains
\bea
&&  
0  = \frac{1}{ Z(\eta,\bar \eta)}\int d\mu_{C} (\bar \varphi, \varphi ) 
\Bigl\{ 
- [C\bar \varphi]^i_{[b]} \, [\eta C^\dag]^j_{[a]}  
+  [C\bar \eta]^i_{[b]}\,   [\varphi C^\dag]^j_{[a]} 
\cr\cr
&&
- \lambda\Big[  C^{i0}_{[b][\mu]}[\varphi C^\dag]^j_{[a]} 
[\bar\varphi^1 \varphi^2 \bar\varphi^3]_{[\mu]} 
- \bar C^{j1}_{[a][\nu]} [C\bar\varphi]^i_{[b]} 
[\varphi^0\varphi^2 \bar\varphi^3]_{[\nu]}  \cr\cr
&&
+ C^{i2}_{[b][\mu]}[ \varphi C^\dag]^j_{[a]}
[\varphi^0 \bar\varphi^1 \bar\varphi^3]_{[\mu]} 
- \bar C^{j3}_{[a][\nu]}[C\bar\varphi]^{i}_{[b]}
[\varphi^0\bar\varphi^1 \varphi^2]   \Big] 
 \;-\;  \bar\lambda \{\bar\varphi\}\cr\cr
&& +C^{ii'}_{[b][\mu]}  \varphi^{i'}_{[\mu]} \bar\varphi^j_{[a]} 
- \bar C^{jj'}_{[a][\nu]} \bar\varphi^{j'}_{[\nu]}\varphi^i_{[b]}\Bigl\}
e^{-S^{\inter}( \varphi, \bar\varphi )  +\bar \varphi \eta  +  \bar \eta \varphi 
 }
\eea
which is the bottom line for deriving the WT identities
using a general unitary transformation.

\subsection{1-action}
\label{app:freereduc1action}

We denote (\ref{eq:intermed}) in the following way
(recalling that $[\alpha]$ is a triple index, while $\mu,\nu$ are single indices, and a dot notifies the position of remaning indices which,
below, are integrated)
\bea
&&
0 =  \frac{1}{Z(\eta,\bar\eta)}  \int d\mu_{C} (\bar \varphi, \varphi )  \Bigl(
- \bar \varphi^{0,2}_{\mu.} \eta^{0,2}_{\nu.} +  \bar \eta^{0,2}_{\mu.}  \varphi^{0,2}_{\nu.}  
 - \bar \varphi^{1,3}_{.\mu} \eta^{1,3}_{.\nu} +  \bar \eta^{1,3}_{.\mu}  \varphi^{1,3}_{.\nu}    \crcr
&&
   +\delta_{\varphi^{0,2}_{\nu.}} C^{0,2}_{\mu. \;[h]}
\delta_{\bar \varphi^{0,2}_{[h]} }
 -\delta_{\varphi^{0,2}_{[h]}} 
C^{0,2}_{[h]\; \nu.}
\delta_{\bar \varphi^{0,2}_{\mu.} } 
 + \delta_{ \varphi^{1,3}_{.\nu}} C^{1,3}_{.\mu \;[h]}
\delta_{\bar \varphi^{1,3}_{[h]} }
 - \delta_{ \varphi^{1,3}_{[h]}} 
C^{1,3}_{[h]\;  .\nu}
\delta_{\bar \varphi^{1,3}_{.\mu} } 
 \Bigr) 
e^{-S^{\inter}( \varphi, \bar\varphi ) +\bar \varphi \eta  +  \bar \eta \varphi  }\;.
\eea
Using $C^{ij}_{[a][b]} = \overline C^{ij}_{[b][a]}$, the above
computes to
\bea
&&
0=  \frac{1}{Z(\eta,\bar\eta)}  \int d\mu_{C} (\bar \varphi, \varphi )  \Bigl(
- \bar \varphi^{0,2}_{\mu.} \eta^{0,2}_{\nu.} +  \bar \eta^{0,2}_{\mu.}  \varphi^{0,2}_{\nu.}  
 - \bar \varphi^{1,3}_{.\mu} \eta^{1,3}_{.\nu} +  \bar \eta^{1,3}_{.\mu}  \varphi^{1,3}_{.\nu}    \crcr
&&
   +\delta_{\varphi^{0,2}_{\nu.}} C^{0,2}_{\mu. \;[h]}
\delta_{\bar \varphi^{0,2}_{[h]} }
 -\delta_{\bar \varphi^{0,2}_{\mu.} } 
\bar C^{0,2}_{\nu.\;[h]}
\delta_{\varphi^{0,2}_{[h]}} 
 + \delta_{ \varphi^{1,3}_{.\nu}} C^{1,3}_{.\mu \;[h]}
\delta_{\bar \varphi^{1,3}_{[h]} }
 - \delta_{\bar \varphi^{1,3}_{.\mu} } 
\bar C^{1,3}_{.\nu\;[h]}
\delta_{ \varphi^{1,3}_{[h]}} 
 \Bigr) 
e^{-S^{\inter}( \varphi, \bar\varphi ) +\bar \varphi \eta  +  \bar \eta \varphi  } \crcr
&& 0=  \frac{1}{Z(\eta,\bar\eta)}  \int d\mu_{C} (\bar \varphi, \varphi )  \Bigl(
- \bar \varphi^{0,2}_{\mu.} \eta^{0,2}_{\nu.} +  \bar \eta^{0,2}_{\mu.}  \varphi^{0,2}_{\nu.}  
 - \bar \varphi^{1,3}_{.\mu} \eta^{1,3}_{.\nu} +  \bar \eta^{1,3}_{.\mu}  \varphi^{1,3}_{.\nu}    \crcr
&&
   +\delta_{\varphi^{0,2}_{\nu.}} 
[C \delta_{\bar \varphi}]^{0,2}_{\mu.}
 -\delta_{\bar \varphi^{0,2}_{\mu.} } 
[\delta_{\varphi}   C^\dag]^{0,2}_{\nu.} 
 + \delta_{ \varphi^{1,3}_{.\nu}} 
[C\delta_{\bar \varphi}]^{1,3}_{.\mu}
 - \delta_{\bar \varphi^{1,3}_{.\mu} } 
[\delta_{ \varphi}  C^\dag]^{1,3}_{.\nu} 
 \Bigr) 
e^{-S^{\inter}( \varphi, \bar\varphi ) +\bar \varphi \eta  +  \bar \eta \varphi  } \;.
\label{eq:reducfree}
\eea
Let us evaluate by Lemma \ref{lem:parts} the functional derivative terms   that we can write
\bea
&&
  \frac{1}{Z(\eta,\bar\eta)}  \int d\mu_{C} (\bar \varphi, \varphi )  \Bigl(
[C \delta_{\bar \varphi}]^{0,2}_{\mu.}
\delta_{\varphi^{0,2}_{\nu.}} 
 -
[\delta_{\varphi}   C^\dag]^{0,2}_{\nu.} 
\delta_{\bar \varphi^{0,2}_{\mu.} } 
 + 
[C\delta_{\bar \varphi}]^{1,3}_{.\mu}
\delta_{ \varphi^{1,3}_{.\nu}} 
 -
[\delta_{ \varphi}  C^\dag]^{1,3}_{.\nu}  
\delta_{\bar \varphi^{1,3}_{.\mu} } 
 \Bigr) 
e^{-S^{\inter}( \varphi, \bar\varphi ) +\bar \varphi \eta  +  \bar \eta \varphi  } \cr\cr
&&=
\frac{1}{Z(\eta,\bar\eta)}  \int d\mu_{C} (\bar \varphi, \varphi )  \Bigl(
\varphi^{0,2}_{\mu.}
\delta_{\varphi^{0,2}_{\nu.}} 
 -
\bar\varphi^{0,2}_{\nu.} 
\delta_{\bar \varphi^{0,2}_{\mu.} } 
 + 
 \varphi^{1,3}_{.\mu}
\delta_{ \varphi^{1,3}_{.\nu}} 
 -
 \bar\varphi^{1,3}_{.\nu}  
\delta_{\bar \varphi^{1,3}_{.\mu} } 
 \Bigr) 
e^{-S^{\inter}( \varphi, \bar\varphi ) +\bar \varphi \eta  +  \bar \eta \varphi  } \cr\cr
&&= \frac{1}{Z(\eta,\bar\eta)}  \int d\mu_{C} (\bar \varphi, \varphi )  \Bigl\{ \int_{h_{ij}h^{ij}} \crcr
&&
\varphi^{0}_{\mu.}
\Big[ \frac{-\lambda}{\sqrt{\delta^{N}(e)}}\bar\varphi^{1}_{.h_{13}h_{12}} \varphi^{2}_{h_{21}.h_{23}} \bar \varphi^{3}_{h_{32}h_{31}\nu}
+ \bar\eta^{0}_{\nu.}\Big]
+
\varphi^{2}_{\mu.}
\Big[\frac{-\lambda}{\sqrt{\delta^{N}(e)}}\varphi^{0}_{h_{03}.h_{01}}\bar\varphi^{1}_{h_{10}h_{13}\nu}  \bar \varphi^{3}_{.h_{31}h_{30}}
+ \bar\eta^{2}_{\nu.}\Big]\crcr
&&+
\varphi^{1}_{.\mu}
\Big[ 
\frac{-\bar\lambda}{\sqrt{\delta^{N}(e)}}
\bar\varphi^{0}_{h^{03}h^{02}.} 
\bar \varphi^{2}_{\nu h^{20}h^{23}}\varphi^{3}_{h^{32}.h^{30}} 
+  \bar\eta^{1}_{.\nu}\Big]
+
\varphi^{3}_{.\mu}
\Big[\frac{-\bar\lambda}{\sqrt{\delta^{N}(e)}}
\bar\varphi^{0}_{\nu h^{02}h^{01}}\varphi^{1}_{h^{10}.h^{12}}  \bar\varphi^{2}_{
 h^{21}h^{20}.}
+  \bar\eta^{3}_{.\nu}\Big]\crcr
&&
-\bar\varphi^{0}_{\nu.}
\Big[ \frac{-\lambda}{\sqrt{\delta^{N}(e)}}\varphi^{1}_{.h^{13}h^{12}} \bar\varphi^{2}_{h^{21}.h^{23}}  \varphi^{3}_{h^{32}h^{31}\mu}
+ \eta^{0}_{\mu.}\Big]
-
\bar\varphi^{2}_{\nu.}
\Big[\frac{-\lambda}{\sqrt{\delta^{N}(e)}}\bar\varphi^{0}_{h^{03}.h^{01}}\varphi^{1}_{h^{10}h^{13}\mu}  \varphi^{3}_{.h^{31}h^{30}}
+ \eta^{2}_{\mu.}\Big]\crcr
&&-
\bar\varphi^{1}_{.\nu}
\Big[ 
\frac{-\bar\lambda}{\sqrt{\delta^{N}(e)}}
\varphi^{0}_{h_{03}h_{02}.} 
 \varphi^{2}_{\mu h_{20}h_{23}}\bar\varphi^{3}_{h_{32}.h_{30}} 
+ \eta^{1}_{.\mu}\Big]
-
\bar\varphi^{3}_{.\nu}
\Big[\frac{-\bar\lambda}{\sqrt{\delta^{N}(e)}}
\varphi^{0}_{\mu h_{02}h_{01}}\bar\varphi^{1}_{h_{10}.h_{12}}  \varphi^{2}_{
 h_{21}h_{20}.}
+ \eta^{3}_{.\mu}\Big]\crcr
&&
 \Bigr\} 
e^{-S^{\inter}( \varphi, \bar\varphi ) +\bar \varphi \eta  +  \bar \eta \varphi  } 
\label{integrals}
\eea
that will be shortly denoted by 
\bea
&&
 \frac{1}{Z(\eta,\bar\eta)}  \int d\mu_{C} (\bar \varphi, \varphi )  \Bigl( \varphi^{0,2}_{\mu.}\bar\eta^{0,2}_{\nu.}
- \lambda \;\varphi^{0,2}_{\mu.} \cdot [\bar\varphi^{1}\varphi^{2,0}\bar \varphi^{3}]_{\nu.} 
+
\varphi^{1,3}_{.\mu}\bar\eta^{1,3}_{.\nu}
-\bar\lambda\; \varphi^{1,3}_{.\mu}\cdot [\bar\varphi^{0}\bar \varphi^{2}\varphi^{3,1}]_{.\nu}
\crcr
&&
-\bar\varphi^{0,2}_{\nu.} \eta^{0,2}_{\mu.}
+\lambda\;
\bar\varphi^{0,2}_{\nu.}
\cdot [\varphi^{1}\bar\varphi^{2,0} \varphi^{3}]_{\mu.}
-
\bar\varphi^{1,3}_{.\nu}\eta^{1,3}_{.\mu}
+
\bar\lambda \;\bar\varphi^{1,3}_{.\nu}
\cdot[ 
\varphi^{0} \varphi^{2}\bar\varphi^{3,1}]_{.\mu}
 \Bigr) 
e^{-S^{\inter}( \varphi, \bar\varphi ) +\bar \varphi \eta  +  \bar \eta \varphi  } \;.
\eea
Inserting this last relation into (\ref{eq:reducfree}), we get
\bea
 0&=&  \frac{1}{Z(\eta,\bar\eta)}  \int d\mu_{C} (\bar \varphi, \varphi )    \Bigl[ 
- \bar \varphi^{0,2}_{\mu.} \eta^{0,2}_{\nu.} +  \bar \eta^{0,2}_{\mu.}  \varphi^{0,2}_{\nu.}  
 - \bar \varphi^{1,3}_{.\mu} \eta^{1,3}_{.\nu} +  \bar \eta^{1,3}_{.\mu}  \varphi^{1,3}_{.\nu}    
+\varphi^{0,2}_{\mu.}\bar\eta^{0,2}_{\nu.}
-\bar\varphi^{0,2}_{\nu.} \eta^{0,2}_{\mu.}
+
\varphi^{1,3}_{.\mu}\bar\eta^{1,3}_{.\nu}
-
\bar\varphi^{1,3}_{.\nu}\eta^{1,3}_{.\mu} \crcr
&&
- \lambda \;\varphi^{0,2}_{\mu.} \cdot [\bar\varphi^{1}\varphi^{2,0}\bar \varphi^{3}]_{\nu.} 
-\bar\lambda\; \varphi^{1,3}_{.\mu}\cdot [\bar\varphi^{0}\bar \varphi^{2}\varphi^{3,1}]_{.\nu}
+\lambda\;
\bar\varphi^{0,2}_{\nu.}
\cdot [\varphi^{1}\bar\varphi^{2,0} \varphi^{3}]_{\mu.}
+
\bar\lambda \;\bar\varphi^{1,3}_{.\nu}
\cdot[ 
\varphi^{0} \varphi^{2}\bar\varphi^{3,1}]_{.\mu}
 \Bigr]
e^{-S^{\inter}( \varphi, \bar\varphi ) +\bar \varphi \eta  +  \bar \eta \varphi  } \crcr
&&
\label{eq:reducfree2}
\eea
where repeated colors are summed as well as arguments 
in dot.  Equation (\ref{eq:reducfree2}) is the starting equation for deriving the WT identities in the case of unitary 1-action.


\begin{thebibliography}{99}


\bibitem{GFT} 
 D.~V.~Boulatov,
  ``A Model of three-dimensional lattice gravity,''
  Mod.\ Phys.\ Lett.\  A {\bf 7}, 1629 (1992)
  [arXiv:hep-th/9202074].

\bibitem{GFT2} 
 H.~Ooguri,
  ``Topological lattice models in four-dimensions,''
  Mod.\ Phys.\ Lett.\  A {\bf 7}, 2799 (1992)
  [arXiv:hep-th/9205090].


\bibitem{laurentgft}
  L.~Freidel,
  ``Group field theory: An overview,''
  Int.\ J.\ Theor.\ Phys.\  {\bf 44}, 1769 (2005)
  [arXiv:hep-th/0505016].

\bibitem{oriti}
  D.~Oriti,
  ``The group field theory approach to quantum gravity,''
  arXiv:gr-qc/0607032.

\bibitem{quantugeom2}
  D.~Oriti,
  ``The group field theory approach to quantum gravity: some recent results,''
  [arXiv:0912.2441 [hep-th]].

\bibitem{oriti2}
D. Oriti (ed.), 
``Approaches to quantum gravity: Towards a new understanding of space, time and matter,'' 
Cambridge Univ. Press., Cambridge (2009).

\bibitem{Gielen:2011dg}
  S.~Gielen and D.~Oriti,
  ``Discrete and continuum third quantization of Gravity,''
  arXiv:1102.2226 [gr-qc].

\bibitem{color}
 R.~Gurau,
  ``Colored Group Field Theory,''
Commun.\ Math.\ Phys.\  {\bf 304}, 69 (2011),
  [arXiv:0907.2582 [hep-th]].

\bibitem{PolyColor}
  R.~Gurau,
  ``Topological Graph Polynomials in Colored Group Field Theory,''
  Annales Henri Poincare {\bf 11}, 565 (2010)
  [arXiv:0911.1945 [hep-th]].

\bibitem{sing}
 R.~Gurau,
  ``Lost in Translation: Topological Singularities in Group Field Theory,''
  Class.\ Quant.\ Grav.\ {\bf 27}, 235023 (2010)
  arXiv:1006.0714 [hep-th].

\bibitem{Gurau:2010ba}
  R.~Gurau,
  ``The 1/N expansion of colored tensor models,''
  Annales Henri Poincare {\bf 12}, 829 (2011)
  [arXiv:1011.2726 [gr-qc]].

\bibitem{Gurau:2011aq}
  R.~Gurau and V.~Rivasseau,
  ``The 1/N expansion of colored tensor models in arbitrary dimension,''
 Europhys.\ Lett.\  {\bf 95}, 50004 (2011)
  [arXiv:1101.4182 [gr-qc]].

\bibitem{Gurau:2011xq}
  R.~Gurau,
  ``The complete 1/N expansion of colored tensor models in arbitrary dimension,''
  arXiv:1102.5759 [gr-qc].

\bibitem{Bonzom:2011zz}
  V.~Bonzom, R.~Gurau, A.~Riello and V.~Rivasseau,
  ``Critical behavior of colored tensor models in the large N limit,''
   Nucl.\ Phys.\  B {\bf 853}, 174 (2011)
  [arXiv:1105.3122 [hep-th]].

\bibitem{Gurau:2011tj}
  R.~Gurau,
  ``A generalization of the Virasoro algebra to arbitrary dimensions,''
  arXiv:1105.6072 [hep-th].

\bibitem{Girelli:2010ct}
  F.~Girelli and E.~R.~Livine,
  ``A Deformed Poincare Invariance for Group Field Theories,''
  Class.\ Quant.\ Grav.\  {\bf 27}, 245018 (2010)
  [arXiv:1001.2919 [gr-qc]].

\bibitem{gftnoncom}
  A.~Baratin and D.~Oriti,
  ``Group field theory with non-commutative metric variables,''
  Phys.\ Rev.\ Lett.\  {\bf 105}, 221302 (2010)
  [arXiv:1002.4723 [hep-th]].

\bibitem{Baratin:2011tg}
  A.~Baratin, F.~Girelli and D.~Oriti,
  ``Diffeomorphisms in group field theories,''
  Phys.\ Rev.\  D {\bf 83}, 104051 (2011)
  [arXiv:1101.0590 [hep-th]].

\bibitem{Carrozza:2011jn}
  S.~Carrozza and D.~Oriti,
  ``Bounding bubbles: the vertex representation of 3d Group Field Theory and the suppression of pseudo-manifolds,''
  arXiv:1104.5158 [hep-th].


\bibitem{Geloun:2011cz}
  J.~Ben~Geloun,
  ``Classical Group Field Theory,''
  arXiv:1107.3122 [hep-th].


\bibitem{Geloun:2011cy}
  J.~Ben~Geloun and V.~Bonzom,
  ``Radiative corrections in the Boulatov-Ooguri tensor model: The 2-point function,'' 
 Int.\ J.\ Theor.\ Phys.\  {\bf 50}, 2819 (2011)
  [arXiv:1101.4294 [hep-th]].

\bibitem{Rivasseau:2011xg}
  V.~Rivasseau,
  ``Towards Renormalizing Group Field Theory,''
  PoS C {\bf NCFG2010}, 004 (2010)
  [arXiv:1103.1900 [gr-qc]].


\bibitem{FreiGurOriti} 
  L.~Freidel, R.~Gurau and D.~Oriti,
  ``Group field theory renormalization - the 3d case: power counting of divergences,''
  Phys.\ Rev.\  D {\bf 80}, 044007 (2009)
  [arXiv:0905.3772 [hep-th]].

\bibitem{sefu1}
  J.~Magnen, K.~Noui, V.~Rivasseau and M.~Smerlak,
  ``Scaling behavior of three-dimensional group field theory,''
  Class.\ Quant.\ Grav.\  {\bf 26}, 185012 (2009)
  [arXiv:0906.5477 [hep-th]].

\bibitem{sefu2}
  J.~Ben~Geloun, J.~Magnen and V.~Rivasseau,
  ``Bosonic Colored Group Field Theory,''
  Eur.\ Phys.\ J.\  C {\bf 70}, 1119 (2010)
  [arXiv:0911.1719 [hep-th]].

\bibitem{sefu3}
   J.~Ben~Geloun, T.~Krajewski, J.~Magnen and V.~Rivasseau,
  ``Linearized Group Field Theory and Power Counting Theorems,''
  Class.\ Quant.\ Grav.\  {\bf 27}, 155012 (2010)
  [arXiv:1002.3592 [hep-th]].

\bibitem{Bonzom:2010ar}
  V.~Bonzom and M.~Smerlak,
  ``Bubble divergences from cellular cohomology,''
  Lett.\ Math.\ Phys.\  {\bf 93}, 295 (2010)
  [arXiv:1004.5196 [gr-qc]].

  V.~Bonzom and M.~Smerlak,
  ``Bubble divergences: sorting out topology from cell structure,''
  arXiv:1103.3961 [gr-qc].


\bibitem{Geloun:2010vj}
  J.~Ben~Geloun, R.~Gurau and V.~Rivasseau,
  ``EPRL/FK Group Field Theory,''
  Europhys.\ Lett.\  {\bf 92}, 60008 (2010)
  [arXiv:1008.0354 [hep-th]].


\bibitem{Grosse:2004yu}
  H.~Grosse and R.~Wulkenhaar,
  ``Renormalization of phi**4 theory on noncommutative R**4 in the matrix base,''
  Commun.\ Math.\ Phys.\  {\bf 256}, 305 (2005)
  [arXiv:hep-th/0401128].


\bibitem{Disertori:2006nq}
  M.~Disertori, R.~Gurau, J.~Magnen and V.~Rivasseau,
  ``Vanishing of Beta Function of Non Commutative Phi**4(4) Theory to all
  orders,''
  Phys.\ Lett.\  B {\bf 649}, 95 (2007)
  [arXiv:hep-th/0612251].

\bibitem{Geloun:2008zk}
  J.~Ben Geloun, R.~Gurau and V.~Rivasseau,
  ``Vanishing beta function for Grosse-Wulkenhaar model in a magnetic field,''
  Phys.\ Lett.\  B {\bf 671}, 284 (2009)
  [arXiv:0805.4362 [hep-th]].

\bibitem{Geloun}
J. Ben Geloun, {\it in preparation}. 

\bibitem{Grosse:2009pa}
  H.~Grosse and R.~Wulkenhaar,
  ``Progress in solving a noncommutative quantum field theory in four
  dimensions,''
  arXiv:0909.1389 [hep-th].






\end{thebibliography}
\end{document}